\documentclass[acmtog]{acmart}
\acmSubmissionID{}

\usepackage{booktabs} %
\usepackage{subcaption}

\usepackage{tabularx}
\usepackage{paralist}
\usepackage{units}
\usepackage{mathtools}
\usepackage{bm}

\definecolor{darkgreen}{RGB}{1, 128, 1}

\newcommand{\hl}[1]{{#1}} %

\newcommand{\edt}[1]{{#1}}  %

\newcommand{\figref}[1]{Figure~\ref{#1}}

\newcommand{\etal}{\textit{et al}. }

\newcommand{\systemname}{\emph{MagPen}}

\newcommand{\RmagtopenBold}{\mathbf{r_{mp}}}
\newcommand{\RmagtopenBoldt}{\mathbf{\Tilde{r}_{mp}}}
\newcommand{\Rmagtopen}{r_{mp}}

\newcommand{\mpBold}{\mathbf{m_p}}
\newcommand{\mpBoldt}{\mathbf{\Tilde{m}_{p}}}
\newcommand{\mmBold}{\mathbf{m_m}}
\newcommand{\posm}{\mathbf{p_m}}
\newcommand{\posp}{\mathbf{p_p}}
\newcommand{\posst}{\mathbf{s}(\theta)}
\newcommand{\ed}{\mathbf{e_d}}
\newcommand{\et}{\mathbf{e_t}}
\newcommand{\ez}{\mathbf{e_z}}
\newcommand{\Rtheta}{\mathbf{r_{\theta}}}
\newcommand{\Cost}{\mathcal{C}}
\newcommand{\stheta}{\sin{\angt}}
\newcommand{\ctheta}{\cos{\angt}}
\newcommand{\sphi}{\sin{\angp}}
\newcommand{\cphi}{\cos{\angp}}
\newcommand{\angt}{\beta}
\newcommand{\angp}{\gamma}

\DeclarePairedDelimiterX{\norm}[1]{\lVert}{\rVert}{#1}

\setcopyright{rightsretained}

\acmYear{2019}
\copyrightyear{2019}
\usepackage[ruled,vlined]{algorithm2e}
\usepackage{algorithmicx,algpseudocode}
\renewcommand{\comment}[2][.5\linewidth]

\graphicspath{{../Figures/}}
\begin{document}
\title{Dynamic Drawing Guidance via Electromagnetic Haptic Feedback}

\author{Thomas Langerak}
\affiliation{%
  \institution{ETH Z\"urich}
  \department{Department of Computer Science}
  \country{Switzerland}
}

\author{Juan Zarate}
\affiliation{%
  \institution{ETH Z\"urich}
  \department{Department of Computer Science}
    \country{Switzerland}
}

\author{Velko Vechev}
\affiliation{%
  \institution{ETH Z\"urich}
  \department{Department of Computer Science}
    \country{Switzerland}
}

\author{Daniele Panozzo}
\affiliation{%
  \institution{New York University}
  \department{Courant Institute of Mathematical Sciences}
    \country{USA}
}

\author{Otmar Hilliges}
\affiliation{%
  \institution{ETH Z\"urich}
  \department{Department of Computer Science}
    \country{Switzerland}
}
\renewcommand{\shortauthors}{T. Langerak et al.}

\begin{teaserfigure}
    \includegraphics[width=\textwidth]{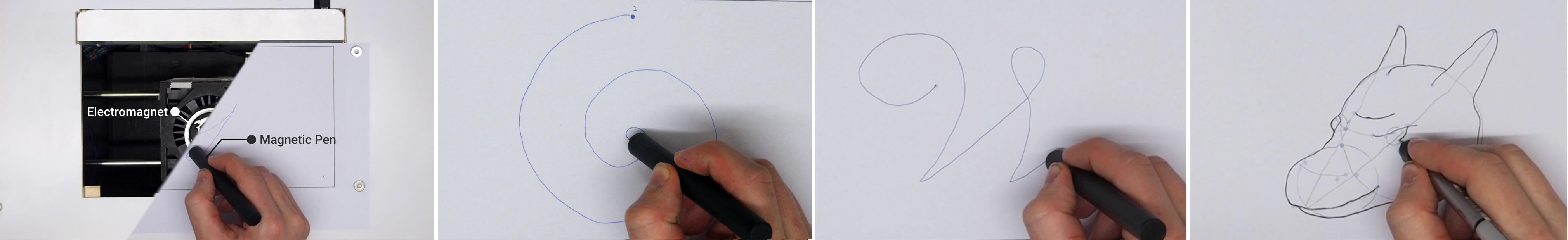}
    \captionof{figure}{Left-to-right: an electromagnet moving on a bi-axial linear stage underneath a high-speed pressure sensitive tablet delivers dynamically adjustable haptic feedback to a minimally instrumented pen. This allows for in-situ haptic feedback aiding users in drawing primitive shapes, or in writing characters, and gives guidance in more complex sketching tasks. We experimentally show that our approach increases accuracy and may help in skill acquisition.
    }\label{fig:teaser}
\end{teaserfigure}

\begin{abstract}
We propose a system to deliver dynamic guidance in drawing, sketching and handwriting tasks via an electromagnet moving underneath a high \hl{refresh rate} pressure sensitive tablet. The system allows the user to move the pen at their own pace and style and does not take away control. 
The system \hl{continously and} iteratively measures the pen motion and adjusts magnet position and power according to the user input in real-time via a receding horizon optimal control formulation. The optimization is based on a novel approximate electromagnet model that is fast enough for use in real-time methods, yet provides very good fit to experimental data. Using a \hl{closed-loop} time-free approach allows for error-correcting behavior, gently pulling the user back to the desired trajectory rather than pushing or pulling the pen to a continuously advancing setpoint.
Our experimental results show that the system can control the pen position with a very low dispersion of \unit[2.8]{mm} ($\pm$\unit[0.8]{mm}). An initial user study indicates that it significantly increases accuracy of users drawing a variety of shapes and that this improvement increases with complexity of the shape.

\end{abstract}


\maketitle

\section{Introduction}
Sketches and handwritten text have been a primary form of communication for centuries. Given their importance for society at large and the arts and design in particular, it is not surprising that a large number of digital pen interfaces exist that aim to combine the expressiveness and flexibility of pens and pencils with the advantages of digital representations via standalone tablets (e.g., Apple iPad, Microsoft Surface) or specialized digitizers (e.g., Wacom). 
Several sketch \cite{xing2015autocomplete, simo2016learning, limpaecher2013real, su2014ez} and digital ink \cite{Aksan:2018:DeepWriting,zitnick2013handwriting} beautification approaches exist. However, these typically improve results a-posteriori and hence do not provide real-time haptic feedback. Comparatively little attention has been devoted to improving sketching in-situ.

\begin{figure}[t]
    \centering
    \includegraphics[width=\columnwidth]{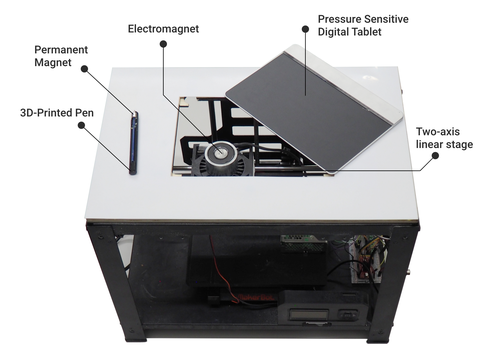}
    \caption{Hardware overview.}
    \label{fig:hardware}
\end{figure}

We propose a haptic feedback system \hl{(Fig. \ref{fig:hardware})} to provide variable strength guidance onto the tip of a minimally instrumented ballpoint pen. \edt{The feedback is delivered via an electromagnet moving on a bi-axial linear stage below a touch and pressure sensitive digital tablet, governed by a closed-loop optimal control algorithm} (see \figref{fig:teaser}). Our approach allows users to perceive different levels of feedback while drawing on the tablet. \edt{Importantly, the variable strength of force feedback is crucial in allowing user autonomy, while maintaining the ability to provide guidance. We show several use casses where this is desirable such as stylization of drawings.}

Previous approaches to pen-based haptic feedback rely on permanent Neodymium magnets (e.g., \cite{yamaoka2013depend}). However, due to the steep increase in magnetic force as the pen approaches the magnet, this fully controls the pen, removing agency of the user. In contrast we propose a novel optimization scheme, inspired by model predictive contouring control (MPCC) \cite{lam2010model}, to position and regulate an electromagnet such that it provides dynamically adjustable in-plane magnetic forces to the pen tip. 
This is challenging due to: 
\begin{inparaenum}[i)]
  \item the quadratic increase in magnetic force as a function of magnet-pen distance \edt{(i.e. a small spatial offset can significantly increase the perceived force)},
  \item the fast pen motion compared to the speed of the linear stage, and
  \item the hard to predict behavior of the user.
\end{inparaenum}
These challenges enforce a very tight computational budget from pen motion to magnet actuation, \edt{that is, we need an efficient numerical solution to design a system with a low overall latency}.

\edt{To this end} we propose an approximate, yet accurate, model of the electromagnetic force field that can be evaluated analytically and is hence suitable for online control. This model is combined with an MPCC-like optimization scheme that iteratively predicts the pen motion and adjusts magnet position and power accordingly. In contrast to simpler control schemes such as MPC \cite{Faulwasser:2009} or PID \cite{aastrom1995pid} control, our approach does not require a timed reference and hence allows users to draw at their desired speed. Furthermore, the optimization scheme allows for error-correcting force feedback, gently pulling the user back to the desired trajectory rather than pushing or pulling the pen to a continuously advancing setpoint on the trajectory. 

To assess the proposed feedback mechanism and control algorithm, we performed an user-study with 12 participants. In our experiment we focus on drawing primitives such as circles, spirals, and more complex curves. Our results indicate that the haptic feedback increases both accuracy  and precision quantitatively, reduces drift (i.e., error over time), and that users qualitatively appreciate the system. We furthermore, illustrate a number of potential use-cases for the proposed method such as a \edt{support} tool for hand drawn sketching and writing, or as in-situ feedback tool for inking, under- and overpainting.

To foster adoption of our technique, and to encourage industrial miniaturization of our device, we will release our reference software implementation and hardware blueprints.

\section{Related Work}

Haptics in touch and pen-based interaction is widely researched. We provide a brief overview of the most closely related work, covering sketching, pen-based interfaces, and magnetic actuation.

\subsubsection*{Sketching Guidance and Beautification}
A number of systems have been proposed to provide varying levels of support during sketching, ranging from basic visual guidance to automatic stroke-refinement and vectorization.
ShadowDraw guides users by continually updating a shadow underneath their sketch \cite{lee2011shadowdraw}. The shadow is generated by first edge-extracting images from a large database, and then analyzing and combining the closest matches to the current sketch. \citet{limpaecher2013real} improve sketches by replacing strokes automatically by strokes extracted from drawings stemming from a large database. The authors report that this reduces undo and other correction operations. 
When tracing images using a technique called \emph{underpainting}, \citet{su2014ez} leverage the underlying image to perform \emph{stroke-refinement} in real-time to better fit the gradient information underneath. They perform optimization both locally on a single stroke, and semi and fully globally by considering the interaction between multiple strokes when no close matches are found. %
\citet{xing2015autocomplete} perform global similarity analysis across multiple frames to assist during animation tasks and to suggest strokes that beautify single frames based on past frames.

\emph{Inking} refers to the process of simplifying sketches and line drawings for improved clarity. Several approaches leverage vectorization \cite{hilaire2006robust, favreau2016fidelity}, stroke aggregation \cite{liu2018strokeaggregator}, and more recently, convolutional neural networks \cite{simo2016learning, simo2018real} and generative adversarial networks \cite{simo2018mastering} for this task. Our system is orthogonal and complementary to these approaches, providing \emph{active physical support} during sketching. We argue that our system is a prime candidate to be combined with such existing systems to re-enforce the intended or corrected stroke of the user, so that the user not only attains a beautified stroke, but also feels the dynamics of creating it.

\subsubsection*{Haptic Feedback for Pen-Based Interfaces}
Haptic feedback has been studied in the domain of stylus and tablet interfaces. \citet{poupyrev2004haptic} used a display instrumented with piezoelectric actuators to provide tactile feedback and report that users were significantly more precise in continuous interaction tasks such as sketching. \citet{lee2004haptic} propose an active stylus containing embedded actuators to provide personalized haptic feedback in collaborative settings. \citet{withana2010impact} embed a linear actuator inside the pen to convey a sense of depth of the screen.  Digital rubbing employs a similar system, however, for the purpose of tracing over digital images on real paper \cite{kim2008digital}. The technique requires a solenoid to activate when the pen passes over a part of the underlying sketch. Because of the delay between measurement and actuation, a simple dead-reckoning movement predictor was applied resulting in significantly better alignment between digital and analog tracings. \hl{Some combined haptic feedback with visual guidance \cite{portillo2005haptic}. This also promoted research into evaluating visuohaptic systems \cite{yang2008validating}.}

Haptic interfaces have also been used to increase the accessibility of GUIs by providing Braille-like feedback on the side of the pen during interaction tasks \cite{kyung2008haptic, kyung2009precise}. 
RealPen \cite{cho2016realpen} increases the realism of sliding over surfaces by mimicking audio-tactile properties of materials like paper.
Pen based input was also combined with a 3-DOF device (Phantom Omni) as a handwriting aid for stroke patients \cite{mullins2005haptic}.  However, since the user writes in air, the realism of writing on paper is lost. \hl{Similarly, there has been some work on tools \cite{zoran2013freed, zoran2014wise, peng2015d}.}

While the benefits of haptic feedback in tablet and pen-based interfaces have been demonstrated \cite{cho2016realpen, poupyrev2004haptic, kyung2009precise}, we argue that a tight adaptive control loop is necessary to improve perception, accuracy and utility of such approaches (cf. \cite{kim2008digital}). We therefore propose a MPCC-like optimization approach to provide programmable real-time haptic feedback to the user with the aim of increasing accuracy and aesthetic output, without removing user agency.

\subsubsection*{Magnetic Actuation}
Providing magnetically-driven haptic feedback on tabletops is desirable as the force can be exerted through the surface without affecting the display. A common approach is using arrays of controllable electromagnets, combined with permanent magnets embedded in objects on the surface \cite{pangaro2002actuated, yoshida2006proactive, weiss2011fingerflux}. Fingerflux by Weiss \etal~\shortcite{weiss2011fingerflux} provide near-surface haptic feedback before the finger touches the screen to guide users to appropriate screen locations. \citeauthor{pangaro2002actuated} \cite{pangaro2002actuated} model the force-field of each electromagnet and combine these using standard aliasing techniques, allowing directed movement of multiple objects on the surface. However, moving objects smoothly across the surface is problematic due to the low resolution of the grid, and the interaction of forces between multiple electromagnets. 
Furthermore magnets are modeled using a simplification of the Gilbert model, considering attraction between single point poles.
In sketching and writing tasks, accuracy is of the utmost importance, and thus, we employ a printer-like setup, allowing for smooth movement across a 2D plane and contribute a more accurate EM model based on oriented dipoles.

dePENd by \citet{yamaoka2013depend} is perhaps the closest prior research to our system. They move a permanent Neodymium magnet on a two-axis setup to control the pen of a user. They make use of the ferromagnetic feature of the metal tip of regular ballpoint pens to attract it. Our work differs significantly in the type of magnet, its mathematical model and the control strategy\edt{, resulting in a different user experience and system capabilities}. The neodymium magnet in their work ``forces'', rather than guides, the user to follow a predefined path. In \citet{yamaoka2013depend} \edt{the pen is not tracked which results necessitates an open-loop control strategy and little to no user autonomy}. The user is allowed to deviate from the stroke only slightly by lifting or moving the pen. However, this input does not alter the behavior of the magnet. In comparison, our system allows the user to move at their own pace through a drawing, and reacts in real-time to user input by altering the position and strength of the magnet to compensate for user input, thus providing \edt{an haptic guidance of the location of the reference path, without prescribing a set velocity along the guidance path and not entirely controlling the user motion}. 

\subsubsection*{Online Path Following} 
Optimal reference following given real world influences is studied in depth in the control theory literature. Methods like MPC \cite{Faulwasser:2009} optimize the reference path and the actuator inputs simultaneously based on the system state. MPC is wildly applied to many robotics (e.g., \cite{Mueller2013}) and graphics applications \cite{dasilva:2008:mpc}. However, \cite{AGUIAR2008} show that the tracking error for following timed-trajectories can be larger than if following a geometric path only. To address this issue Model Predictive Contouring Control (MPCC) \cite{lam2013model} has been proposed to follow a time-free reference, optimizing system control inputs for time-optimal progress. MPCC approaches have been successfully applied in industrial contouring \cite{lam2013model} and RC racing \cite{Liniger2014} and in drone cinematography \cite{Naegeli:2017:MultiDroneCine}. We also pose our optimization problem in the MPCC framework. However, to the best of our knowledge we are the first to do so in the context of haptic feedback systems where one has to consider both a controllable (i.e., the linear stage) and non-controllable (i.e., the user) system. Furthermore, we contribute a fast approximate electromagnet (EM) model, that gives good experimental fit, for use in iterative optimization schemes.

\section{Overview}
The goal of our work is to provide an integrated software-hardware solution that can provide dynamically adjustable force feedback to a regular ballpoint or digital pen. Importantly, we argue that user agency is crucial. \edt{This is defined as the user staying in control of their actions and the system playing only a supportive role}. \edt{Hence the system should never control the user's motion but only provide feedback. In particular, the user may maintain personal speed and style of drawing}.%

\SetKwProg{Fn}{Function}{}{}
\begin{algorithm}[t]
\SetAlgoLined
\DontPrintSemicolon
 \Fn{MPCC($\mathbf{x0}, \mathbf{w},\posp, \mathbf{parameters}$)}{
 $[\mathcal{C}_l, \mathcal{C}_c, \mathcal{C}_\theta, \mathcal{C}_{\dot{\theta}}]\leftarrow$compute lag and contour error \Comment{Sec \ref{sc:lag_cont_err}}\;
 $[\mathcal{C}_f, \mathcal{C}_d, \mathcal{C}_\alpha]\leftarrow$compute force error \Comment{Sec \ref{sc:hardware} \& Sec \ref{sc:em_costs}}\;
 $J_k \leftarrow $ sum($\mathcal{C}_l, \mathcal{C}_c, \mathcal{C}_\theta, \mathcal{C}_{\dot{\theta}},\mathcal{C}_f, \mathcal{C}_d, \mathcal{C}_\alpha$)\;
 $[\mathbf{x}, \mathbf{u}]\leftarrow$minimize($J_k$) \Comment{Sec \ref{sc:Jk}}\;
 }
  \Return $[\mathbf{x}, \mathbf{u}]$\;
  \;
 
 $\mathbf{x_0}$, $\mathbf{w} \leftarrow \ $initialize\;
\While{drawing not finished}{
  $\posp \leftarrow$ Measure pen position\;
  ${\posp}_{,k} \leftarrow$ $KalmanFilter(\posp) $\Comment{Sec \ref{sc:dyn_model}}\;
  $\mathbf{x_0} \leftarrow$ update system states, from sensor data\;
  $\mathbf{parameters} \leftarrow$ update MPCC parameters \;
  $[\mathbf{x_{t=1..n}},\mathbf{u_{t=1...n}}] \leftarrow$ $MPCC(\mathbf{x_0},\mathbf{w},\mathbf{p},\mathbf{params})$ \Comment{Sec \ref{sc:control}}\;
  $\mathbf{x_0}\leftarrow \mathbf{x_1}$\;
  apply($\mathbf{u_0})$
 }
 \caption{Closed Loop Haptic Feedback Control.} \label{alg:optimization-overview}
\end{algorithm}

We propose the \systemname{} system, shown in \figref{fig:hardware}. \systemname{} consists of a high speed pressure sensitive tablet under which an electromagnet is moved on a bi-axial linear stage. The electromagnet delivers adjustable force feedback to a mostly unmodified pen -- we only attach a small permanent magnet to an otherwise passive ballpoint pen.
Our proposed optimization scheme (summarized in Alg. \ref{alg:optimization-overview}) allows us to adjust the magnet position and strength such that it gently pulls the pen tip towards a desired stroke, while allowing users to draw at their desired speed and without fully taking over control. 
Here we assume that the user traces a known trajectory. This already enables a number of applications such as practice support in writing of characters or sketching, or as active support in inking or underpainting. We leave integration with a full predictive model (e.g., \cite{Aksan:2018:DeepWriting}) for future work.

At each time step, we minimize a cost functional over a receding time horizon in order to find optimized values for system states $\mathbf{x}$ and inputs $\mathbf{u}$. The cost function, here given as high-level abstraction, 
\begin{equation}
    \underset{\mathbf{x},\mathbf{u}}{\text{minimize}} \sum \underbrace{\mathcal{C}_{
    \text{path}}(\mathbf{x},\mathbf{u})}_{\text{Eq. \ref{eq:errL} \& \ref{eq:errC}}} + \underbrace{\mathcal{C}_{\text{progress}}(\mathbf{x},\mathbf{u})}_{\text{Eq. \ref{eq:err_theta} \& \ref{eq:err_theta_dot}}} + \underbrace{\mathcal{C}_{\text{force}}(\mathbf{x},\mathbf{u})}_{\text{Eq. \ref{eq:err_F}, \ref{eq:err_d} \& \ref{eq:err_alpha} }}  \label{eq:min1},
\end{equation}
serves three main purposes:  
1) ensuring that the user stays close to the desired path, 2) makes progress along it and 3) the user perceives haptic feedback of dynamically adjustable force.

We first introduce the hardware platform in Sec. \ref{sc:hardware}, including a novel approximate model of the electromagnet that can be evaluated in closed form and hence is usable within an iterative optimization method. We then introduce our closed-loop control formulation in Sec. \ref{sc:control} that leverages the force behavior model to compute control inputs for the bi-axial linear stage and for the electromagnet.

\section{Hardware}\label{sc:hardware}

In designing the \systemname{} hardware (Figure \ref{fig:hardware}) we balanced several important aspects. First, the magnetic force must be controlled in a fine-grained manner to allow for user agency, essentially ruling out the use of permanent Neodynium magnets. Second, very small pen displacement can lead to undesired snapping of the pen tip unless the magnet is adjusted almost instantaneously, due to the quadratic rise in magnet attraction as a function of the pen magnet distance \edt{(see accompanying video for an illustration)}. This requires low-latency sensing hardware and control software. We chose a display-less pressure-sensitive tablet since it provides a high framerate. Finally, modeling electromagnetic fields is highly involved and typically requires FEM simulation which would not lend itself for use in iterative optimization schemes. %
This problem is intensified if EM fields overlap and interact, such as in a grid of electromagnets (cf. \cite{pangaro2002actuated}). Therefore we move a single electromagnet on a bi-axial linear stage and contribute an approximate model of the EM field. We leave miniaturization into a tablet form-factor for future work but argue that the current form-factor already is an interesting solution for tracing tables or digital whiteboards.       

\subsection{Sensing and Actuation}\label{sc:sensing_actuation}
One of our design goals is to provide users with an as unencumbered as possible experience, staying close to the experience of sketching on pen and paper but allowing for in-situ haptic force feedback. 

\begin{figure}[tb]
    \centering
    \includegraphics[width=\columnwidth]{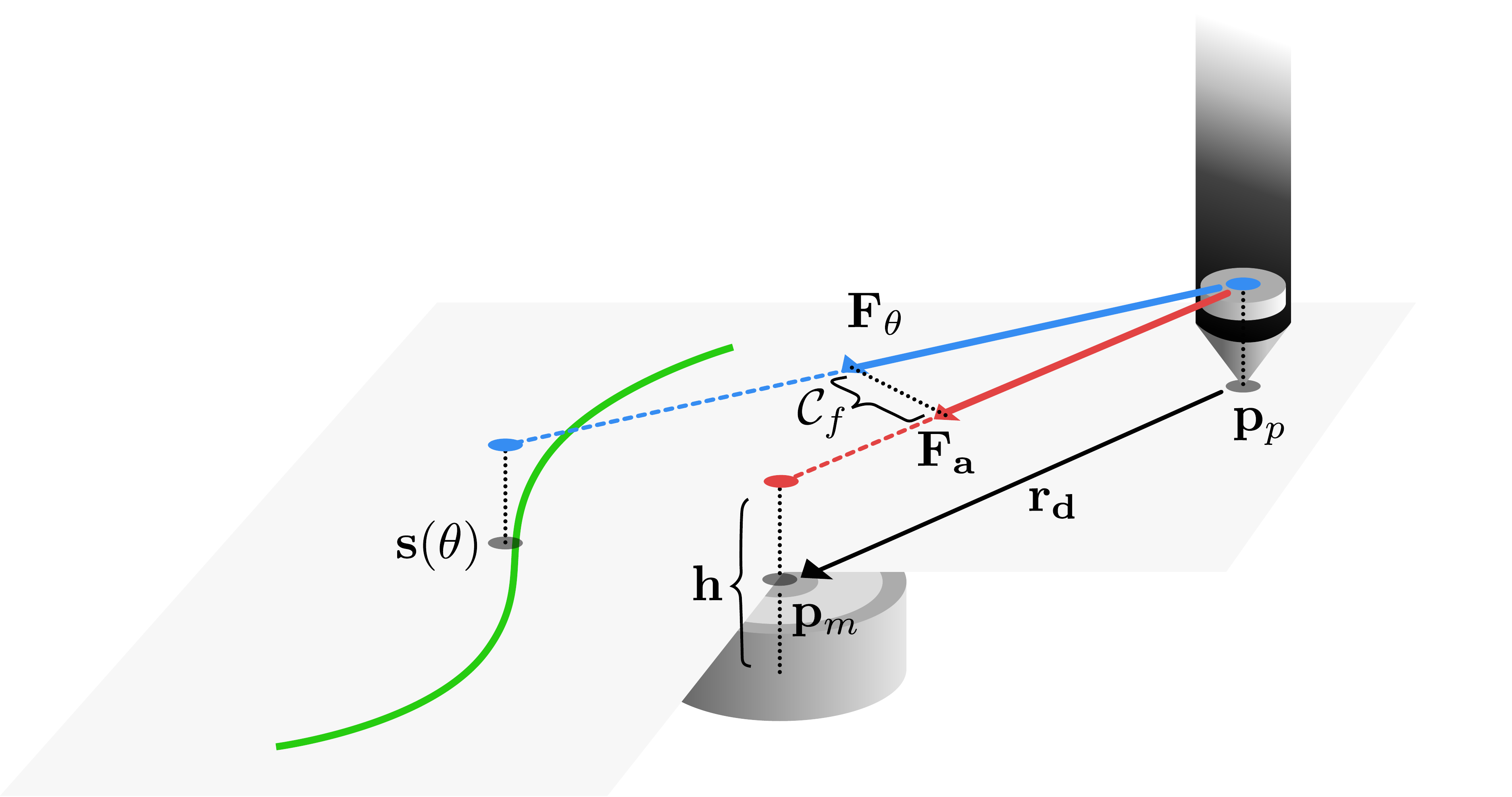} %
    \caption{Illustration of actuation force $\mathbf{F_a}$, desired force $\mathbf{F_{\theta}}$, and the force cost-term $\Cost_f$ associate with the difference between those two forces. 
    }
    \label{fig:em_model}
\end{figure}

In \systemname{} users draw on normal paper with a minimally modified ballpoint pen. The strokes are recorded by a Sensel Morph (\url{https://sensel.com/}) pressure sensitive touch pad. We chose the Sensel board for it's high spatial resolution (\unit[6502]{DPI}), high speed (\unit[500]{Hz}) and low latency (\unit[2]{ms}). Since the board is designed to be used in combination with different overlays, the sketching surface does not interfere with the input recognition. Users draw with a 3D printed ballpoint pen with a permanent ring magnet mounted in the shaft (see \figref{fig:em_model}).

To deliver haptic feedback to the pen, we move a programmatically controlled electromagnet on a bi-axial linear stage directly underneath the input sensor. Our implementation leverages the bi-axial system and motors from a Makerbot Replicator 2X 3D printer. We replaced the printer head with an electromagnet (Intertec ITS-MS-5030-12VDC, diameter = \unit[5]{cm}, height=\unit[3]{cm}, \unit[12]{V} \unit[11]{W}) mounted on a heatsink and cooled by a fan. The choice of the electromagnet is non-trivial: we need a strong enough electromagnet, fitting in our hardware setup which limits size and heat dissipation, while not being too heavy for the linear stage.
We employ FEM analysis to make an informed choice with regard to the electromagnetic characteristics. Our system allows up to \unit[488]{mN} of lateral force (at 11 W), with only \unit[2.8]{mm} of point dispersion (see Sec \ref{sc:results-pos-dispersion})

The stepper motors are controlled via a Sparkfun EasyDriver motor-shield and an Arduino Uno, allowing for micro stepping (a full step is \unit[.2]{mm}, we use quarter stepping) of the motors which increases the smoothness and accuracy of the magnet motion. The electromagnet is controlled via pulse-width modulation (PWM) and an H-Bridge. While allowing for both repulsive and attractive forces, we only use attraction for simplicity.  

\subsection{Electromagnetic Force Model}\label{sc:em_model}
With the hardware platform in place, we require a mathematical model of the electromagnet to compute the attraction force between magnet and pen as a function of their distance. Here we face a significant challenge: modeling the full EM field, \hl{with the help of a FEM analysis}, is not computational feasible in real-time; yet we require a physically accurate description of the force behavior at every iteration of our optimization procedure. One of our main contributions is an approximate model that is efficient to evaluate and provides a very good fit to empirical data. Here we briefly discuss its derivation and we refer to Appendix \ref{sc:ap.dipole.eq} and \ref{ap:sc.EM.model} for additional details and for a full validation of the model. 

\edt{The electromagnet core is made of a non-linear ferromagnetic material. Thus, }calculating the physical correct force behavior would involve pre-computation of a volumetric map of the EM field $\mathbf{B_m}$ via FEM simulation for all levels of the electrical current. The actuation force on the pen $\mathbf{F_p}$ is then given by integrating over the volume of the permanent magnet in the pen:
\begin{equation}
    \mathbf{F_p} = \iiint \nabla \left( \mathbf{M_p} \cdot  \mathbf{B_m}(\cdot)\right) dxdydz , \label{eq:gradB2}
\end{equation}
\noindent where $\mathbf{M_p}$ is the magnetization of pen magnet and $\mathbf{B_m(\cdot)}$ is the EM field evaluated at the pen position. Intuitively this can be read as the force response corresponding to the gradient of the EM field evaluated within the volume of the pen magnet. Since this is too costly to evaluate in real-time, we propose an approximate model that is consistent with the underlying physical phenomena. 

We make the following two assumptions in our derivations:
\begin{inparaenum}[1)]
    \item the electromagnet and the permanent magnet can be approximated as dipoles (i.e., oriented point magnets), and
    \item \edt{for the smaller dipole (the permanent magnet in the pen) the out-of-plane vector component is much larger than the in-plane counterpart. This allows us to use only the vertical component in the calculation of the force.} 
\end{inparaenum}
Note that these simplifications lead only to a small approximation error. %
Compared to an angle dependent formulation (see Appendix \ref{ap:sc.EM.model} Eq. \ref{eq:ap.Fd}), a tilt of up to $\angt = 30^{\circ}$ leads to a max error in our model (Eq. \ref{eq:Fa}) equivalent to shifting the distance $d$ by $\pm 3$ [mm] (see Figure \ref{fig:approx_error}). This uncertainty in $d$ is comparable with the in-plane positioning error of our overall system (Sec. \ref{sc:results-pos-dispersion}).
Furthermore, we found that users can not perceive a difference in strength when tilting the pen in-place.
\begin{figure}[!t]
    \centering
    \includegraphics[width=\columnwidth]{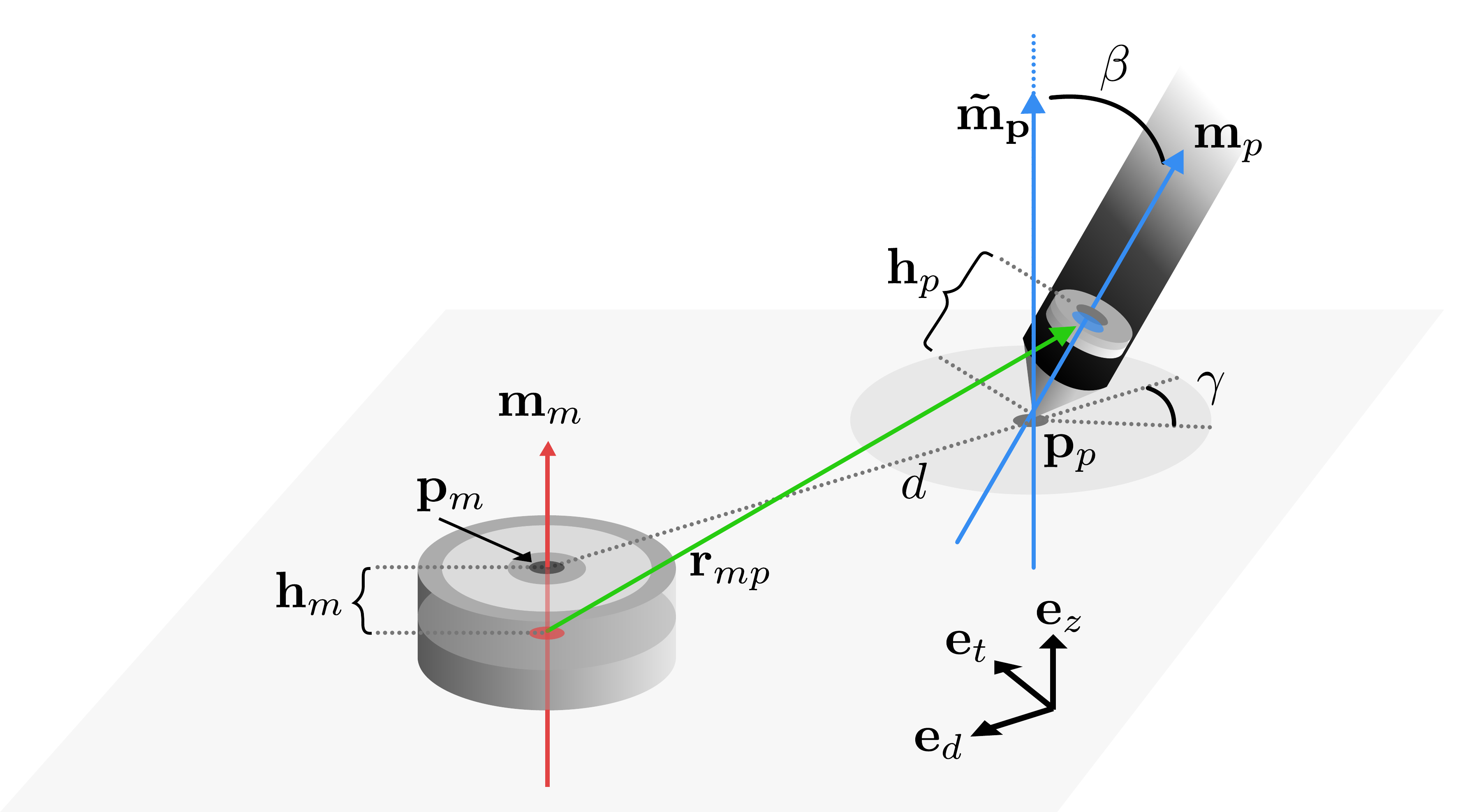} \\
    \caption{Illustration of the model to compute the force $\mathbf{F_p}$ on dipole $\mpBold$ due to dipole $\mmBold$ (see Eq. \ref{eq:F21-dip}). In Eq. \ref{eq:Fa} we describe the actuation force under the upright pen approximation ($\angt = 0$), while in Appendix \ref{ap:sc.EM.model} we report a more general model for small tilt angles ($\angt \leq 30^\circ$).}
    \label{fig:dipole_dipole}
\end{figure}

The first approximation allows us to follow the formulation in \cite{yung1998analytic} and to model the force $\mathbf{F_p}$ exerted by the electromagnet dipole $\mmBold$ onto the pen dipole $\mpBold$ (see \figref{fig:dipole_dipole}) as: 
\begin{multline}
   \mathbf{F_p} = {\dfrac  {3\mu _{0}}{4\pi \Rmagtopen^{5}}}
   \left [ \left(\langle\mpBold,\RmagtopenBold\rangle \right) \mmBold + 
   \left(\langle\mmBold,\RmagtopenBold\rangle\right) \mpBold \right . +
   \\
   \left(\langle\mpBold,\mmBold\rangle\right) \RmagtopenBold - 
    \left . {\dfrac{5\left(\langle\mpBold,\RmagtopenBold\rangle\right)
    \left(\langle\mmBold,\RmagtopenBold\rangle\right)}{\Rmagtopen^{2}}} \RmagtopenBold \right ] \ , \label{eq:F21-dip}
\end{multline}
where $\mu_0$ is a constant (see Table \ref{tab:em_model}) and  $\RmagtopenBold$ is the 3D vector between \hl{the centers of the electromagnet and pen dipoles.}
\hl{The electromagnet can be seen as magnetic dipole with variable strength, controlled by the dimensionless scalar $\alpha \in \left[0,1\right]$.}

\edt{The three vectors needed to compute} Eq. \ref{eq:F21-dip} \hl{can be expressed in the coordinate system of} Figure \ref{fig:dipole_dipole} as 
\begin{eqnarray}
 \mmBold &=& \alpha \ m_m \ \ez \label{eq:m1}, \\
 \mpBold &=& - (m_p \stheta \cphi) \ \ed \nonumber \\
          && + (m_p \stheta \sphi) \ \et \nonumber \\
          && + (m_p \ctheta) \ \ez \label{eq:mp}, \\
 \RmagtopenBold &=& - (d+h_p \stheta \cphi) \ \ed \nonumber \\
          && + (h_p \stheta \sphi) \ \et \nonumber \\
          && + (h - (1-\ctheta) h_p) \ \ez. \label{eq:r21a} 
\end{eqnarray}

\begin{figure}[!t]
    \centering
    \includegraphics[width=\columnwidth]{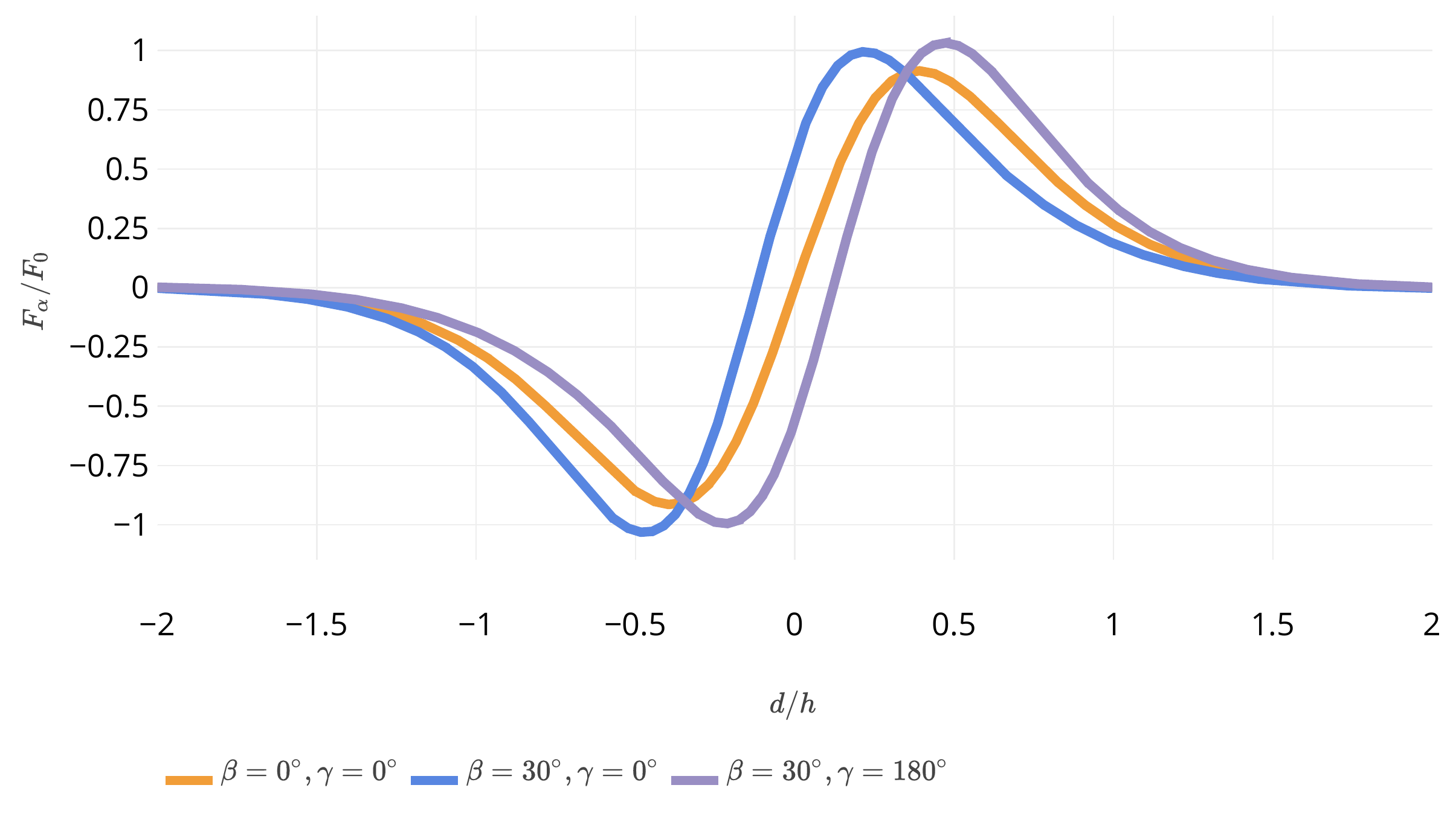} \\
    \caption{In-plane magnetic force as function of position. The horizontal displacement between curves (each denoting a different pen-tilt) is the approximation error induced by the upright pen (purple) assumption (angles defined in \protect\figref{fig:dipole_dipole}).}
    \label{fig:approx_error}
\end{figure}

\edt{While providing us with an analytic and differentiable expression, this leads to an equation for the actuation force $\mathbf{F_p}$ that depends on the tilt of the pen through the angles $\angt$ and $\angp$. With our hardware these data are non-trivial to attain.} %
\edt{We now leverage our second assumption by rewriting Eq. }\ref{eq:F21-dip} \edt{ with an equivalent pen dipole $\mpBoldt$, obtained by applying the small tilting angle approximation ($\ctheta \simeq 1$ and $\stheta \simeq 0$) to Eq.} \ref{eq:mp}, %
\begin{equation}
    \mpBoldt = m_p \ \mathbf{e_z} \label{eq:m1t} \ ,
\end{equation} 
\noindent \hl{where the scalar magnetization is given by $m_p = B_r V/\mu_0$. $B_r$ is the residual magnetization of the permanent magnet and $V$ its volume and  $\mathbf{e_z}$ is the $z$-unit vector. This approximation removes the requirement for tracking the pen tilt. More importantly it drastically simplifies the force equation since both dipoles now only have a $z$ component and thus the actuation only depends on the distance $d$ between pen and magnet (not on $\angt$ nor $\angp$).}
\edt{This provides a simplified version of the 3D distance vector,} 
\begin{equation}
    \RmagtopenBoldt = - d  \ \ed + h  \ \mathbf{e_z} , \label{eq:r21b} 
\end{equation}
\noindent \edt{where the vertical distance, $h = h_m + h_p$, is constant. Note that the in-plane distance $d = \norm{\posp - \posm}$ is one of the variables we seek to control, given the projections of the pen position ($\posm$) and the electromagnet position ($\posp$) onto the paper plane.} %

\edt{The electromagnet dipole ($\mmBold$) is mounted in a fixed upright position. Therefore it can be expressed via Eq.} \ref{eq:m1}, \edt{without incurring any approximation error.}
The magnetization value of the \hl{full-strength} dipole $m_m$, which approximates the electromagnet, can be derived experimentally. For this purpose we scan the magnetic field generated \hl{by the electromagnet, setting $\alpha = 1$ and using} a hall sensor and adjust the parameters of EM field equation to give a good fit. The full procedure and derivations can be found in Appendix \ref{sc:ap.dipole.eq}.
Table \ref{tab:em_model} \edt{reports the values of $m_m$, $m_p$ and $h$ that were used in our experiments.}

\begin{table}[t]
  \caption{List of electromagnet model and hardware parameters}
  \label{tab:em_model}
  \begin{tabular}{lll}
    \toprule
    Name&Value&Description\\
    \midrule
    $\mu_0$ & $4\pi \ 10^{-7}$ [H/m] & Vacuum permeability \\
    $B_r$ & 1.3 [T] & Pen magnet type (NIB N42) \\
    $V$ & 0.66 [cm$^3$] & Pen magnet volume \\
    $m_p$ & 0.683 [A m$^2$]& pen dipole ($=B_r V / \mu_0$)\\
    $m_m$ & 1.286 [A m$^2$]& electromagnet dipole (see App. \ref{sc:ap.dipole.eq})\\
    $h$ & 2.71 [cm] & z-distance $\mmBold$ to $\mpBold$ (see App. \ref{sc:ap.dipole.eq})\\
    $h_p$ & 1.40 [cm] & height pen-tip to magnet (see Fig. \ref{fig:dipole_dipole})\\
    $F_0$ & 0.488 [N] & force factor in Eq. \ref{eq:Fa}), $F_0 = \frac{3 \mu_0 m_p m_m}{4 \pi h^4}$ \\
  \bottomrule
\end{tabular}
\end{table}

The total force acting on the pen (Eq. \ref{eq:F21-dip}) can now be decomposed into the in-plane and vertical force components: 
\begin{equation}
    \mathbf{F_p} = F_a \ \mathbf{e_d} + F_z  \ \mathbf{e_z} \ . \label{eq:Fp_decomp}
\end{equation}

\noindent Here $\mathbf{F_a} = F_a \ \mathbf{e_d}$ represents the quantity we seek to control. 
By substituting the results form Eq. \ref{eq:m1}, \ref{eq:m1t} and \ref{eq:r21b} into Eq. \ref{eq:F21-dip} and maintaining only the in-plane contributions ($\mathbf{e_d}$ direction), we obtain the expression for the actuation force as function of pen-magnet separation:
\begin{equation}
    \mathbf{F_a} = \alpha \ F_0 \ \left( \frac{d \left(4 - \frac{d^2}{h^2}\right)}{h \left(1 + \frac{d^2}{h^2}\right)^\frac{7}{2}} \right)  \ \mathbf{e_d} , \label{eq:Fa}
\end{equation}
where $F_0$ is a constant force parameter given by the expression,
\begin{equation}
 F_0 = \frac{3 \ \mu_0 \ m_p \ m_m}{4 \ \pi \ h^4} \ . \label{eq:F0}
\end{equation}

\edt{Fig. }\ref{fig:approx_error} \edt{illustrates how the dimensionless ratio within parentheses in Eq.} \ref{eq:Fa} governs the force strength as function of distance $d=\norm{\mathbf{r_d}}$. 
The actuation force $F_a$ is zero if the two magnets are aligned with one another ($d=0$), it has a maximum $F_a^{max} = 0.9 F_0$ at $d=0.39h$, and we can assume there is no more attraction for distances $d>2h$. In Table \ref{tab:em_model} \hl{we report the value of $F_0$ we obtained for our setup.}

We note that the vertical force component $F_z$ from Eq. \ref{eq:Fp_decomp} pulls the pen downwards. However, during our experiments there was no significant change in ink thickness or perceived friction when comparing the drawings with and without electromagnet (i.e., with or without $F_z$). For this reason we do not actively optimize for $F_z$ in our optimization. Finally, we provide an angle dependent formulation of our model in Appendix \ref{ap:sc.EM.model} for future use in cases where the pen angle is tracked.

\section{Closed-Loop Control}\label{sc:control}

With the force behavior model from Eq. \ref{eq:Fa} we can now derive our control strategy, such that a force $F_a$ of desired strength is exerted onto the pen, in order to keep the user close to the desired path. 
The known path $\mathbf{s}$ is parametrized by $\theta \in[0,L]$, where $L$ is the length of the path. Note that we do not want to prescribe how fast the user draws the shape and hence for each given pen position $\posp$ we first need to establish the closest position on the path parametrized by $\mathbf{s}(\theta)$. Furthermore, we seek to find optimized values for the electromagnet intensity $\alpha$ and the in-plane electromagnet position $\mathbf{p}_{m}$. We phrase this problem in the MPCC framework \cite{lam2013model}. Solving the error functional given in Eq. (\ref{eq:J_k}) at each timestep, yields optimized values for system states $\mathbf{x}$ and inputs $\mathbf{u}$.  

\begin{figure}[!t]
    \centering
    \includegraphics[width=\columnwidth]{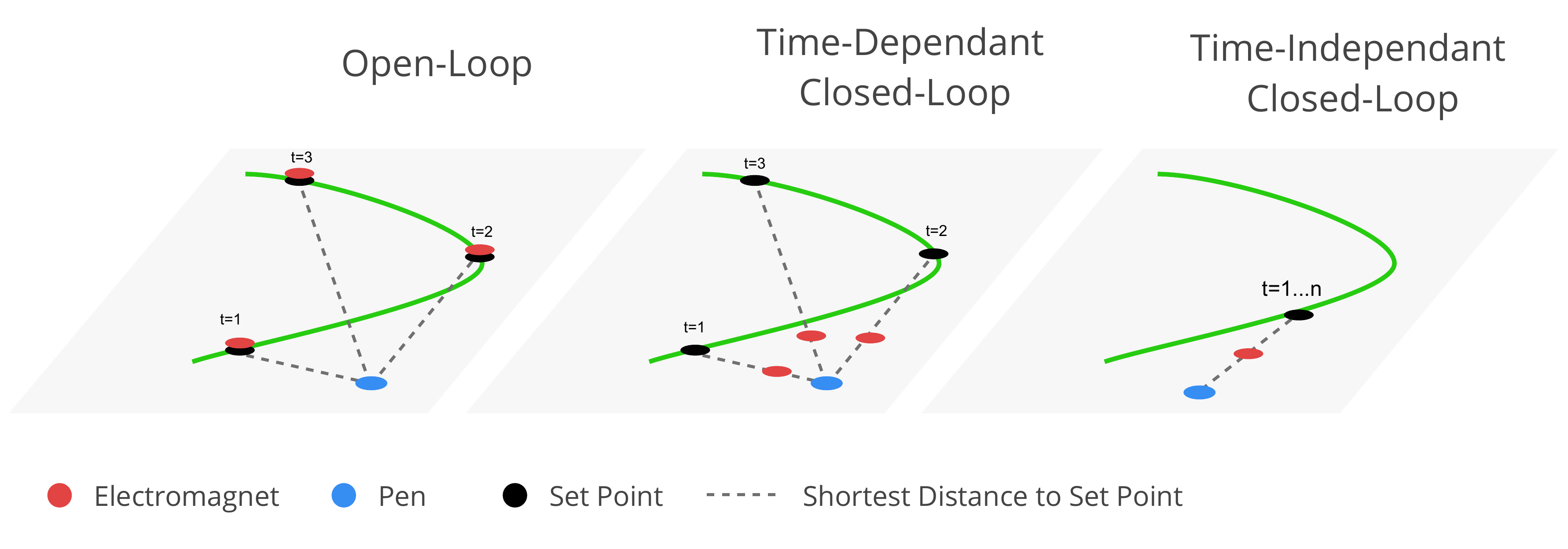}
    \caption{\hl{Overview of different control strategies and their theoretical behavior. For sake of simplicity the user is kept at a constant position for all time-steps. In the case of Open-Loop, the position of the electromagnet is identical to the setpoint. The setpoint is defined per timestep. For MPC the setpoint is still defined per timestep, however the EM is at an optimized position between the setpoint and the pen. MPCC, on the other hand, also optimizes the setpoint based on the pen position (hence, in this stationary case, the setpoint is also stationary). In a time-dependent setting the electromagnet might not be close to the pen (open-loop) or in the wrong direction (time-dependant closed loop). These problems are not present in a time-independent closed loop approach.}}
    \label{fig:control}
\end{figure}

\edt{From a high level perspective, Model Predictive Contour Control is a closed-loop} \emph{time-independent} \edt{control strategy that minimizes a cost-function over a fixed receding horizon. As is commonly done in MPC(C), the system is initialized from measurements at $t=0$. The system state is then propagated over the horizon with the help of the system dynamics (Eq} \ref{eq:model}). \hl{The state vector x contains only variables that are controlled by the algorithm (cf. Eq.} \ref{eq:states}). \hl{Only the first of the optimized inputs ($u_0$) is then applied to the physical system, transitioning the system state to $x_1$, before iteratively repeating the process. This allows for iterative correction of noisy predictions due to modelling errors.}  

\edt{There are several advantages in using a time-independent closed-loop controls strategy, such as MPCC, over open-loop or time-dependent strategies. First, closed-loop control allows the system to react to user-input, whereas open-loop control removes all user agency. Both MPC and MPCC are closed-loop control strategies. The main difference is that MPC tracks a timed reference, prescribing a fixed velocity, whereas MPCC follows a time-free trajectory, which allows the user to progress at the desired speed.} \figref{fig:control} \edt{illustrates the expected behavior for a closed-loop versus a timed and time-free strategy respectively, given that the user decides to slow down or stop moving the pen.
}

\subsection{Dynamics Model}\label{sc:dyn_model}
To control linear stage and electromagnet we require a model $f(\mathbf{x},\mathbf{u})$ describing the system dynamics given its states $\mathbf{x}$ and inputs $\mathbf{u}$.  
\begin{equation}
\label{eq:model}
    \dot{\mathbf{x}} = f(\mathbf{x}, \mathbf{u}) .
\end{equation}

We model the electromagnet with its position $\mathbf{p}_{m}$ and magnet intensity $\alpha$ and include a path progress $\theta$.
\begin{equation}
\label{eq:states}
    \mathbf{x} = [\mathbf{p}_{m},\dot{\mathbf{p}}_{m}, \alpha, \theta] \in \mathbb{R}^6 .
\end{equation}

The inputs to the system consist of the in-plane electromagnet accelerations $\ddot{\mathbf{p}}_{m}$, and velocities $\dot{\alpha}$ and $\dot{\theta}$ for magnet intensity and the spline progress respectively: 
\begin{equation}
    \mathbf{u} = [\ddot{\mathbf{p}}_{m}, \dot{\alpha}, \dot{\theta}]  \in \mathbb{R}^4 .
\end{equation}
Note that we empirically found that magnet accelerations yield smoother motion than using velocities. \edt{The state and inputs in the model are in SI-units. However, the acceleration is converted to stepper motor increments before they are send to the EM controller.}

The system model is given by the non-linear ordinary differential equations using first and second derivatives as inputs:
\begin{equation}
  \ddot{\mathbf{p}}_{m} = v_{m}, \quad \dot{\alpha} = v_{\alpha} \quad \text{and} \quad \dot{\theta} = v_{\theta} ,  
\end{equation}
where $v_{\left(\cdot\right)}$ are the external inputs. The continuous dynamics model $\dot{\mathbf{x}} = f(\mathbf{x}, \mathbf{u})$ is discretized using a standard forward Euler approach: $\mathbf{x}_{t+1} = f(\mathbf{x}_t, \mathbf{u}_t)$ \cite{gibbs2011advanced}.

We derive the sets of admissible states $\bm{\chi}$ and inputs $\bm{\zeta}$ empirically to conform to the physical hardware constraints of the linear stage and EM specifications (e.g. max voltage). \hl{These will be used in the contrained optimization problem solved in Eq. }\ref{eq:mpcc-formulation}.

The pen position is propagated via a standard linear Kalman filter \cite{gibbs2011advanced}. While not an accurate user model, it works well in practice since the states are recalculated at every timestep. 

\subsection{Path Following} \label{sc:lag_cont_err}
We continuously optimize the EM parameters with \edt{the} goal of keeping the distance between the desired path and the pen minimal. However, we cannot rely on a timed trajectory as is commonly done for example in MPC formulations (e.g., \cite{Faulwasser:2009}), since we want to give the user freedom in deciding their drawing speed. To achieve this trade-off, we first need to find the reference point $s(\theta)$ itself. Finding the closest point on the path is an optimization problem itself and hence can not be used within our optimization. Similar to recent work in MAV trajectory generation \cite{Naegeli:2017:MultiDroneCine,Gebhardt:2018} we decompose the distance to the closest point into a contouring and lag error. 

\begin{figure}[]
    \centering
    \includegraphics[width=\columnwidth]{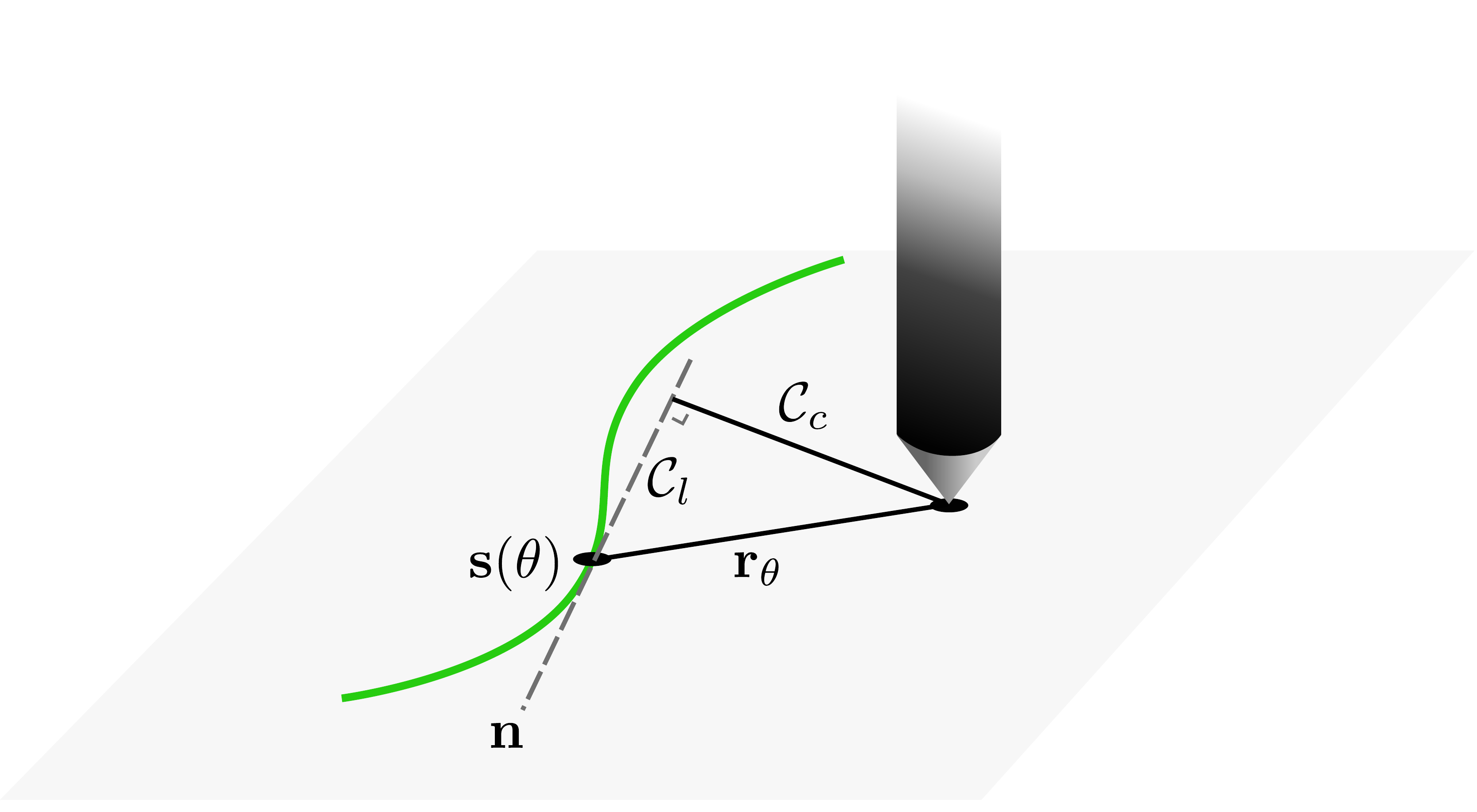}
    \caption{Illustration of lag- and contouring error decomposition.}
    \label{fig:elc}
\end{figure}

\begin{figure*}[t]
    \begin{subfigure}[t]{0.33\textwidth}
    \includegraphics[width=\columnwidth]{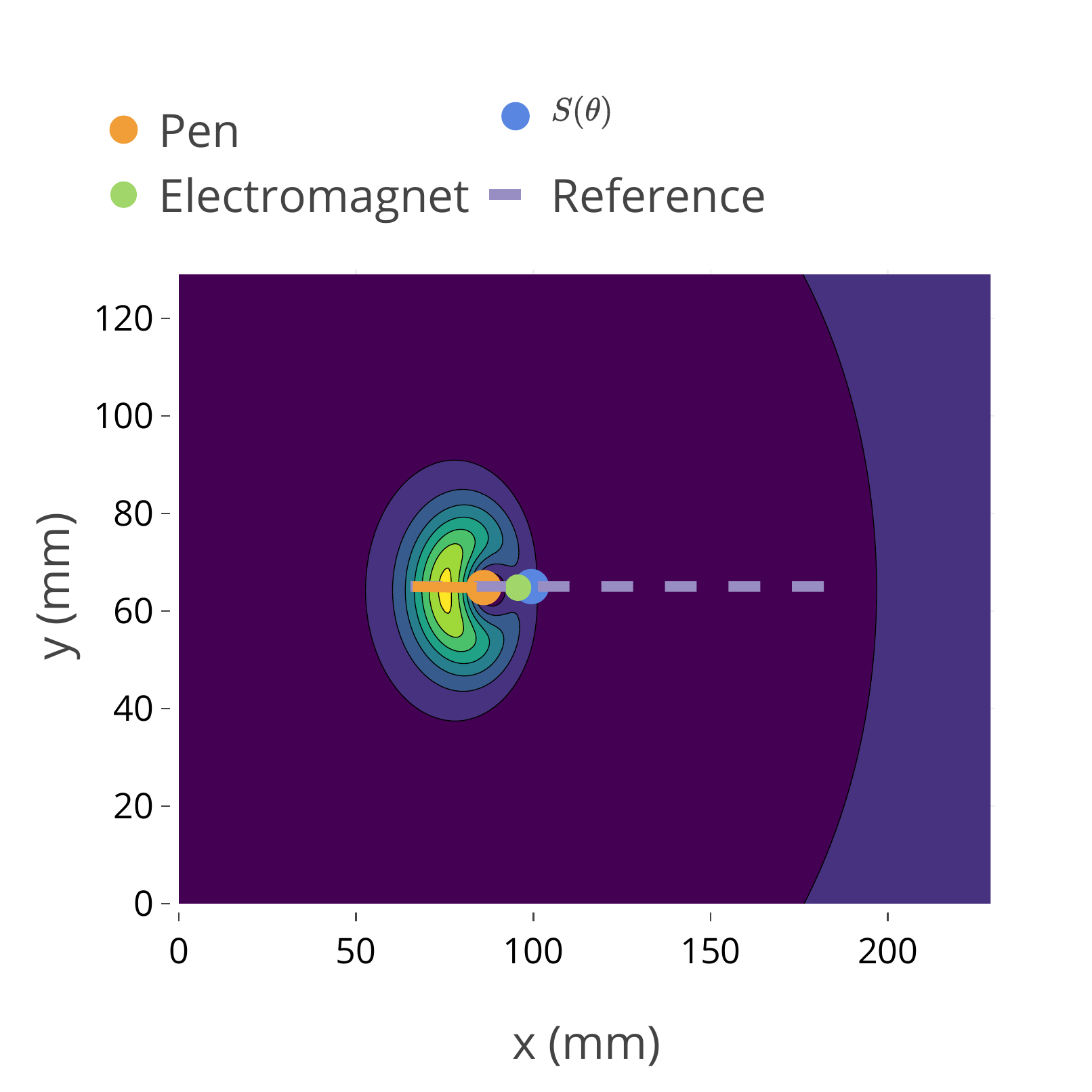}
    \caption{On-path behavior.}\label{fig:error_correction_a}
    \end{subfigure}
    \begin{subfigure}[t]{0.33\textwidth}
     \includegraphics[width=\columnwidth]{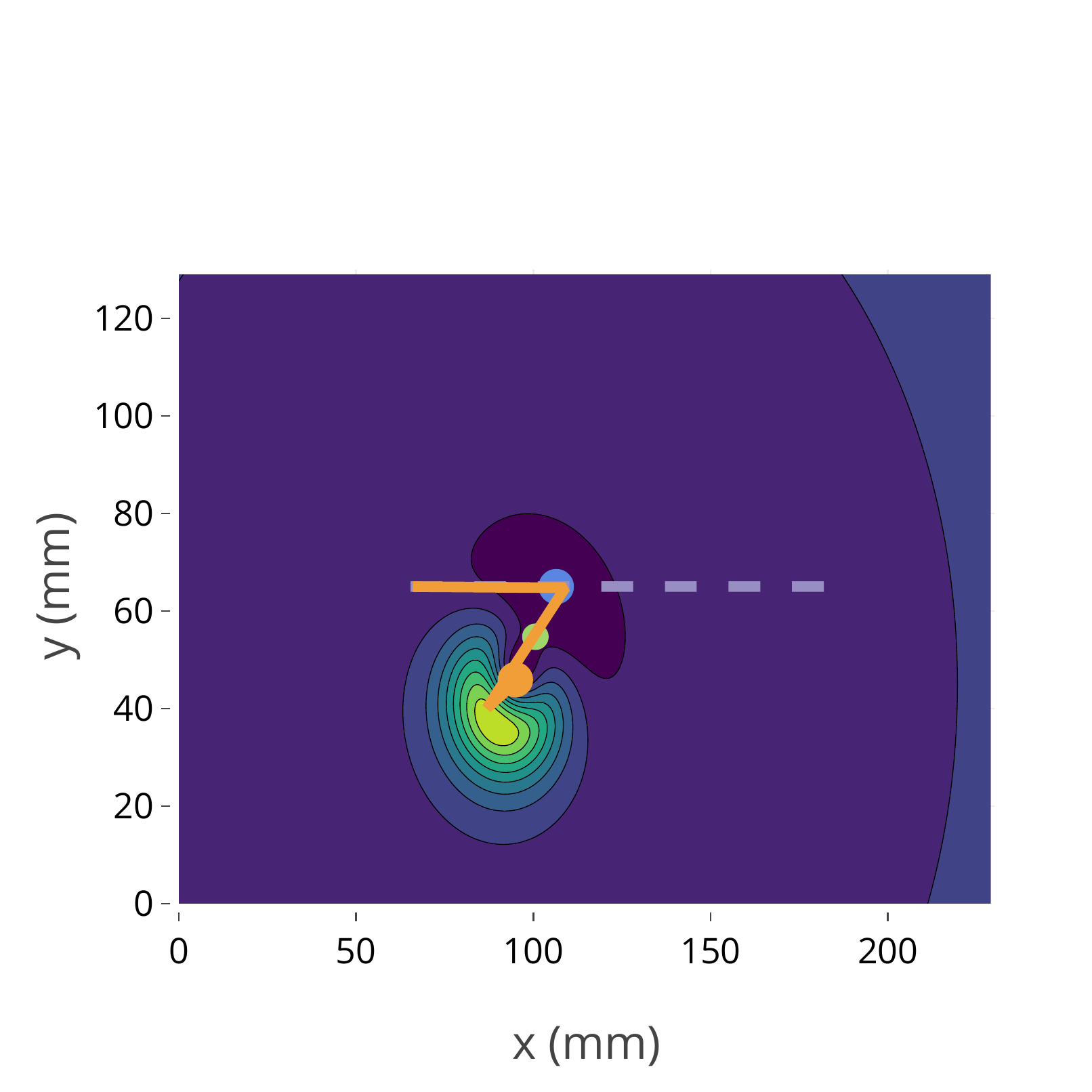}
     \caption{Sudden jump in pen position.}
     \label{fig:error_correction_b}
    \end{subfigure}
    \begin{subfigure}[t]{0.33\textwidth}
        \includegraphics[width=\columnwidth]{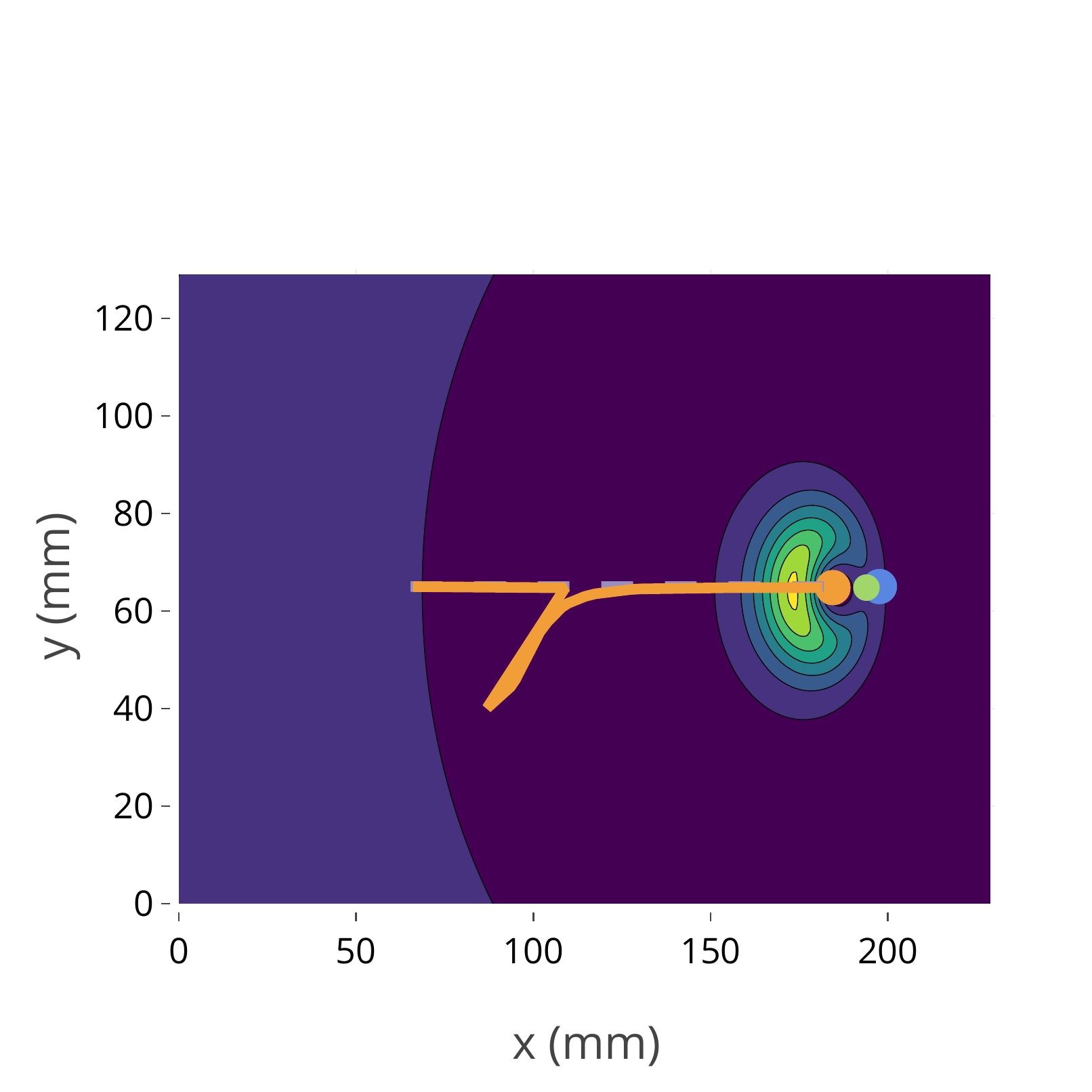}
        \caption{Smooth correction.}
        \label{fig:error_correction_c}
    \end{subfigure}
    \centering
    \caption{Illustration of error correcting behavior. Left-to-right: (\protect\subref{fig:error_correction_a}) a simulated user is close to the desired path. (\protect\subref{fig:error_correction_b}) A sudden jump in pen-position causes the magnet to move towards the pen and to increase the magnet strength, (\protect\subref{fig:error_correction_c}) which pulls the pen smoothly back to the trajectory . See Sec. \protect\ref{sc:results-error-correcting}. The background is a plot of the cost function; in which purple is the lowest.}
    \label{fig:error_correction}
\end{figure*}

We define $\Rtheta$ as the distance between the pen $\posp$ and a point $\mathbf{s}(\theta)$ on the spline, and $\mathbf{n}$ as the normalized tangent vector to the spline at that point:
\begin{align}
    \Rtheta &= \posst -\posp \ , \label{eq:r_theta} \\
    \mathbf{n} &= \frac{\mathbf{s'}}{\norm{\mathbf{s'}}}, \label{eq:n}
\end{align}
\noindent  with $\mathbf{s'} = \frac{\partial \mathbf{s} (\theta)}{\partial\theta}$. The vector $\Rtheta$ can now be decomposed into a lag error and a contour error (see \figref{fig:elc}). %
The lag-error $\Cost_l$ is computed as the projection of $\Rtheta$ on the tangent of $\posst$%
, while the contour-error $\Cost_c$ is the component of $\Rtheta$ orthogonal to the normal:%
\begin{align}
\Cost_l (\posp, \theta) &= \norm{\langle\Rtheta,\mathbf{n}\rangle}^2 \label{eq:errL}, \\
\Cost_c (\posp, \theta) &= \norm{ \Rtheta - \left( \langle\Rtheta,\mathbf{n}\rangle \right) \mathbf{n} }^2.\label{eq:errC} 
\end{align}

Separating lag from contouring error allows us, for example, to differentiate how we penalize a deviation outside the path ($\Cost_c$), from encouraging the user to progress forward ($\Cost_l$). %
We furthermore include cost terms to ensure the magnet stays ahead of the pen ($\Cost_{\theta}(\theta)$) and to encourage smooth progress ($\Cost_{\dot{\theta}}(\dot{\theta})$): 
\begin{align}
\Cost_{\theta}(\theta) &= - \theta \label{eq:err_theta},\\
\Cost_{\dot{\theta}}(\dot{\theta}) &= (\dot{\theta}_t-\dot{\theta}_{t-1})^2 \label{eq:err_theta_dot}.
\end{align}

\edt{Finally, we optimize the progress variable $\theta$ so that $\mathbf{s}(\theta)$ is a combination between the closest point and ensuring that the magnet eventually does progress to indicate to the user in which direction to continue.}

\subsection{EM Force Control} \label{sc:em_costs}
Leveraging our physical model for the electromagnetic force $F_a$ (Eq. \ref{eq:Fa}), we propose a residual for the desired force $F_{\theta}$ (see Figure \ref{fig:em_model}). %
The desired force, $F_{\theta}$ pulls the pen towards the target point $\mathbf{s}(\theta)$. The cost is modeled with a spring-like behaviour, so that it linearly increases with the distance $r_{\theta}$: 
\begin{equation}
     \mathbf{F}_{\theta} (\Rtheta) = c \ F_0 \ \frac{r_{\theta}}{h} \ \mathbf{e_{r_{\theta}}} \ , \label{eq:Fd}
\end{equation}
where $c$ is a scalar that regulates the stiffness of the spring, $F_0$ a scaling of the EM force and $h$ the distance between dipoles in $z$ (set manually). Although simple, this formulation ensures that the haptic guidance is strong under large deviation from the path while vanishing as the user approaches the target path ($r_{\theta} \to 0$). %
Note that the EM force saturates at $F_a^{max}$.
Using the expressions for the actuation force $\mathbf{F_a}$ (Eq. \ref{eq:Fa}) and desired force (Eq \ref{eq:Fd}) we formulate a quadratic cost term to penalize the difference:
\begin{equation}
    \Cost_f(\posm, \posp, \alpha) = \norm{ \ \mathbf{F}_{\theta}(\Rtheta) \ - \ \mathbf{F_a}(\mathbf{d}) \ }^2. \label{eq:err_F}
\end{equation}

Since the actuation force $\mathbf{F_a}$ declines rapidly with distance, the gradient of $\Cost_f$ goes to 0 for large values of $d$ causing the optimization to become unstable. To counterbalance this issue we encourage the electromagnet to stay close to the pen:
\begin{equation}
    \Cost_d(\posm,\posp) = d^2. \label{eq:err_d}
\end{equation}
Finally we encourage an interplay between magnet position and intensity by penalizing excessive use of magnetic intensity $\alpha$:
\begin{equation}
    \Cost_{\alpha}(\alpha) = \alpha^2. \label{eq:err_alpha}
\end{equation}

\subsection{Problem Formulation}\label{sc:Jk}

We combine the cost terms Eq. \ref{eq:errL}--\ref{eq:err_theta_dot} and Eq. \ref{eq:err_F}--\ref{eq:err_alpha} to form the final stage cost:
\begin{align}\label{eq:J_k}
J_k= \quad & w_l \Cost_l(\mathbf{p}_{p,k}, \theta_k) + 
     w_c \Cost_c(\mathbf{p}_{p,k}, \theta_k) + \nonumber \\
     & w_{\theta} \Cost_{\theta}(\theta_k) + 
        w_{\dot{\theta}} \Cost_{\dot{\theta}}(\dot{\theta}_k) + \nonumber \\
     & w_f \Cost_f(\mathbf{p}_{m,k}, \mathbf{p}_{p,k}, \alpha_k, \theta_k) + \nonumber \\
     & w_d \Cost_d(\mathbf{p}_{m,k}, \mathbf{p}_{p,k}) +
        w_\alpha \Cost_{\alpha}(\alpha_k),
\end{align}
where the scalar weights $w_l,w_c,w_{\theta},w_{\dot{\theta}},w_f,w_d>0$ control the influence of the different cost terms. The values used in our experiments can be found in Table \ref{tab:var} in Appendix \ref{sc:ap.notation}. The states and inputs to the bi-axial linear stage and to the electromagnet are then computed by solving the $N$-step %
finite horizon constrained non-linear optimization problem at time instance $t$. 

\begin{align}
\label{eq:mpcc-formulation}
\underset{\mathbf{x}, \mathbf{u}, \theta}{\text{minimize}}\quad & \sum_{k=0}^{N} w_k\left ( J_k + \mathbf{u}_k^T \mathbf{R} \mathbf{u} \right ) && \\
\text{Subject to:}\quad & \mathbf{x} _{k+1} = f(\mathbf{x_k}, \mathbf{u_k}) & \text{(System Model)} \nonumber\\
                        & \mathbf{x}_0 = \hat{\mathbf{x}}(t) & \text{(Initial State)} \nonumber \\
                        & \theta_0 = \hat{\theta}(t) & \text{(Initial Progress)} \nonumber \\
                        & \theta_{k+1} = \theta_k + \dot{\theta}_k dt & \text{(Progress along path)} \nonumber \\
                        & 0 \leq \theta_k \leq L& \text{(Path Length)} \nonumber \\
                        & \mathbf{x}_k \in \bm{\chi} & \text{(State Constraints)} \nonumber \\
                        & \mathbf{u}_k \in \bm{\zeta} & \text{(Input Constraints)} \nonumber
\end{align}

Here $k$ indicates the horizon stage and the additional weight $w_k$ reduces over the horizon, so that the current timestep has relatively more importance than later timesteps. $\mathbf{R}\in\mathbb{S}_+^{n_u}$ is a positive definite penalty matrix avoiding excessive use of the control inputs. In our implementation we use a horizon length of $N=10$. Experimentally we found that this is sufficient to yield robust solutions to problem instances and longer horizons did not improve results.

\section{Technical Evaluation}
We evaluate our system both from a technical and performance perspective. Furthermore we report quantitative and qualitative findings from an initial user study. 

\subsection{Implementation Details}
We use a standard desktop computer (Intel Core i7-4770 CPU 4 cores at \unit[3.40]{GHz}) running Ubuntu 17.10 in our experiments. The solver is implemented in FORCES Pro \cite{domahidi2014forces}, which produces efficient C-code. The communication with the Sensel Morph tablet and the Arduino board controlling the stepper motors happens asynchronously over a serial connection.

\subsection{Error-correcting behavior} \label{sc:results-error-correcting}
An important design goal is to provide guidance to the user, especially under deviation from the path. Here we analyze how the algorithm responds to various types of erroneous situations. To this end we simulate a cooperative user, always following the guidance, but initialize the pen position $\mathbf{p}_p$ at different off-path positions. We update the pen position via a simple constant velocity model:
\begin{equation}
    \mathbf{p}_{p}^{t+1} = \mathbf{p}_{p}^{t} + v_c\left(w_v\frac{\mathbf{p}_{v}^{t}}{\norm{\mathbf{p}_{v}^{t}}}+w_{m}\ed\right) ,
    \label{eq:user_sim}
\end{equation}
where the pen-position $\mathbf{p}_{p}^{t+1}$ at timestep $t+1$ is updated via the previous position and a weighted sum of the normalized velocity $\mathbf{p}_{v}^{t}$, the exerted force direction $\ed$ and a constant velocity factor $v_c$.

We analyze a typical but difficult to handle path deviation: a y-offset in combination with an offset in the direction of negative $\theta$ (backwards). \figref{fig:error_correction} illustrates that our algorithm handles the error gracefully. In particular the magnet is controlled such that it gently pulls the pen back to the path, rather than pulling directly towards either the closest point on the path or a steadily advancing setpoint, as would be the case in traditional control methods.

\subsection{Positional dispersion}\label{sc:results-pos-dispersion} 
We now analyze the positional accuracy of the system while isolating the user contribution. Note that in the context of drawing the interesting aspect is not the accuracy of the position of the electromagnet (stepper motors are discrete and accurate), but the influence of the magnet on the pen. To assess this we moved the electromagnet to random locations and then always back to the center of the drawing surface. With a pen being held fully upright and the magnet at full strength ($\alpha=1$) and the user following the magnet passively, this allows us to measure the system's positional dispersion. We collected data from 300 repetitions, resulting in a mean offset from the target of \unit[2.8]{mm} with a standard deviation (SD) of \unit[0.8]{mm}, indicating that the system can control the pen well. \edt{One of the factors that contribute to this dispersion is the vanishing of the actuation force $F_a$ as $d\rightarrow0$. This can lead to the pen motion stopping slightly before it reaches the target.}

\subsection{Latency} Due to the steepness of the electromagnetic force $\mathbf{F_p}$ and the potentially fast pen motion, runtime and latency are crucial performance metrics. The optimization algorithm contributes to both, whereas latency is dominated by the hardware and I/O. The mean solve time for a problem instance is \unit[7.4]{ms} ($\pm$ \unit[3.0]{ms}). Since we do not manipulate the system state space and pen input is the only measurement input this can be expected to be mostly constant. The hardware and overall system latency adds to the solve time. We employ a high-speed camera (\unit[1000]{fps}) to establish the motion (pen) to motion (magnet) latency. This yields an approximate latency of \textasciitilde{}\unit[10]{ms}. Given the combined latency envelope of \textasciitilde{}\unit[20]{ms}, we did not experience any abrupt pen snapping in our experiments.

\section{User Evaluation - Control Strategies}
\label{Sc:preliminary_user_evaluation}
\edt{We first conduct a preliminary experiment to validate the choice of a time-free closed-loop optimization strategy. For this purpose we compare our implementation to a simpler MPC variant and an open-loop strategy (our implementation of dePENd \cite{yamaoka2013depend}). The main difference in \emph{ours} is that $s(\theta)$ is an optimization variable itself, whereas MPC used a setpoint that progresses along the reference at a predetermined speed.}

\subsection{Procedure}
\edt{In this experiment we asked participantes ($N=12$) to draw one complex shape (Fig. }\ref{fig:usertest_examples_dog}\edt{) in three different conditions: Open-Loop (\emph{OL}), time-dependent closed loop (\emph{MPC}), and time-free closed loop (\emph{ours}). The order of the trials was counterbalanced via a latin-square design.  
}
 \begin{figure}
     \centering
     \includegraphics[width=\columnwidth]{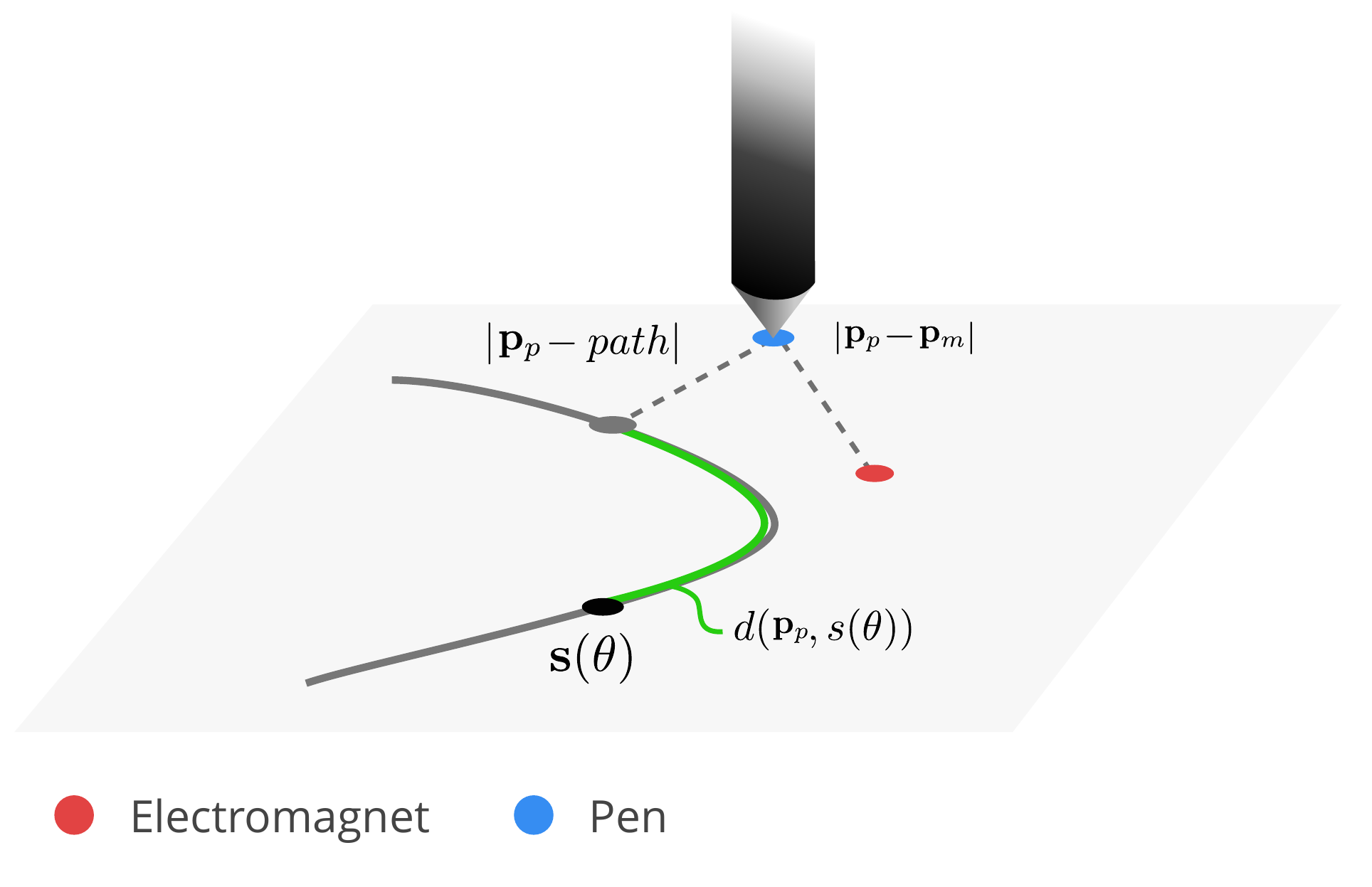}
     \caption{Overview of the different metrics for the preliminary user evaluation.}
     \label{fig:ontrol-strategies-metrics}
 \end{figure}

\subsection{Measures}
\edt{
We analyze three measures; i) The mean distance from the pen to the path, ii) the mean distance from the pen position projected onto the path and} $\mathbf{s}(\theta)$ along the path and iii) \edt{the mean distance from the pen to the electromagnet. Taking the mean of the error terms over subjects we ensure equal numbers of datapoints, accounting for differences in speed. Note that here we assume that the user roughly maintains a constant speed.}\footnote{\edt{In our full implementation and haptic feedback experiments this assumption is not necessary. However, the metric used here assumes time-independent datapoints.}} \edt{ We also gathered qualitative feedback in the form of a semi-structured interview.}

\edt{A one-way ANOVA with Kruskal-Wallis test was performed for data analysis.}

\subsection{Results}
\begin{figure*}[t]%
    \centering
    \begin{subfigure}[b]{.45\textwidth}
    \includegraphics[width=\columnwidth]{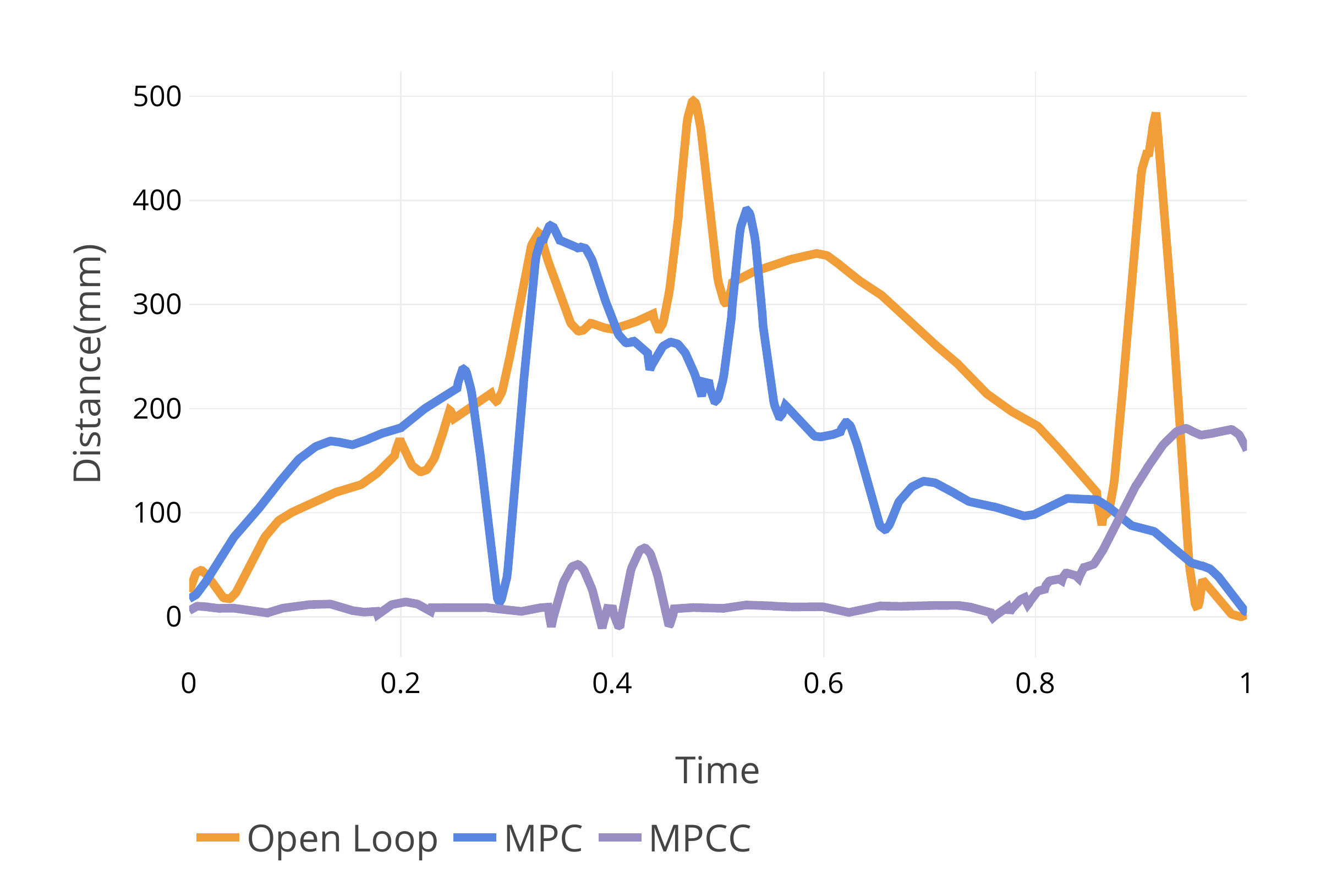}%
    \caption{Path distance pen-$\mathbf{s}(\theta)$}%
    \label{fig:path_pen_s}%
    \end{subfigure}
    \begin{subfigure}[b]{.45\textwidth}
    \includegraphics[width=\columnwidth]{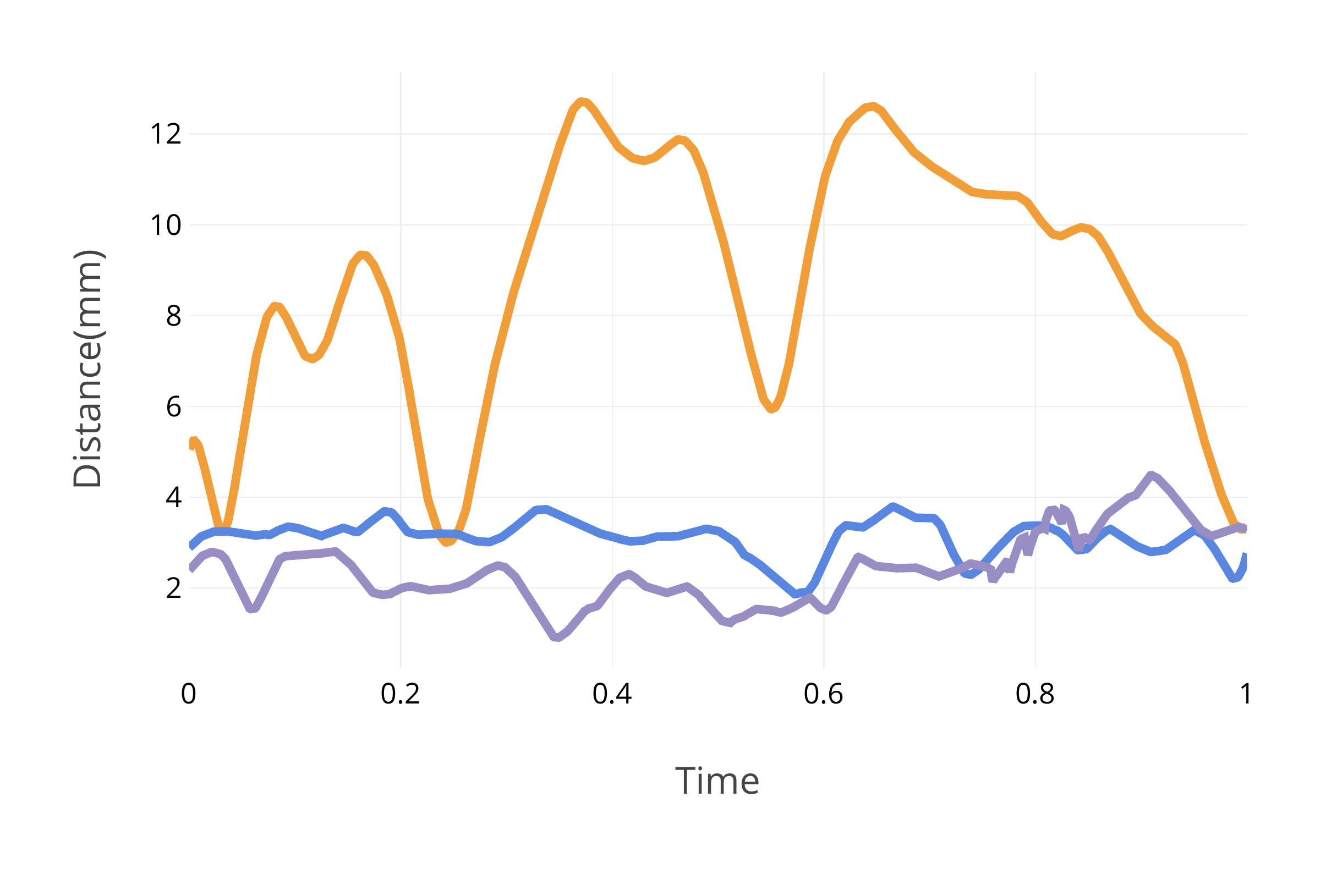}%
    \caption{The euclidean distance from pen-magnet}%
    \label{fig:euc_pen_em}%
    \end{subfigure}\hfill%
    \caption{\hl{Comparison} of error over time for a single participant (P1). In \ref{fig:path_pen_s} there is the inverse u-shape that illustrates that the $\mathbf{s}(\theta$) moves at a different speed than the user for OL and MPC. Sub-figure \ref{fig:euc_pen_em} show how the EM is relatively close and constant to the pen with MPC and MPCC. With OL the magnet moves away from the pen. The data is smoothed to increase readability.} 
    \label{fig:single_user_control}
\end{figure*}

\begin{table}[t]
    \centering
    \caption{\hl{Mean distance in mm from 1) the pen to the closest point on the path, 2) distance from pen to } $\mathbf{s}(\theta)$\hl{ along the path, 3) the pen to the electromagnet and 4) the fraction of the measurements where the electromagnet is more than 15 mm away from the pen (based on Figure} \ref{fig:approx_error}). All units are given in mm.}
    \begin{tabular}{l|ccc}
         &|pen-path|& d(pen, $s(\theta)$) & |pen-em| \\
         \midrule
         OL & $4.1(\pm 0.7)$ & $38.0(\pm 56.9)$  &$38.2(\pm 25.1)$ \\ 
         MPC & $3.9(\pm 1.3)$& $45.0(\pm 50.8)$ & $8.6(\pm 1.6)$ \\ 
         MPCC & \textbf{2.0}$(\pm 0.6 )$& \textbf{6.2}$(\pm 0.8)$ & \textbf{4.6}$(\pm 0.9)$\ 
    \end{tabular}
    \label{tab:strategy_results}
\end{table}

\subsubsection*{Quantitative} \edt{Table} \ref{tab:strategy_results} \edt{summarizes our quantitative findings. Not surprisingly, the distance from the electromagnet to the pen for (\emph{OL}) is significant. Since the force exerted on the pen falls off quadratically with distance, participants often lost any haptic guidance early on, confirmed via user comments such as ``I don't feel anything'' (P3) and ``Is the system on?'' (P6). Similarly, we see that $d(pen, \mathbf{s(\theta)})$ is larger compared to \emph{ours} by a factor of six.} %

\edt{While \emph{MPC} reduces the distance from the pen to the magnet (and hence always provides haptic feedback), it does not optimize for the progress along the path and hence may pull the pen into undesired directions. For example, \emph{MPC} produces extreme corner cutting behavior to catch up to the setpoint.}

 \edt{Finally, \emph{ours} has the highest accuracy (H(2)=20.76, p<.001). Furthermore, the setpoint }$\mathbf{s}(\theta)$\edt{ (H(2)=7.362, p <.05) and the electromagnet (H(2)=27.12, p <.001) are closest to the pen. Thus our time-free formulation overcoms both problems of wrong setpoints (\emph{MPC}) and a run-away electromagnet, as can happen with strategies proposed in prior-work} \cite{yamaoka2013depend}.
 
 \edt{Figure} \ref{fig:single_user_control} \edt{shows that both the distance along the path and the pen-magnet distance accumulate over time if \emph{OL} or \emph{MPC} control strategies are employed. Note that this is not the case with our implementation (MPCC) and both errors remain low over the entire path length.}

\subsubsection*{Standard Deviation}\edt{Note that for \emph{OL} and \emph{MPC} the standard deviation is high. This is likely due to the absence of direct coupling between user feedback and path progress, which makes it possible for the user to lag behind the setpoint significantly (albeit at the cost of reduced force feedback). In our implementation the path progress is adjusted to the user's drawing speed, drastically reducing the standard deviation and in consequence ensuring delivery of force feedback throughout the drawn path. 
}

\subsubsection*{Qualitative} \hl{From our observations we saw that} $\mathbf{s}(\theta)$\hl{ was either in front or behind the user for MPC. This was also confirmed in our interview, where people especially commented on the MPC strategy: ``The system tries to push me in the wrong direction'' (P2) and ``It is counteracting me'' (P11). This also resulted in the MPC being the least preferred option. In contrast with our formulation the magnet remains always slightly ahead of the pen, resulting in users rating the MPCC as the most preferred condition. In the words of one subject this is: ``since I still had control'' (P9). }

\edt{Taking both the quantitative and qualitative results into account, we see that our MPCC formulation performs best overall. Open-Loop causes numerous problems, including users not perceiving any haptic feedback. This is especially troublesome in settings where autonomy is desired. In MPC the haptic feedback is perceived, but can be erroneous. This is especially evident when users do not conform the expected behavior. We therefore only report results from the MPCC formulation in all further evaluations.}

\section{User Evaluation - Haptic Feedback}
\begin{figure}[!t]
    \centering
        \begin{subfigure}[b]{0.45\columnwidth}
        \includegraphics[width=\columnwidth]{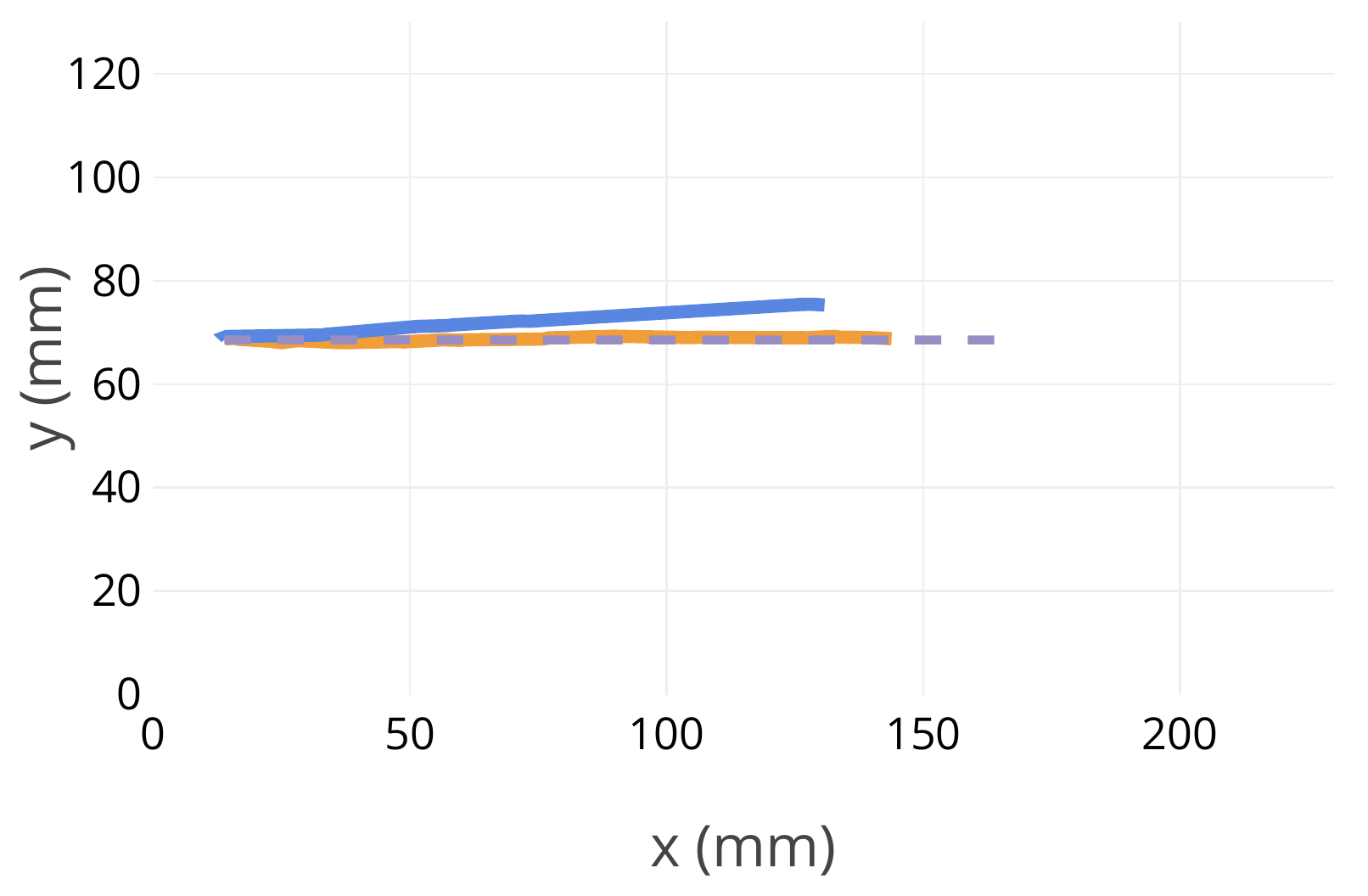}
        \caption{Line}\label{fig:qualitative_results_a}
    \end{subfigure}
        \begin{subfigure}[b]{0.45\columnwidth}
        \includegraphics[width=\columnwidth]{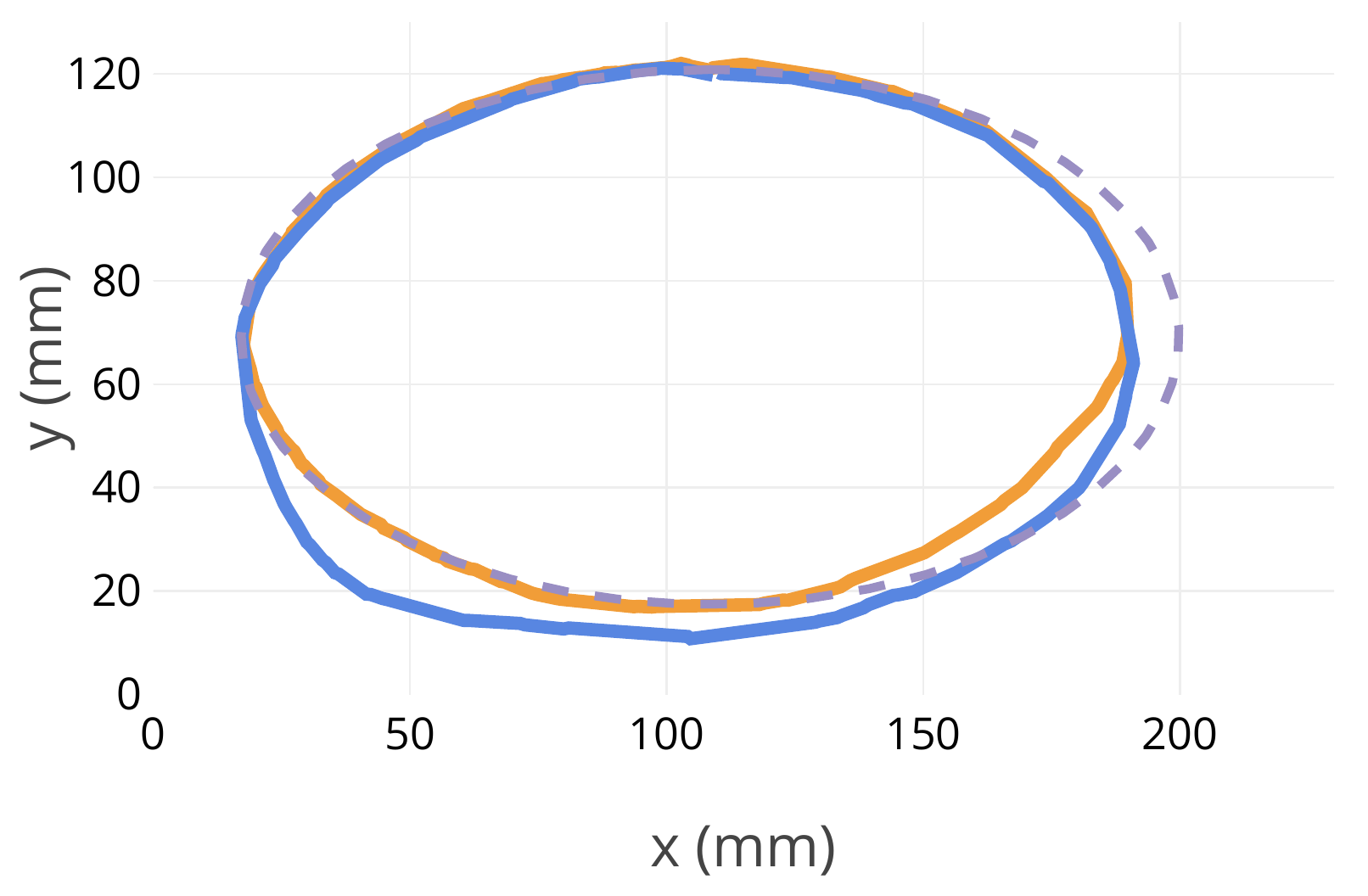}
        \caption{Ellipse}\label{fig:qualitative_results_b}
    \end{subfigure}
    \begin{subfigure}[b]{0.45\columnwidth}
        \includegraphics[width=\columnwidth]{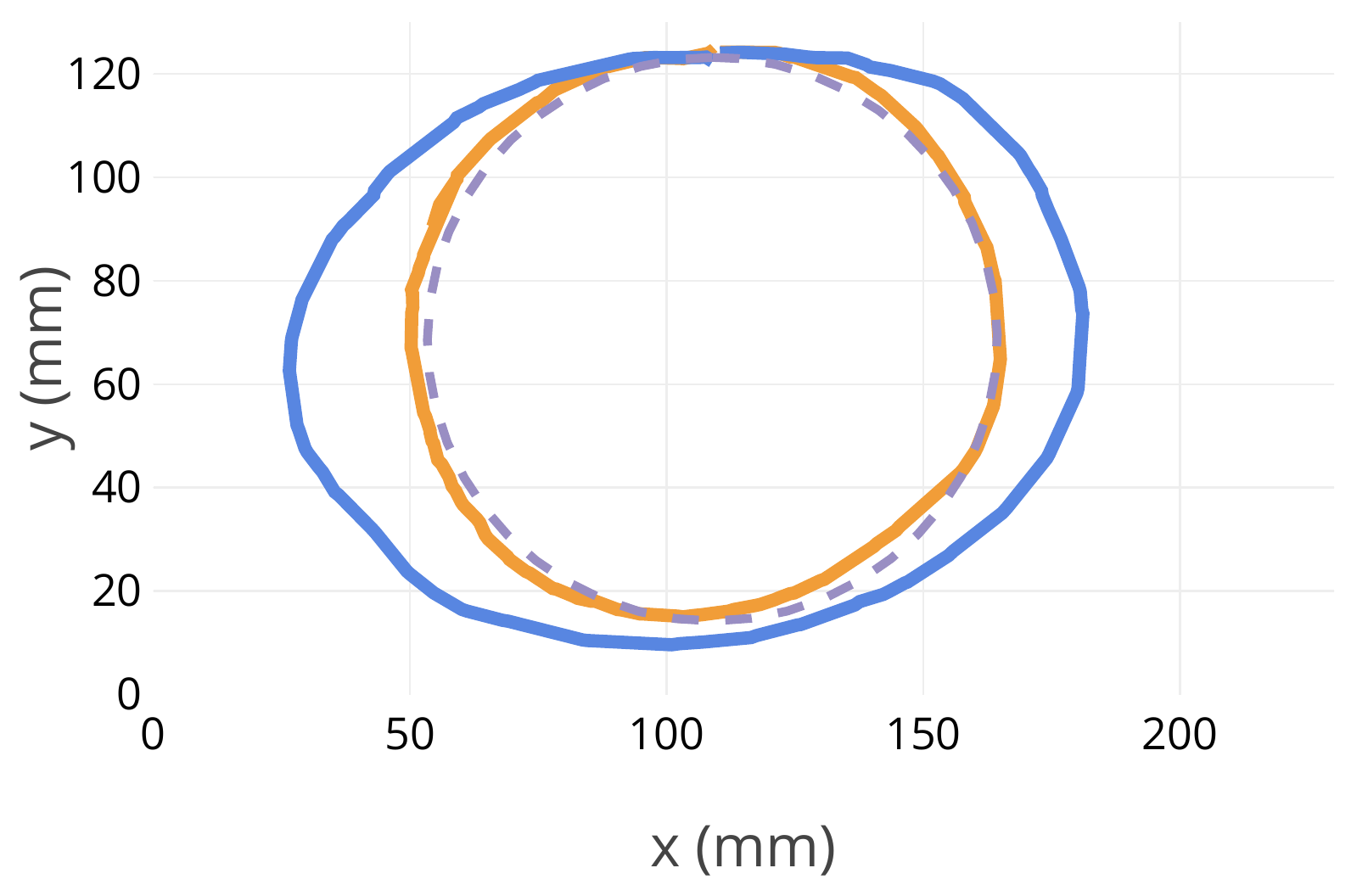}
        \caption{Circle}\label{fig:qualitative_results_c}
    \end{subfigure}
    \begin{subfigure}[b]{0.45\columnwidth}
        \includegraphics[width=\columnwidth]{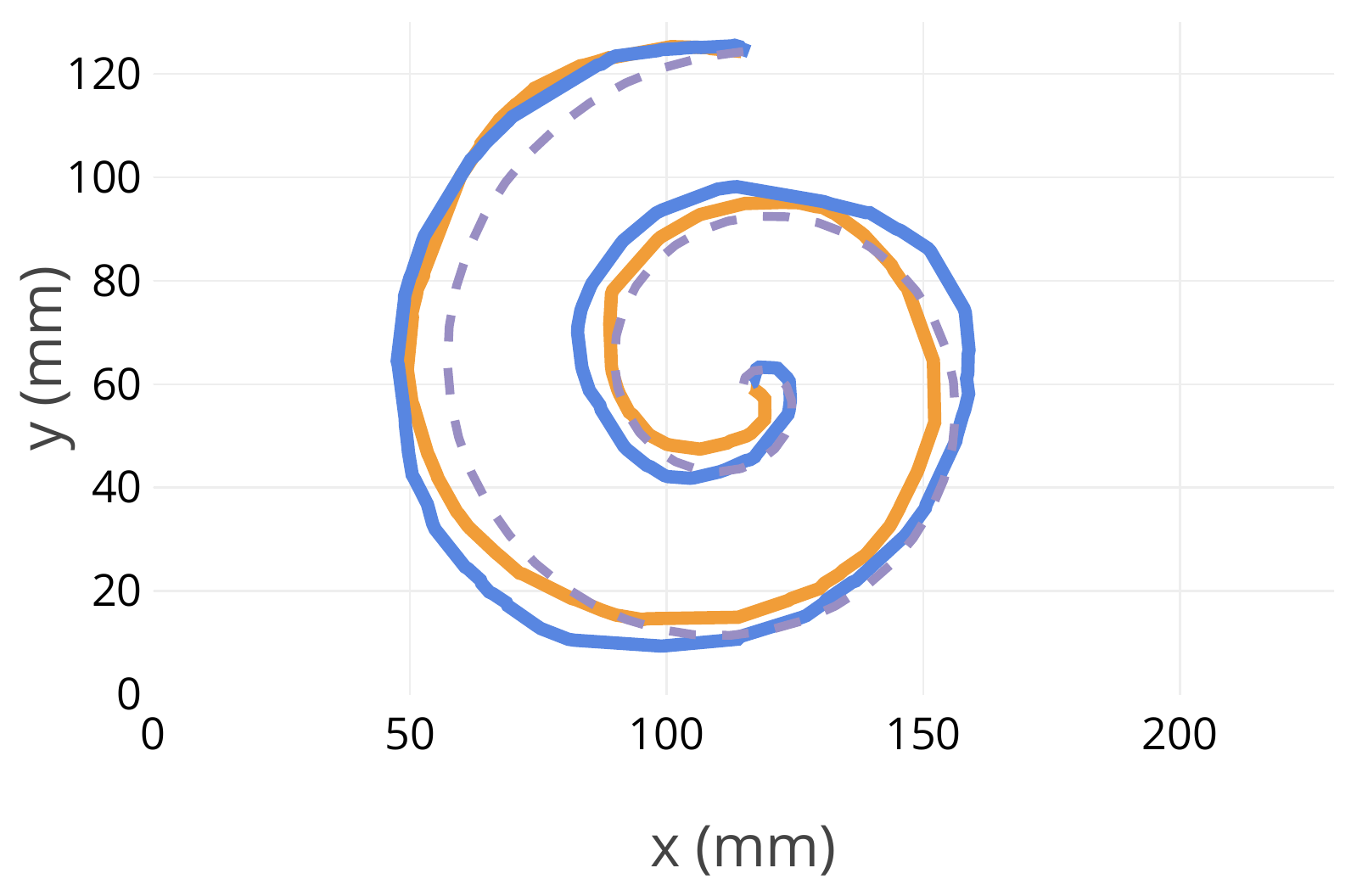}
        \caption{Spiral}\label{fig:qualitative_results_d}
    \end{subfigure}
    \begin{subfigure}[b]{0.45\columnwidth}
        \includegraphics[width=\columnwidth]{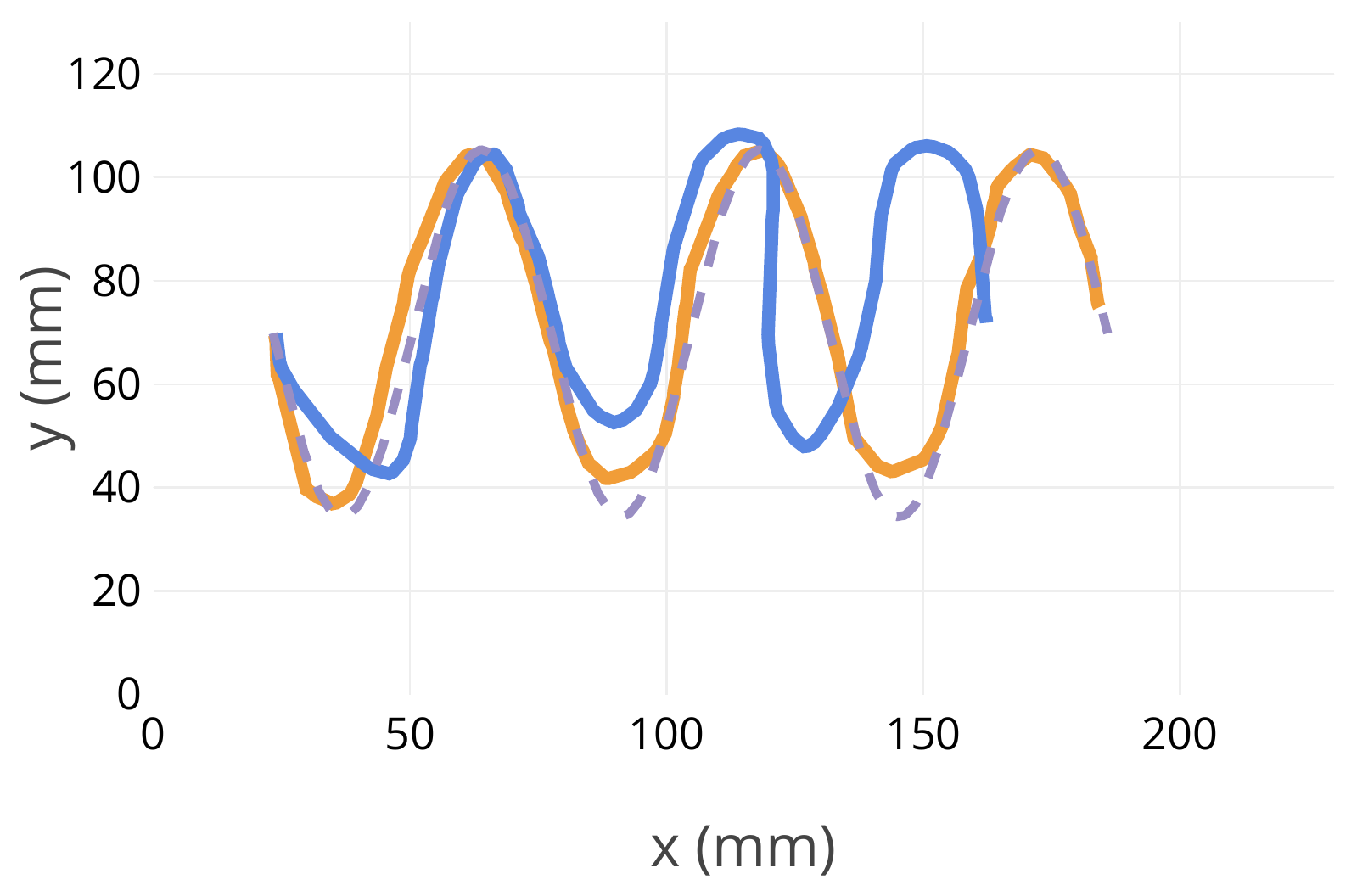}
        \caption{Sinusoidal}\label{fig:qualitative_results_e}
    \end{subfigure}
    \begin{subfigure}[b]{0.45\columnwidth}
        \includegraphics[width=\columnwidth]{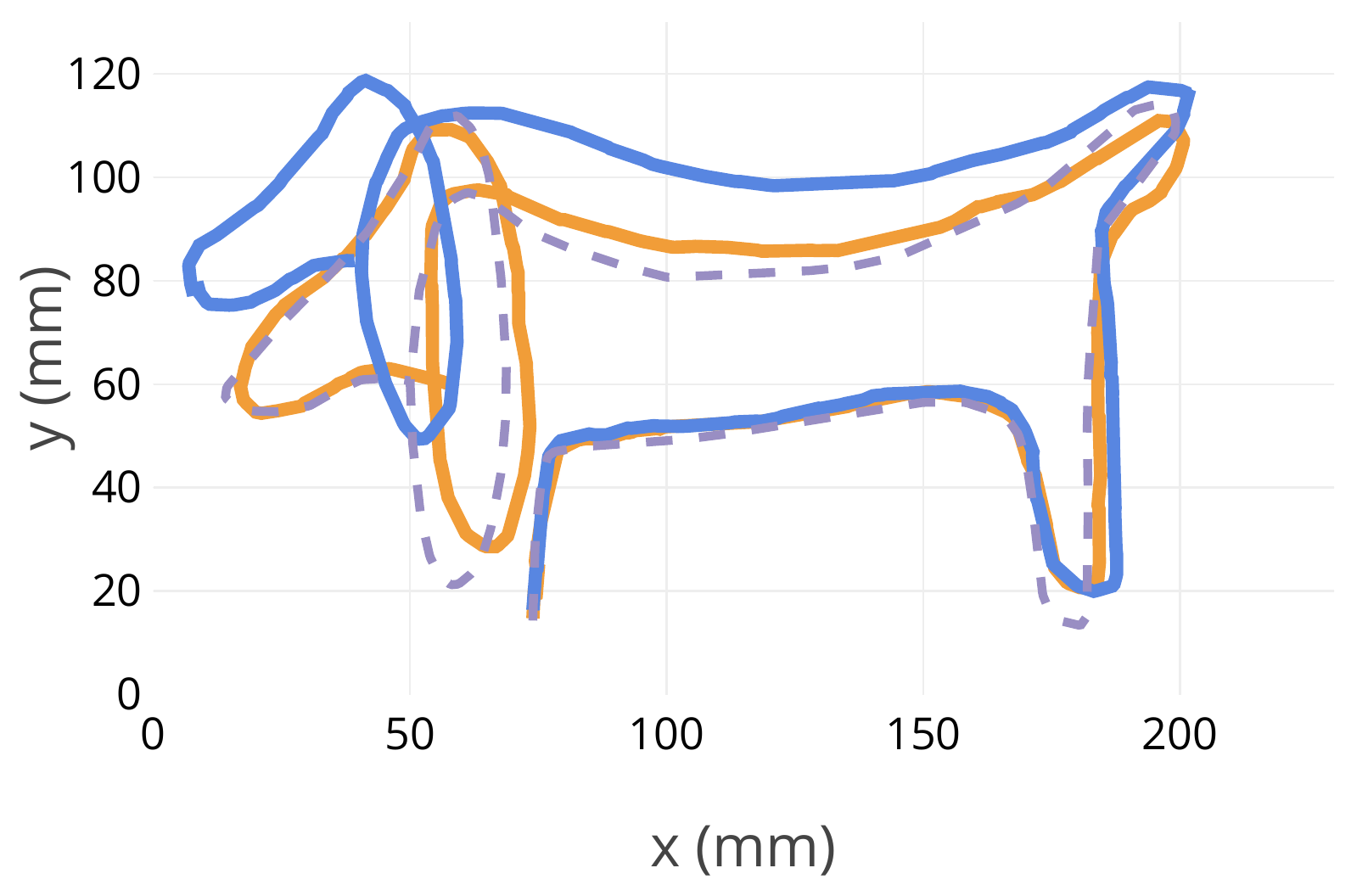}
        \caption{Dog}\label{fig:qualitative_results_f}
    \end{subfigure}
    \caption{Selected experimental results. Each shape drawn by different participants with (orange) and without (blue) guidance, compared to the reference (dotted). Sinusoidal is different from the one in \protect\figref{fig:single_user_a}}
    \label{fig:qualitative_results}
\end{figure}

The impact on task performance and user perception is potentially the most important aspect of any haptic feedback system. To assess these factors we ran an initial controlled user study. \edt{
In this study, we investigate the overall performance of our MPCC formulation in more detail. To this end we conduct experiments with several  different shapes, and compare to a no-feedback baseline.} \hl{A first 
impression of the results can be found in Figure  \ref{fig:qualitative_results}.}

\begin{figure}[!t]
    \centering
    \begin{subfigure}[b]{0.3\columnwidth}
        \includegraphics[width=\columnwidth]{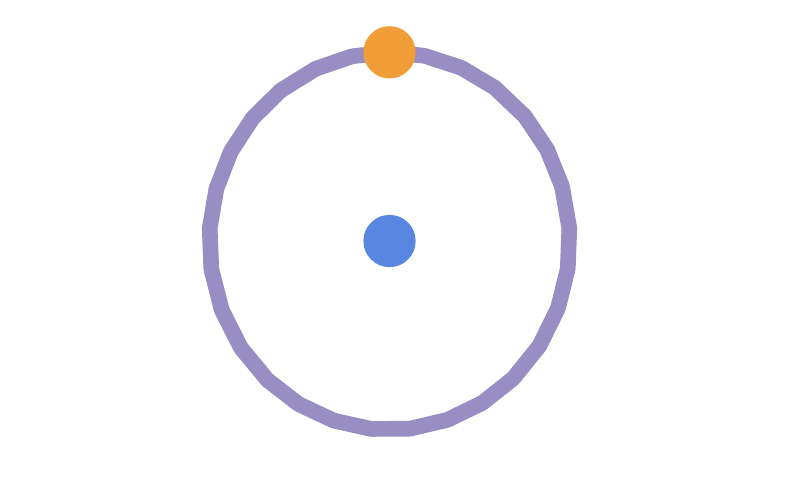}
        \caption{Circle}
    \end{subfigure}
    \begin{subfigure}[b]{0.3\columnwidth}
        \includegraphics[width=\columnwidth]{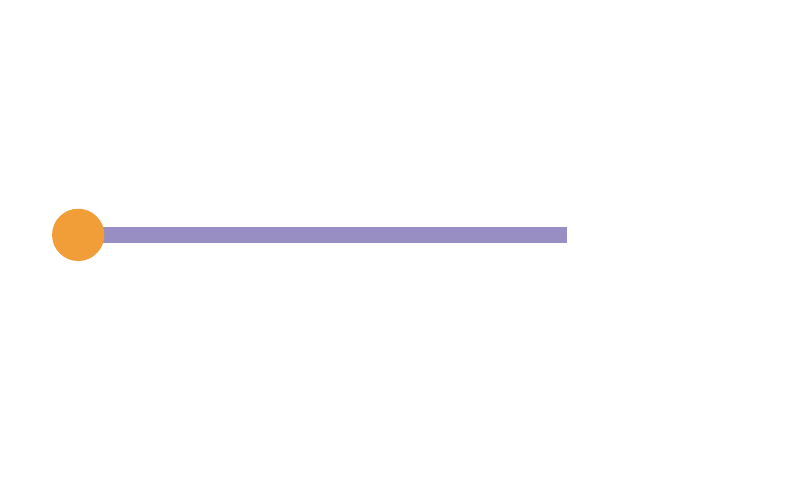}
        \caption{Line}
    \end{subfigure}
    \begin{subfigure}[b]{0.3\columnwidth}
        \includegraphics[width=\columnwidth]{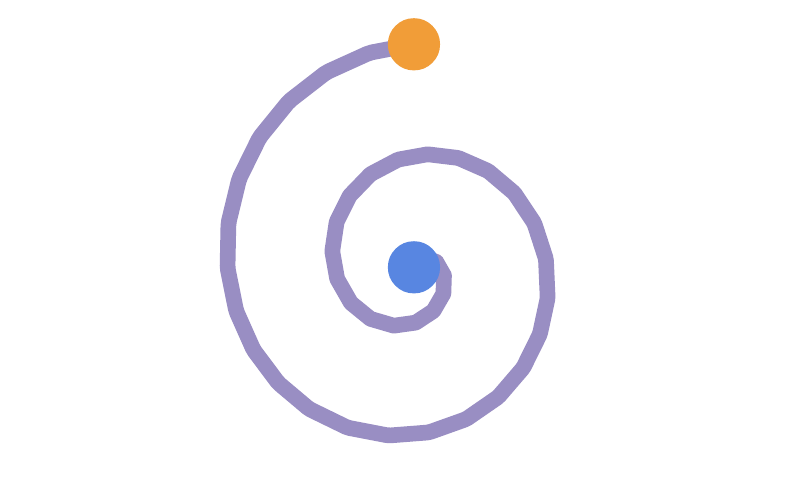}
        \caption{Spiral}
    \end{subfigure}
    \begin{subfigure}[b]{0.3\columnwidth}
        \includegraphics[width=\columnwidth]{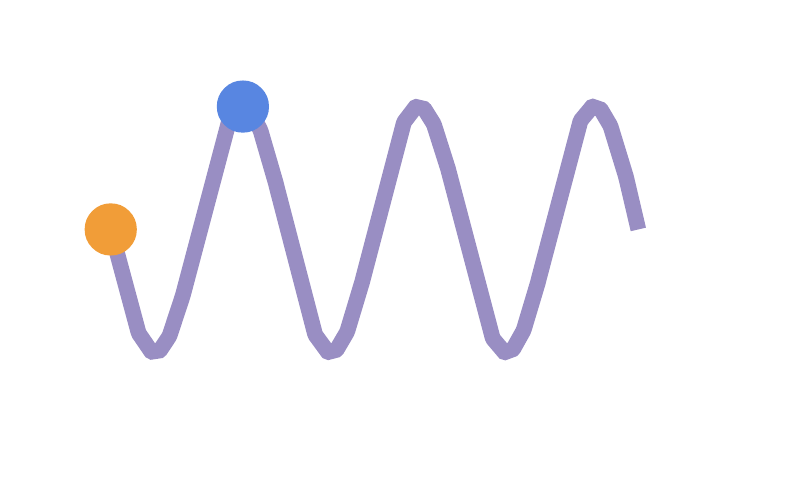}
        \caption{Sinusoidal}
    \end{subfigure}
    \begin{subfigure}[b]{0.3\columnwidth}
        \includegraphics[width=\columnwidth]{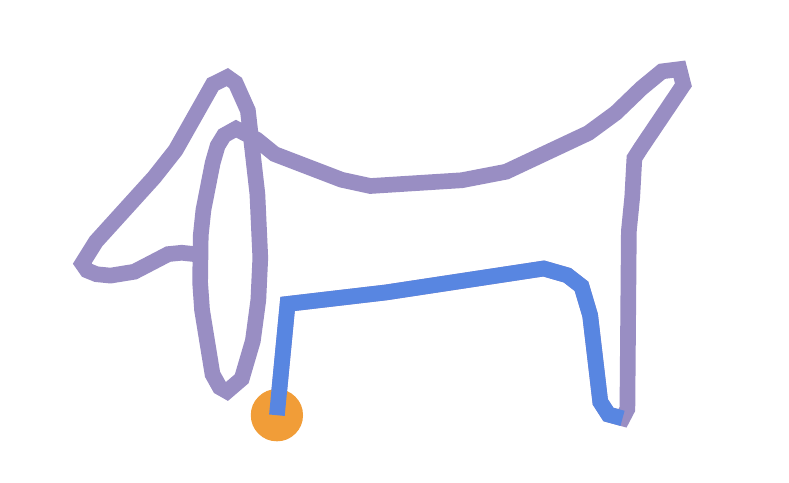}
        \caption{Dog}
        \label{fig:usertest_examples_dog}
    \end{subfigure}
    \begin{subfigure}[b]{0.3\columnwidth}
        \includegraphics[width=\columnwidth]{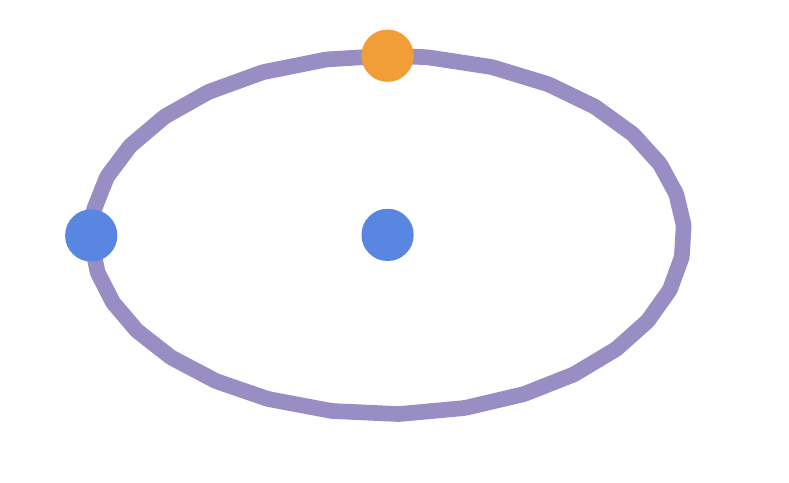}
        \caption{Ellipse}
    \end{subfigure}
    \caption{Shapes used in our user tests. 
    Note that the drawing surface only contained sparse visual references (shown in blue) and starting positions (orange).}
    \label{fig:usertest_examples}
\end{figure}

\subsection{Procedure}
We invited 12 participants ($M$=8; $F$=4, Age=28.2 $\pm$ 2.2) into our lab. All subjects were right-handed and did not have any professional drawing experience. 
Before commencing the experiment, users were given an introduction to the system functionality and got to experience the system in a self-timed training phase. Only once participants were reasonably confident in the system we continued with the experiment. 

During the experiment we asked each participant to draw six basic shapes, illustrated in \figref{fig:usertest_examples}. Each participant drew each shape with and without haptic feedback once. The presentation order of shapes and interface condition was counterbalanced. The drawing surface only contained a starting point and, in the case of more complex shapes, additional visual guidance (shown in red in \figref{fig:usertest_examples}). Furthermore, the participants were shown a scaled version during task execution (scaled to prevent 1:1 copying).

\begin{figure*}[!h]%
    \centering
    \begin{subfigure}[b]{.3\textwidth}
    \includegraphics[width=\columnwidth]{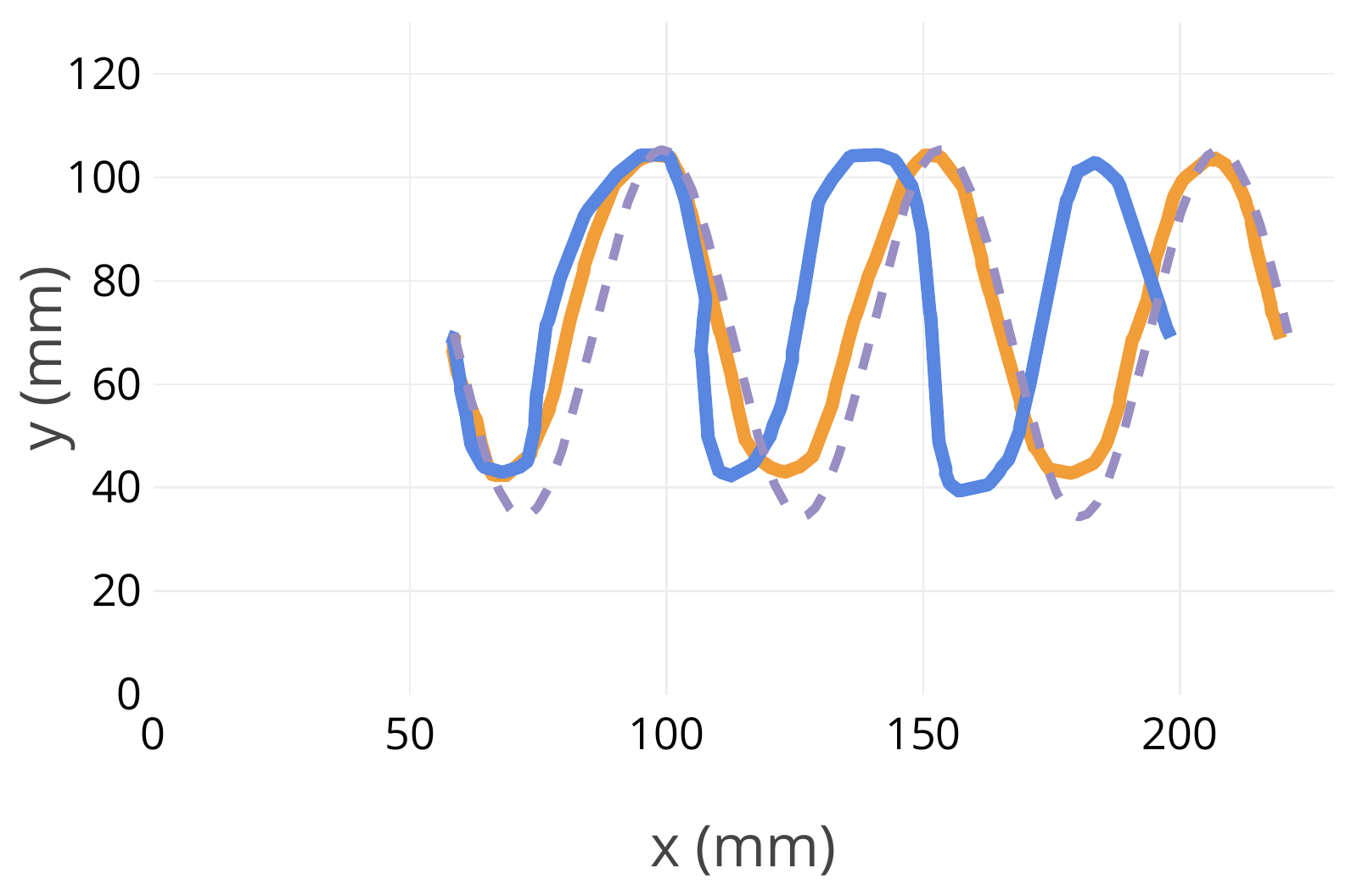}%
    \caption{Sinusoidal}%
    \label{fig:single_user_a}%
    \end{subfigure}\hfill%
    \begin{subfigure}[b]{.3\textwidth}
    \includegraphics[width=\columnwidth]{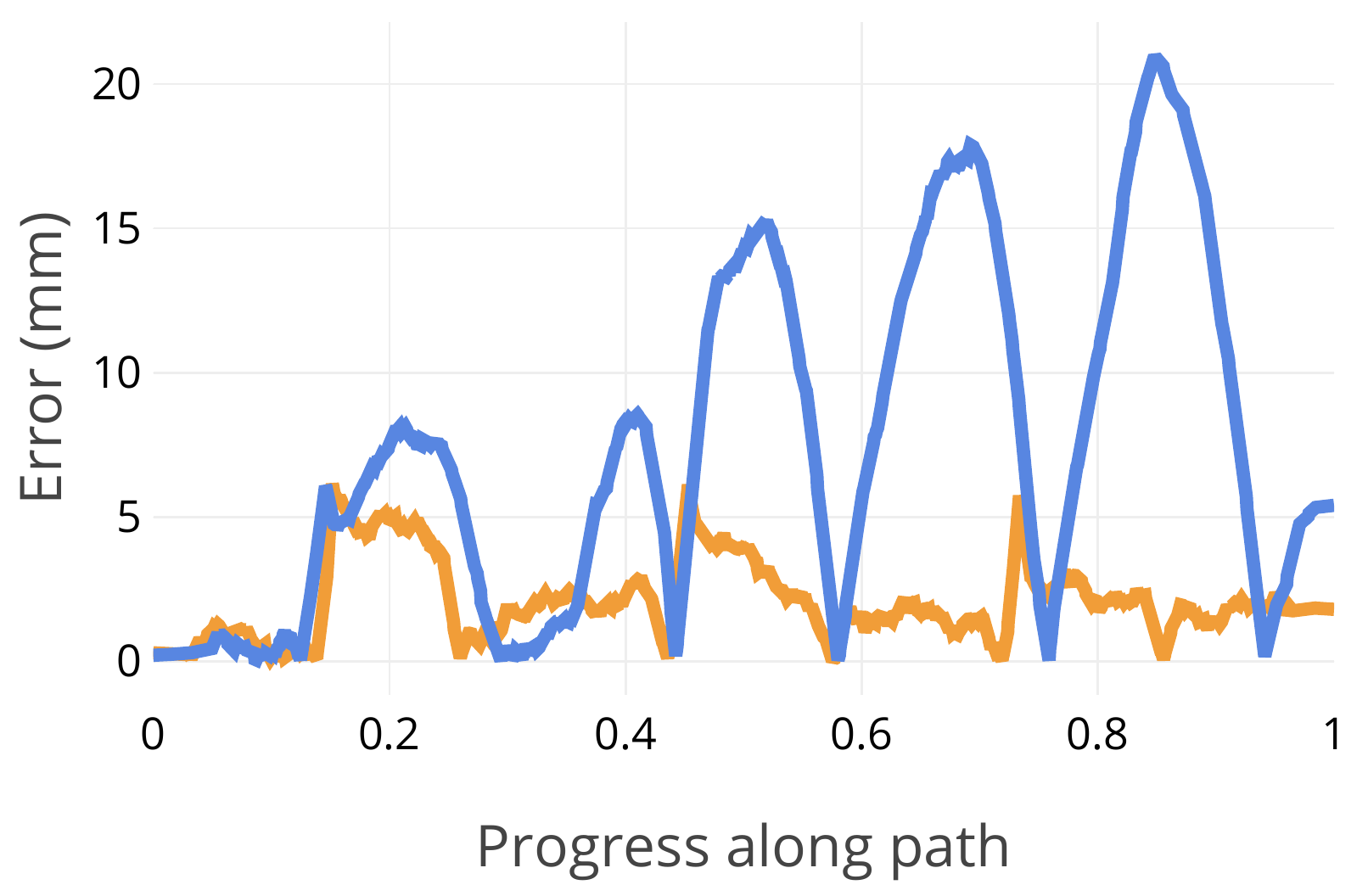}%
    \caption{Error over time}%
    \label{fig:single_user_b}%
    \end{subfigure}\hfill%
    \begin{subfigure}[b]{.3\textwidth}
    \includegraphics[width=\columnwidth]{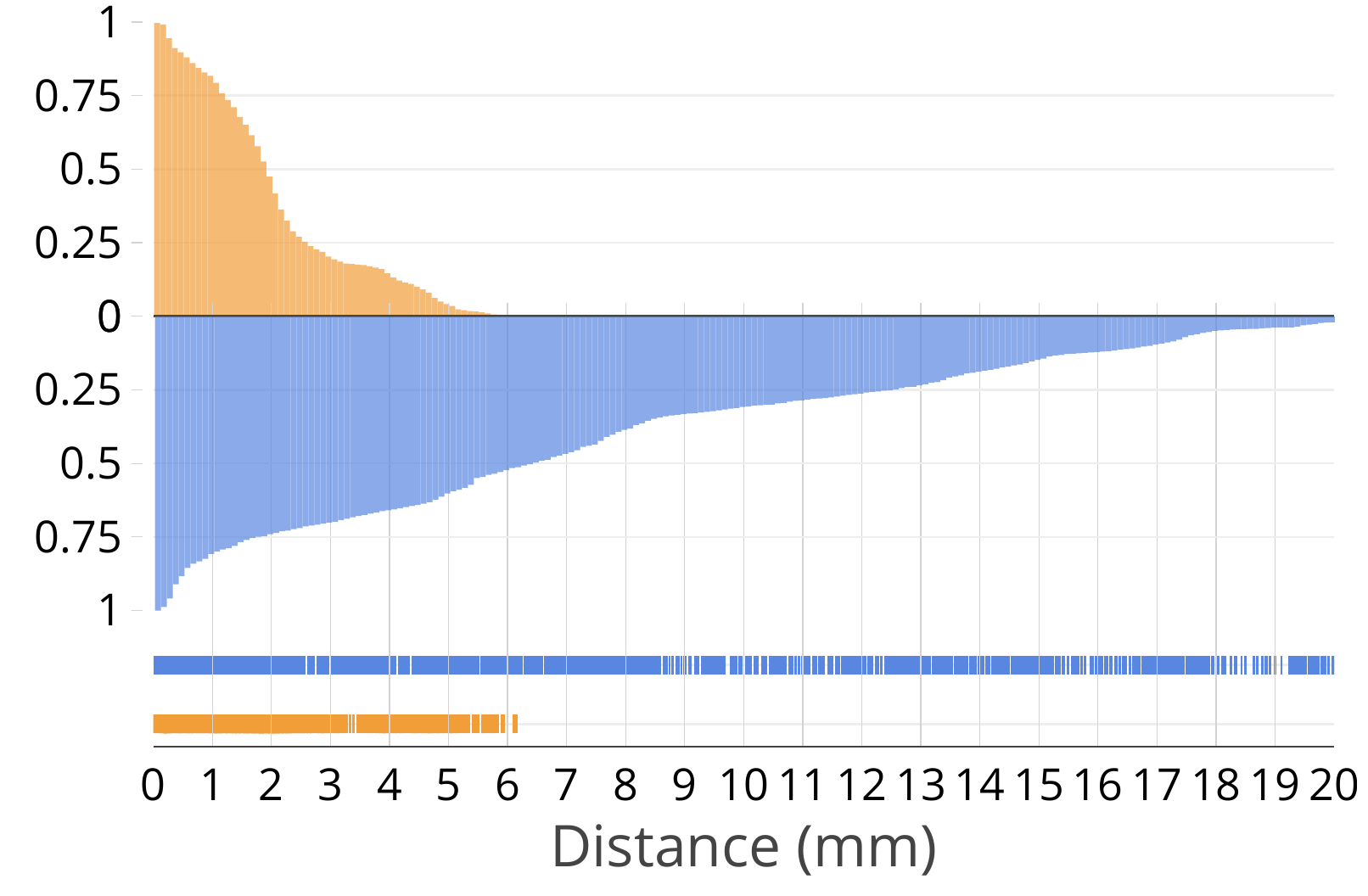}%
    \caption{Pen to reference error histogram.}%
    \label{fig:single_user_c}%
    \end{subfigure}
    \caption{Accuracy comparison for a single participant. \protect\subref{fig:single_user_a} reference shape (dotted line) overlaid by path drawn by the same user with (orange) and without (blue) haptic guidance. The absolute error increases over time without error correction \protect\subref{fig:single_user_b}. Error histogram reveals more compact distribution skewed towards low errors.} 
    \label{fig:single_user}
\end{figure*}

\subsection{Results}
\subsubsection*{Quantitative Results}
\label{sc:quantitative_results}
We first analyze the results quantitatively. We use a Hausdorff-like distance \cite{rockafellar2009variational} as error metric. To make the metric robust to drawing speed, we re-sample the drawn path and the reference equidistantly. We then compute the distance from all re-sampled points to the closest point on the reference. To ensure fairness we also compute the distance from reference to the drawn path. A Kolmogorov-Smirnov test \cite{kolmogorov1933sulla} indicates that both sets are the same and we report uni-directional distances. 

\begin{figure}[!t]
    \centering
    \includegraphics[width=\columnwidth]{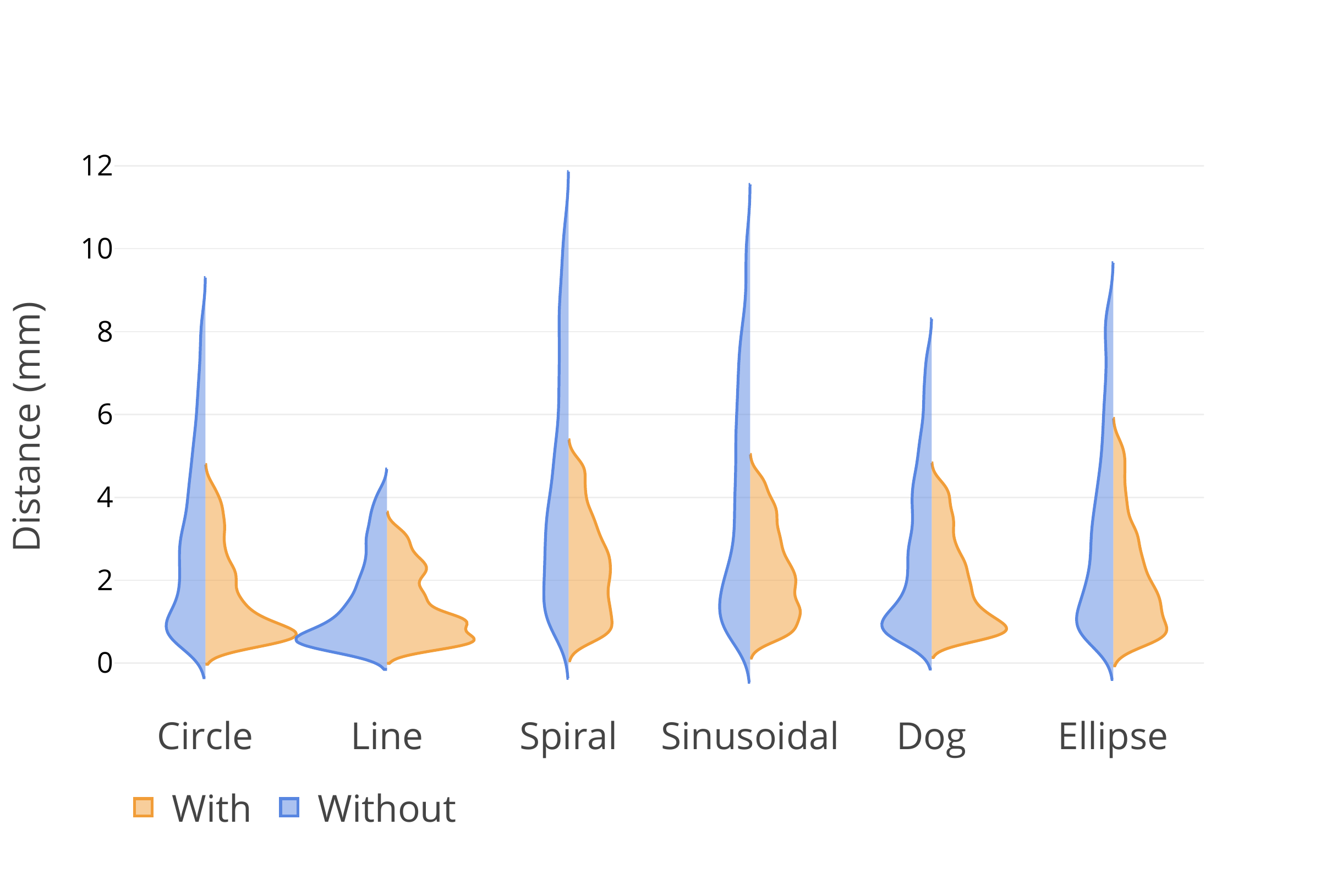}
    \caption{Distribution of pen to reference distances. All subjects combined. The entries have been trimmed for 10\% on both the upper and lower limit in order to increase readability.}
    \label{fig:results_summary}
\end{figure}

\figref{fig:single_user_a} compares the reference (dotted-line) to a sinusoidal drawn without (blue) and with haptic feedback (orange) by one of our participants. The path drawn with guidance clearly stays closer to the reference and drifts less. Plotting the error over time confirms this observation (\figref{fig:single_user_b}), where the error with guidance stays more or less constant and the guidance-free error continuously increases. \figref{fig:single_user_c}, plots the error histogram for both conditions showing a longer tail without haptic assistance. This trend holds for all users as can be seen in \ref{fig:results_summary}, plotting the pen-reference distances for all users. 

We conducted a two-way ANOVA on the mean error (computed over all users), as metric for accuracy. Results show a main effect for the feedback type (F=46.187, p<.001) and for the shapes (F=11.771, p<.001). Post-hoc analysis reveals  that the line is statistically significantly different from all other shapes and we report its results separately. Mean accuracy per shape and significance levels are summarized in in Table \ref{tab:accuracy}, showing that haptic feedback significantly improves accuracy across shapes to \unit[1.871]{mm} (p<.001). While there is no significant difference for the line, even when including it, we attain a significant accuracy improvement by \unit[1.537]{mm} (p<.001).

\begin{table}[h]
\caption{Mean accuracy in mm. Percentage of error: avg(with)/avg(without) -- lower is better. Significance values set to: *p<0.05, **p<0.01, ***p<0.005}
    \begin{tabular}{l|cc|cc|c}
    \multicolumn{1}{c}{} &\multicolumn{2}{c}{With}&\multicolumn{2}{c}{Without}&\multicolumn{1}{c}{}\\
    \midrule
Scenario& Mean & SD & Mean & SD & Err $\%$ \\
 \midrule
 Circle* & \textbf{2.19} & 0.90 & 4.26 & 2.39 & 0.51 \\
 Line  & 1.18 & 0.80 & \textbf{1.03} & 0.84 & 1.15  \\
 Spiral*** & \textbf{2.55} & 0.75 & 4.38 & 1.64 & 0.58  \\
 Sinus***  & \textbf{2.53} & 0.70 & 5.08 & 2.19 & 0.50  \\
 Dog***  & \textbf{2.31} & 0.54 & 3.81 & 1.32 & 0.60 \\
 Ellipse***  & \textbf{2.40} & 0.56 & 3.84 & 1.22 & 0.62 \\
\end{tabular}
\label{tab:accuracy}
\end{table}

\subsubsection*{Qualitative Results}
A brief exit interview (see Table \ref{tab:survey}) shows that users subjectively rate the system favourably in terms of accuracy, speed, force and overall performance on a 5-point Likert scale. 

Finally, we qualitatively show the effect of using haptic feedback in
Figure \ref{fig:qualitative_results} using different shapes drawn by different participants. The more complex the shape, the more pronounced the difference between the conditions (cf. circle, spiral, dog).

\begin{table}[htb]
    \centering
    \caption{Survey results. Likert scale: 1=Not Accurate/Slow/Weak/No Improvement. 5=Very Accurate/Fast/Strong/Much Improvement.}
    \begin{tabular}{l|cccc}
    
    Question & Accuracy & Speed &Force &Improvement \\
    \midrule
    Mean    & 4.33 & 4.00 &3.50 &4.50  \\
    SD   & 0.62 & 0.91 & 0.86 & 0.90 \\
    \end{tabular}
    \label{tab:survey}
\end{table}

\section{Additional Results}
\begin{figure}[t]
    \centering
    \begin{subfigure}[b]{0.45\columnwidth}
        \includegraphics[width=\columnwidth]{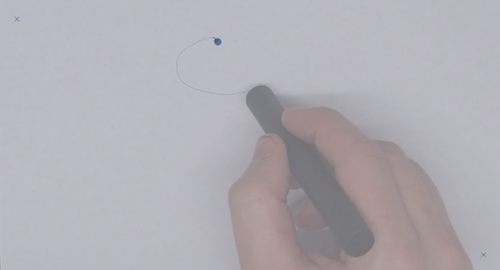}
            \caption{}
    \label{fig:g_start}
    \end{subfigure}
    \begin{subfigure}[b]{0.45\columnwidth}
        \includegraphics[width=\columnwidth]{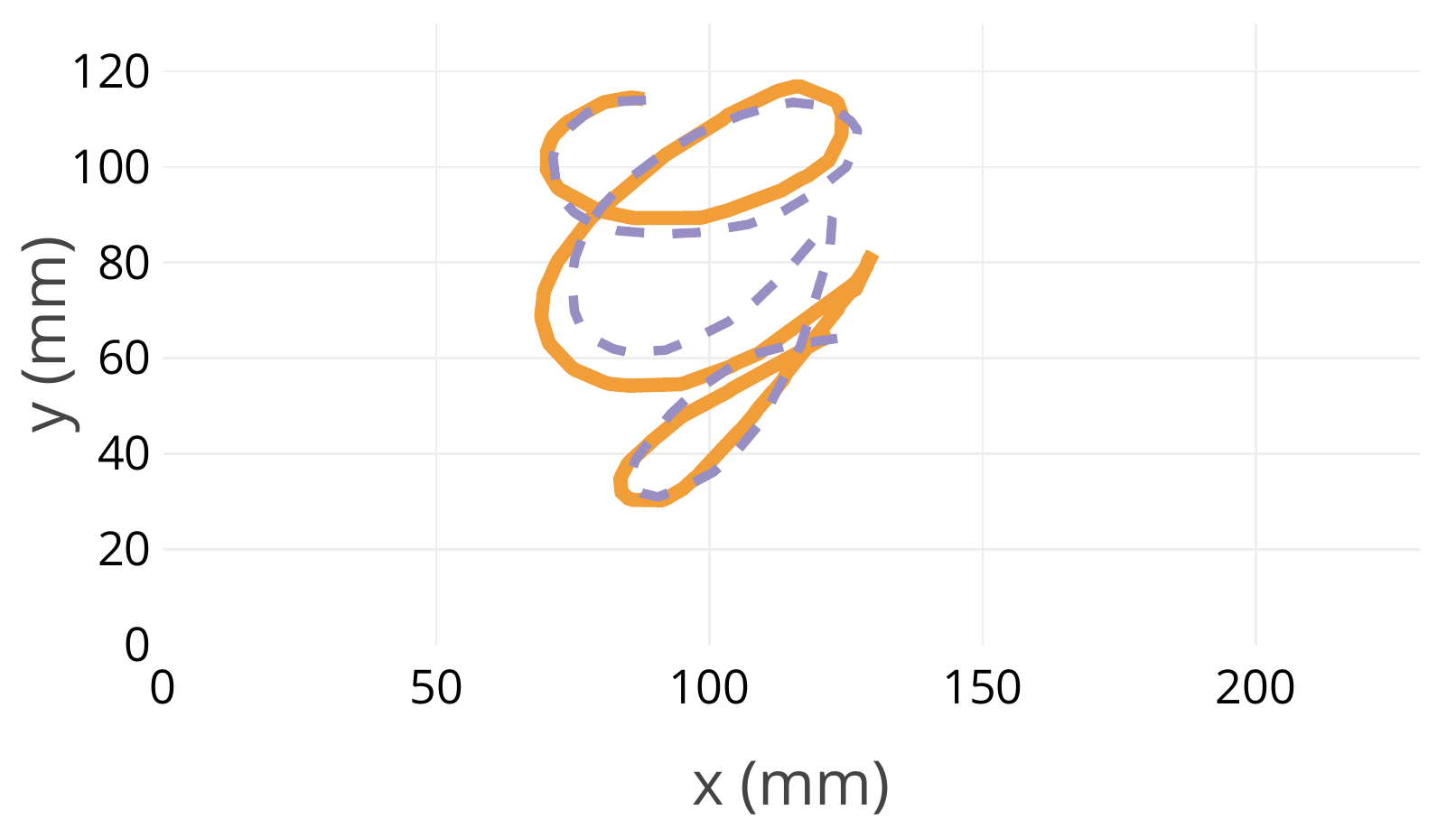}
            \caption{}
    \label{fig:g}
    \end{subfigure}
    \begin{subfigure}[b]{0.45\columnwidth}
        \includegraphics[width=\columnwidth]{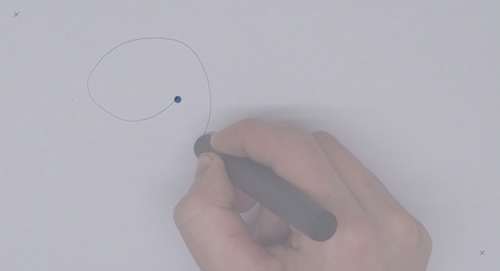}
            \caption{}
    \label{fig:w_start}
    \end{subfigure}
    \begin{subfigure}[b]{0.45\columnwidth}
        \includegraphics[width=\columnwidth]{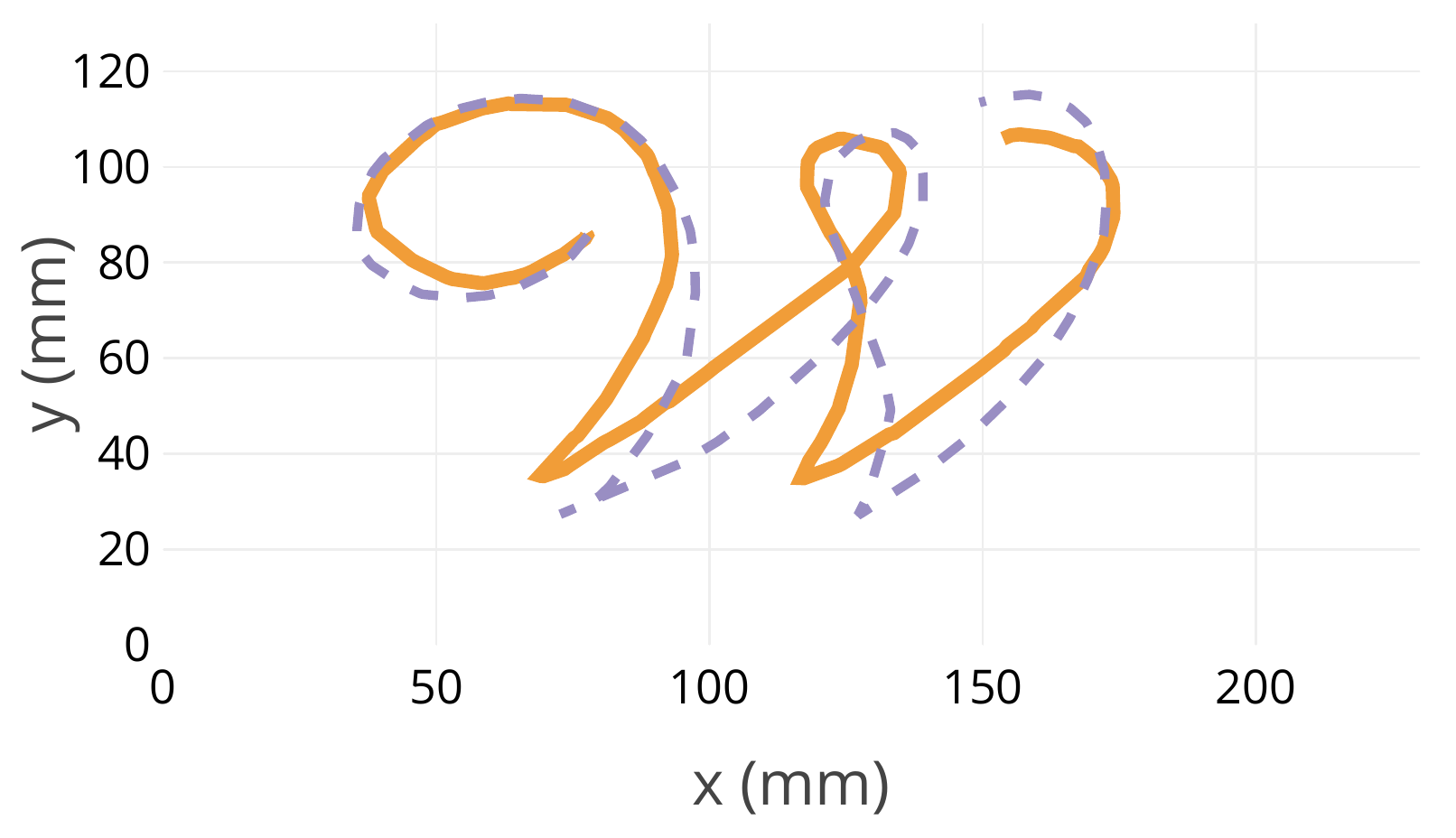}
            \caption{}
    \label{fig:w}
    \end{subfigure}
    \begin{subfigure}[b]{0.45\columnwidth}
        \includegraphics[width=\columnwidth]{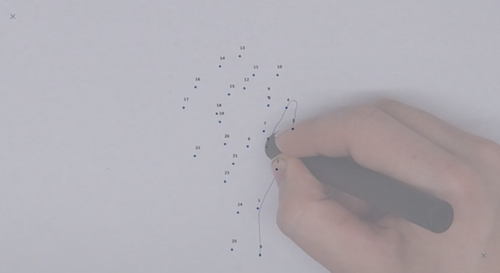}
            \caption{}
    \label{fig:dots_start}
    \end{subfigure}
    \begin{subfigure}[b]{0.45\columnwidth}
        \includegraphics[width=\columnwidth]{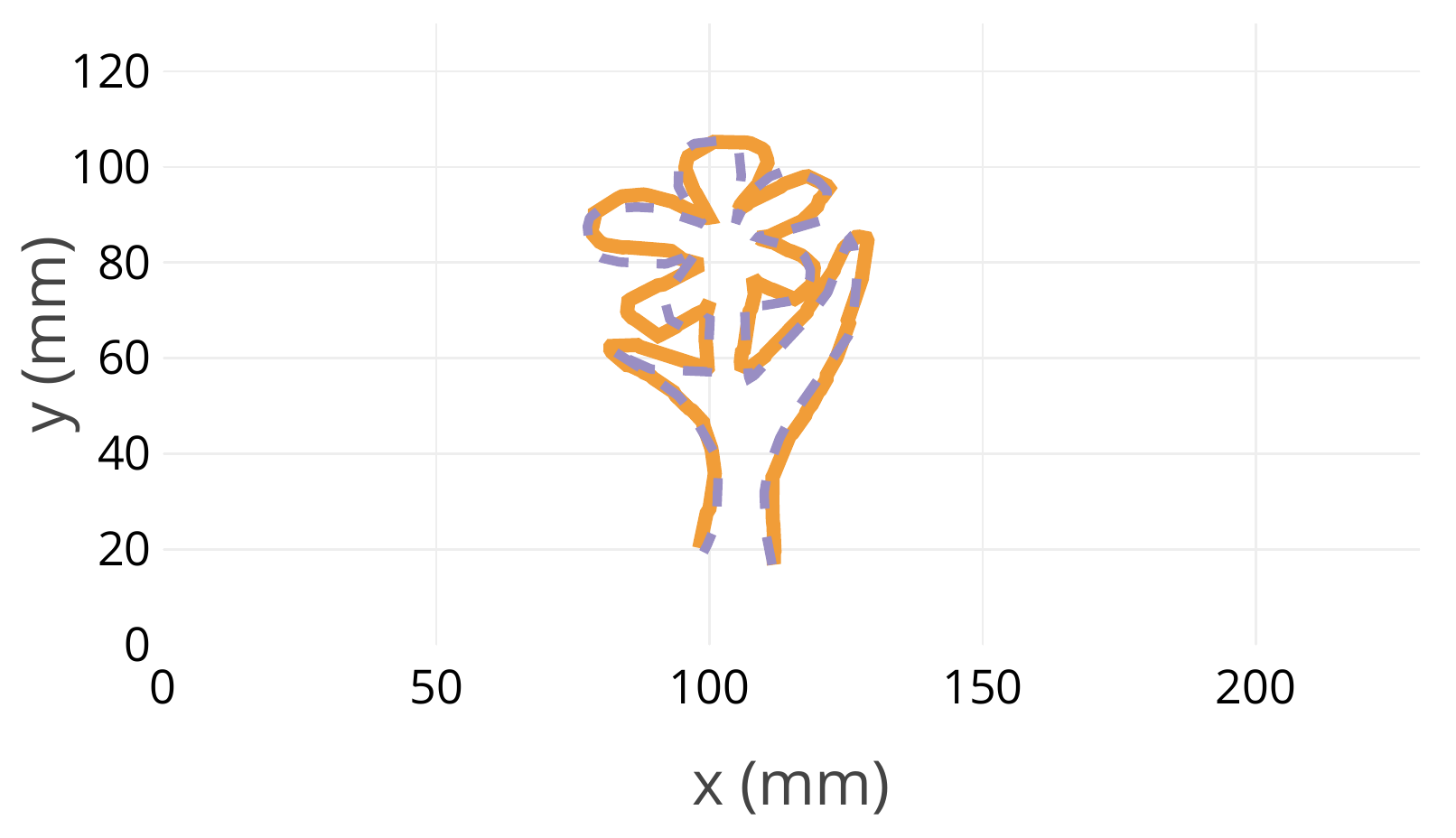}
            \caption{}
    \label{fig:dots}
    \end{subfigure}
    \caption{Overview of use cases: calligraphy (\ref{fig:g} and \ref{fig:w}) and drawing exercises (\ref{fig:dots}). Figures \ref{fig:g_start}, \ref{fig:w_start}, \ref{fig:dots_start} show the guidance given during the experiment.}
    \label{fig:usertest_additional}
\end{figure}
To further demonstrate the capabilities of our system we illustrate potential use-cases including applications in learning to draw, in outlining and in inking.  

\subsubsection*{Calligraphy:}
\figref{fig:g} and \figref{fig:w} illustrate writing of flourished characters, with only minimal visual guidance (single starting point). Although an offset from the reference path remains, the lines are smooth and the overall shape is close to the desired characters. 

\subsubsection*{Drawing teaching aid:}
Connect-the-dots exercises are often used to teach children motor skills as well as stroke ordering. \figref{fig:dots} shows results from a similar exercise performed with our system, yet using much fewer dots as visual guidance than a paper version.

\subsubsection*{Outlining \& inking:}
\begin{figure}[!t]
    \centering
    \begin{subfigure}[b]{0.49\columnwidth}
        \includegraphics[trim={0 50 0 0},clip,trim={0 50 0 0},clip,width=\columnwidth, angle=-90, scale=0.8]{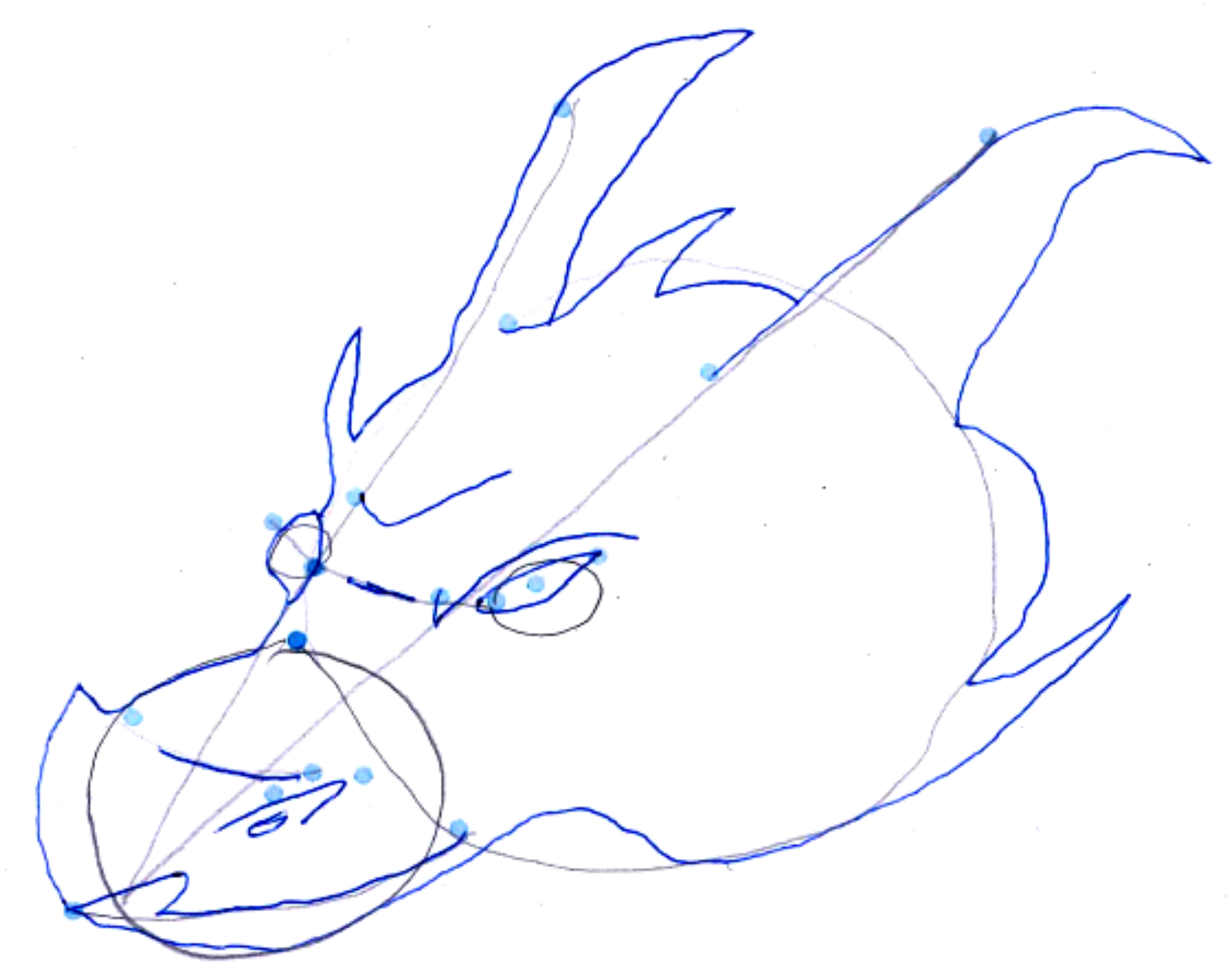}
    \end{subfigure}
    \begin{subfigure}[b]{0.49\columnwidth}
        \includegraphics[trim={0 50 0 0},clip,width=\columnwidth, angle=-90, scale=0.8]{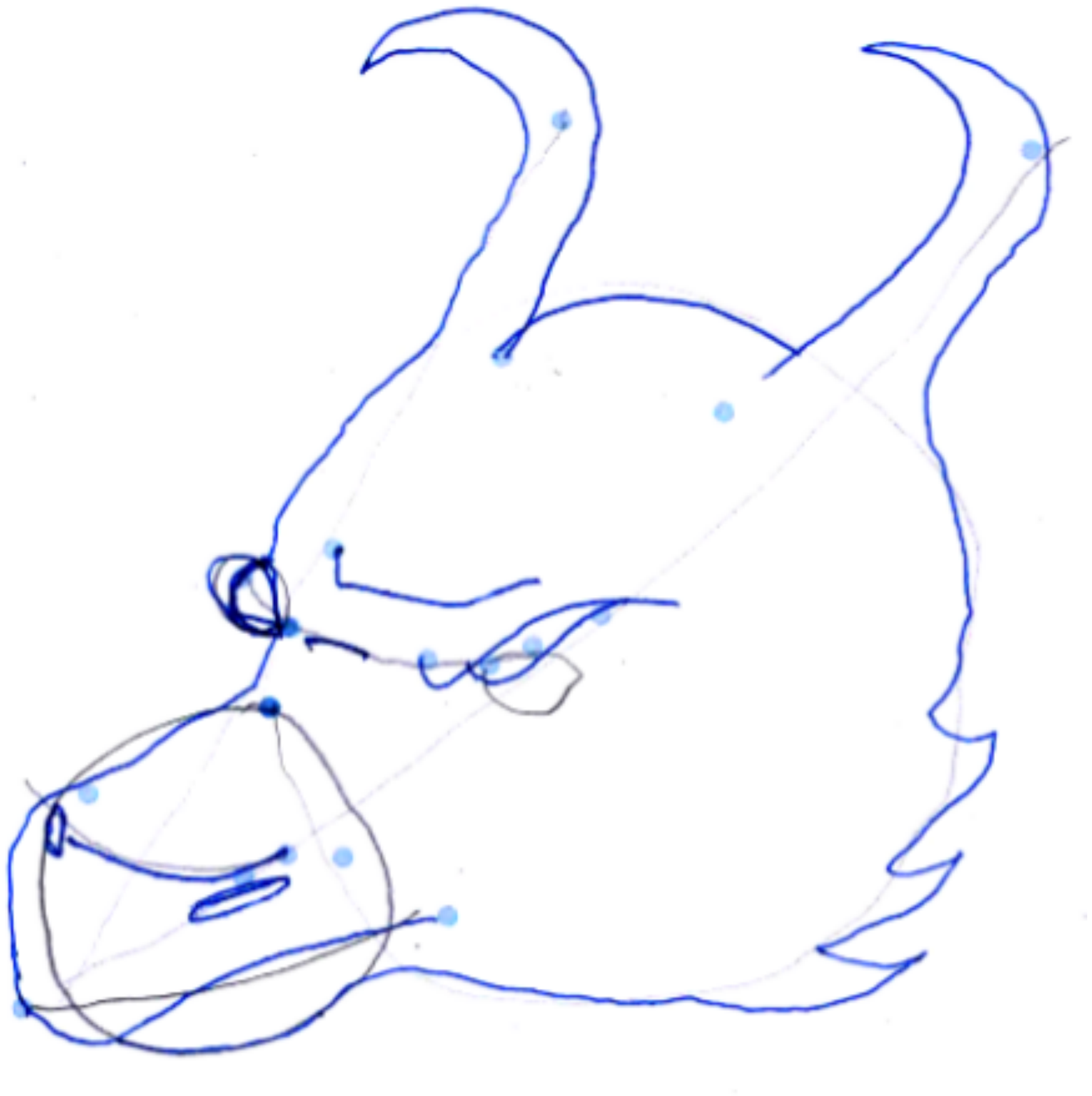}
    \end{subfigure}
    \begin{subfigure}[b]{0.49\columnwidth}
        \includegraphics[trim={0 50 0 0},clip,width=\columnwidth, angle=-90, scale=0.8]{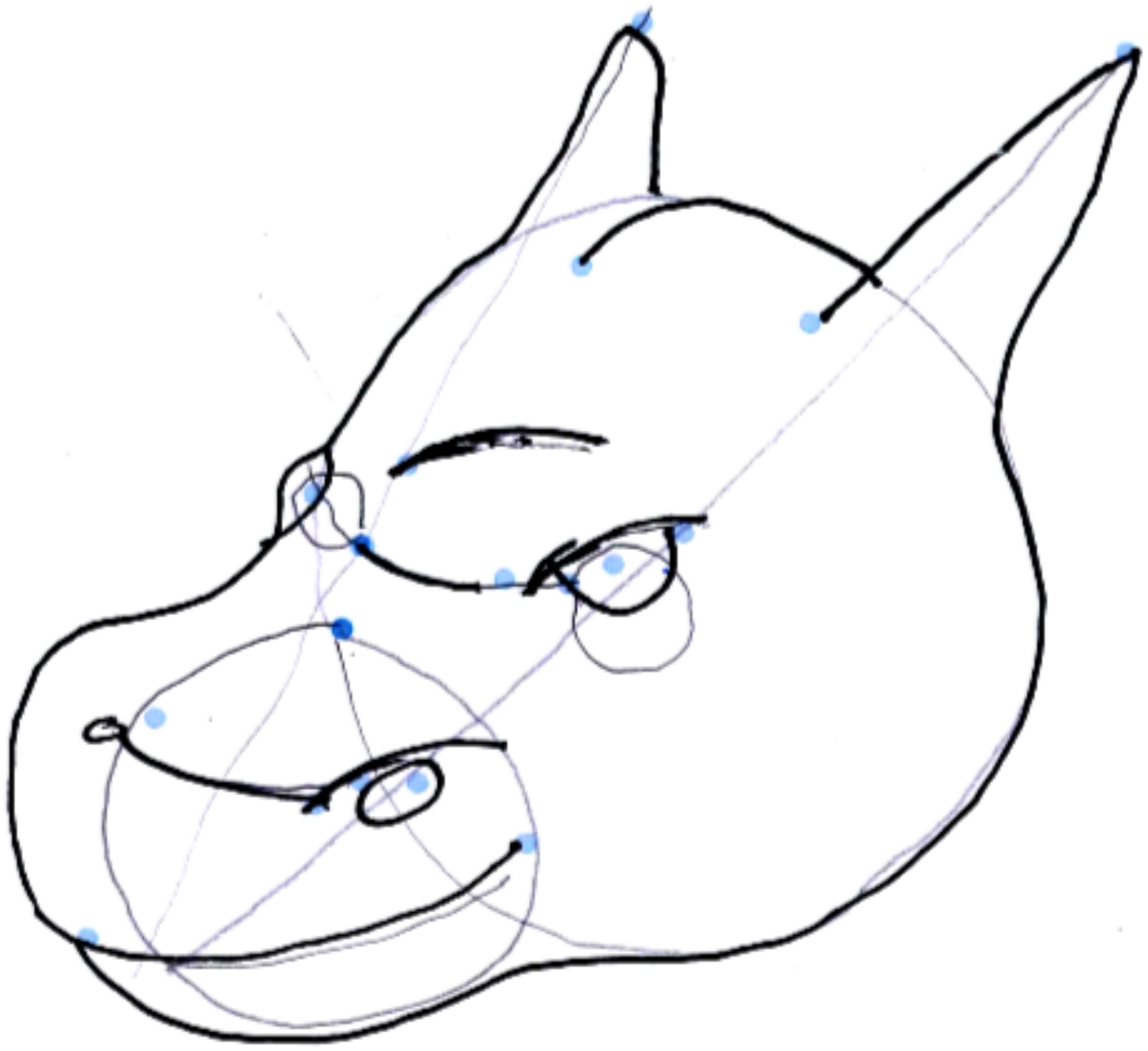}
    \end{subfigure}
    \begin{subfigure}[b]{0.49\columnwidth}
        \includegraphics[trim={0 50 0 0},clip,width=\columnwidth, angle=-90,scale=0.8]{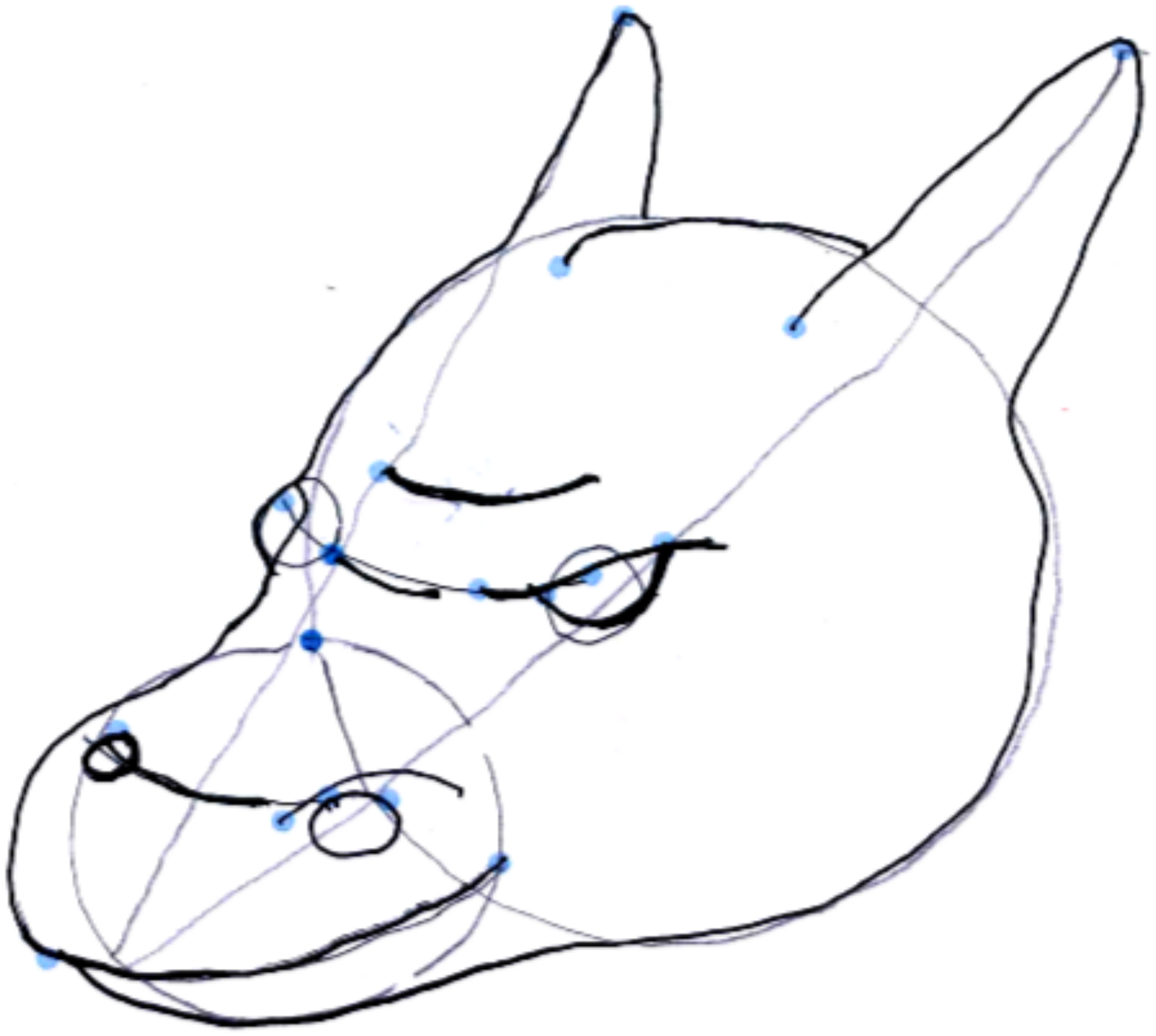}
    \end{subfigure}
    \caption{Different variants of the same dragon, drawn with identical system settings \hl{by a novice}. Each pair of drawings used with different tools. First a pencil for proportions and a fine-liner (top) or pencil (bottom) to ink-in details. \hl{Multi-stroke lines are achieved by approaching each seperate instance as a new figure, the system is trigger by a pen lift.}}
    \label{fig:dragon}
\end{figure}

\figref{fig:dragon} illustrates the effect of two core capabilities of the proposed approach. Here we first outline the proportions of the dragon head and then use different pens to ink-in the details. Note that the system provides haptic guidance but allows the user to draw the shape in different styles and with varying high-frequency detail, while maintaining similarity to the reference shape. \hl{This is a direct consequence of using time-free closed loop control approach, as is alluded to in Sec.} \ref{Sc:preliminary_user_evaluation}. In this case, all four variants were drawn without changes to the system or desired path.

\section{Discussion}
\edt{From a technical perspective we can we can conclude our time-independent closed-loop control formulation, provides qualitatively and quantitatively better results than simpler approaches. Open-loop control might lead to complete loss of feedback due to differences in speed. This is solved in a closed-loop setting. However, time-dependent implementations can maintain haptic feedback but the perceived direction of the  feedback maybe wrong. This is solved by our time-independent implementation, where} $\mathbf{s}(\theta)$ \hl{is part of the optimization problem.}

Our \hl{haptic feedback} experiments indicate that the proposed approach indeed increases accuracy in drawing tasks and that users rate the system favorably. We did not find a significant difference for the straight line. This may be explained by user feedback, that the maximum speed of the linear stage is a limiting factor in the current implementation. We leave faster magnet positioning for future work. 

Two aspects from the exit interviews are noteworthy. First, there is a high standard deviation in how users rated the perceived force. We hypothesize that this is due to the way users operate the pen, with some leveraging the full arm and others rely more on the wrist. We note that our palm rejection implementation is simple and may have contributed to this. Furthermore, some users indicated that they had the feeling that their drawings without feedback were more accurate once they experienced the haptic guidance, indicating the possibility of short-term muscle memory. Long-term learning however is difficult to evaluate experimentally and goes beyond the scope of this paper. 

\begin{figure}[t]
    \centering
    \begin{subfigure}[b]{0.49\columnwidth}
    \includegraphics[trim={0 20 0 80},clip,width=\columnwidth]{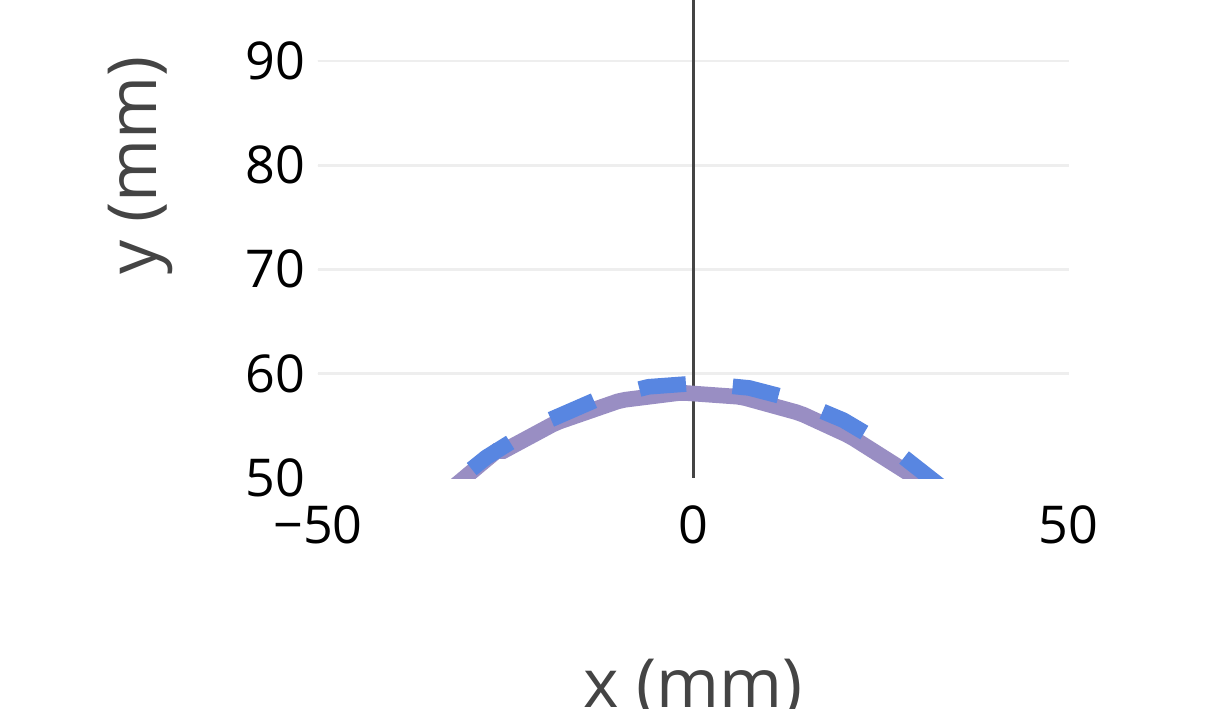}
    \caption{\unit[36.9]{$^\circ$}}\label{subfig:angle_error_a}
    \end{subfigure}
    \begin{subfigure}[b]{0.49\columnwidth}
    \includegraphics[trim={0 20 0 80},clip,width=\columnwidth]{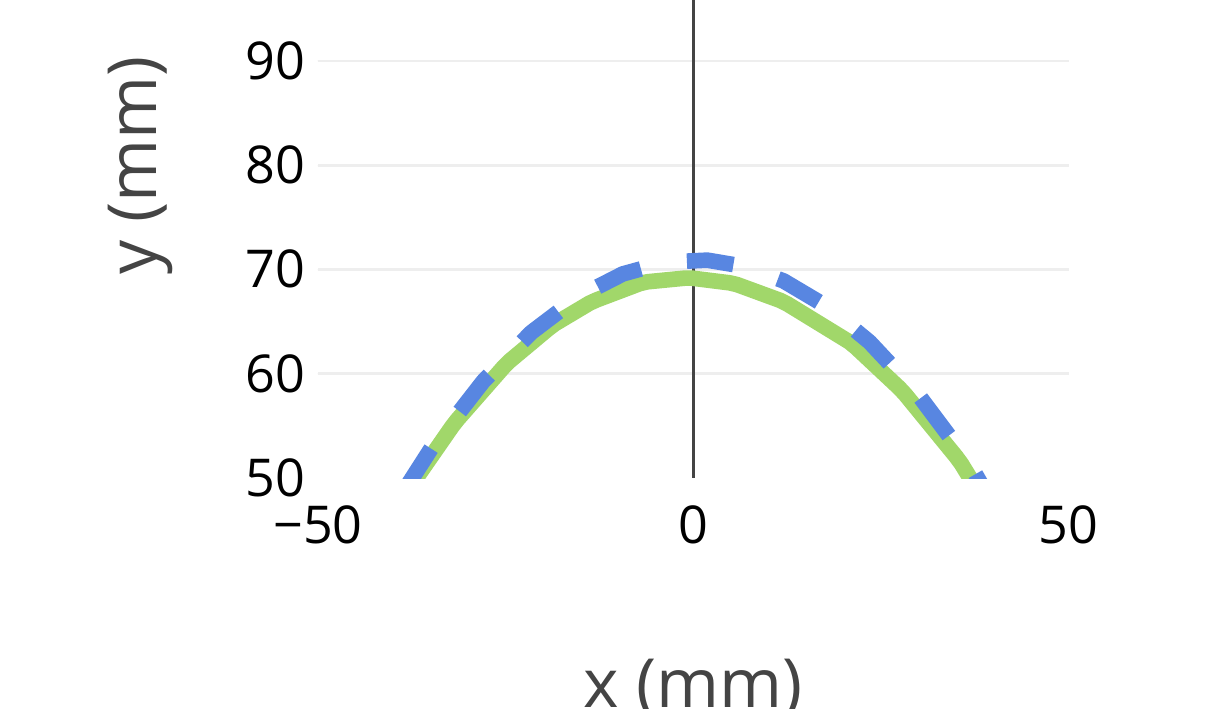}
    \caption{46.1$^\circ$}\label{subfig:angle_error_b}
    \end{subfigure}
    \begin{subfigure}[b]{0.49\columnwidth}
    \includegraphics[trim={0 20 0 80},clip,width=\columnwidth]{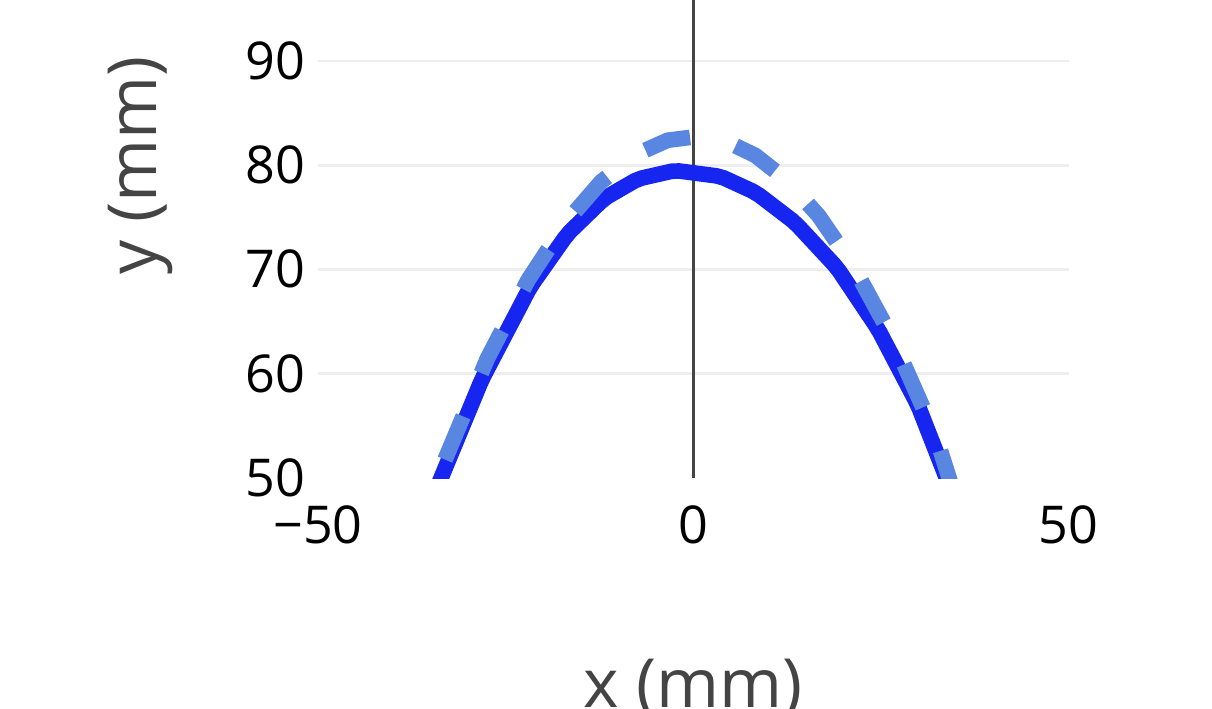}
    \caption{57.1$^\circ$}\label{subfig:angle_error_c}
    \end{subfigure}
    \begin{subfigure}[b]{0.49\columnwidth}
    \includegraphics[trim={0 20 0 80},clip,width=\columnwidth]{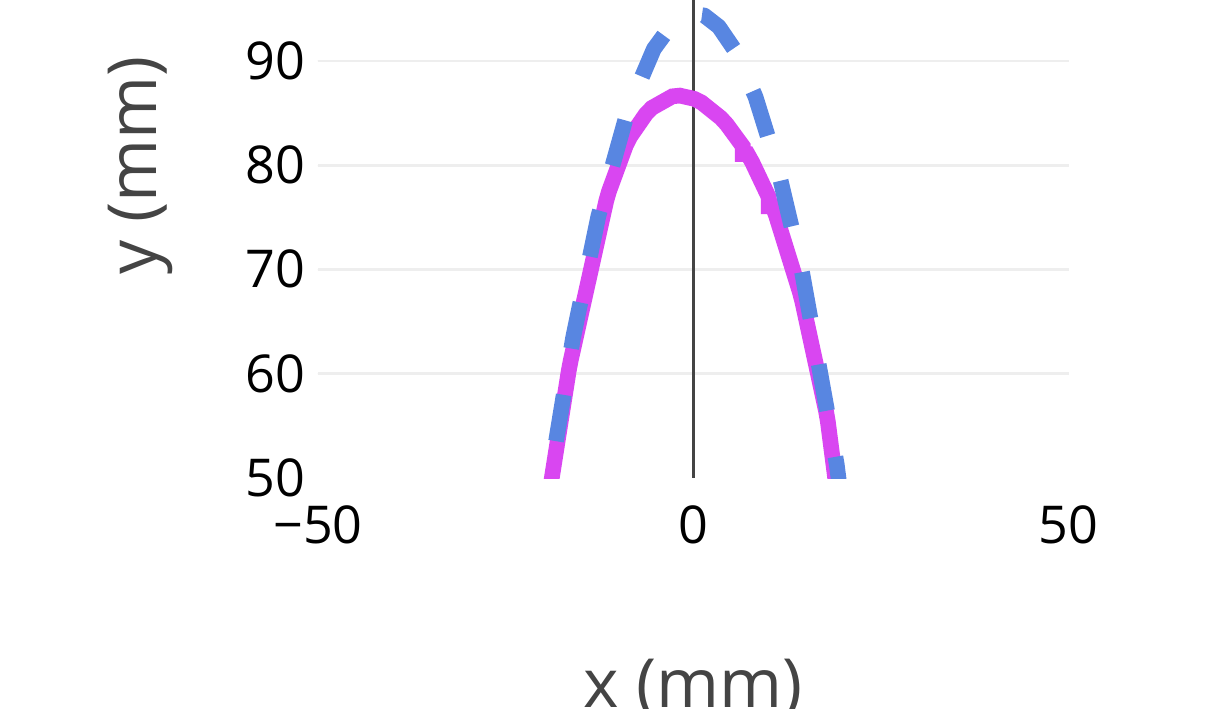}
    \caption{73.3$^\circ$}\label{subfig:angle_error_d}
    \end{subfigure} \\
    \begin{subfigure}[b]{\columnwidth}
    \includegraphics[trim={0 20 0 0},clip,width=\columnwidth]{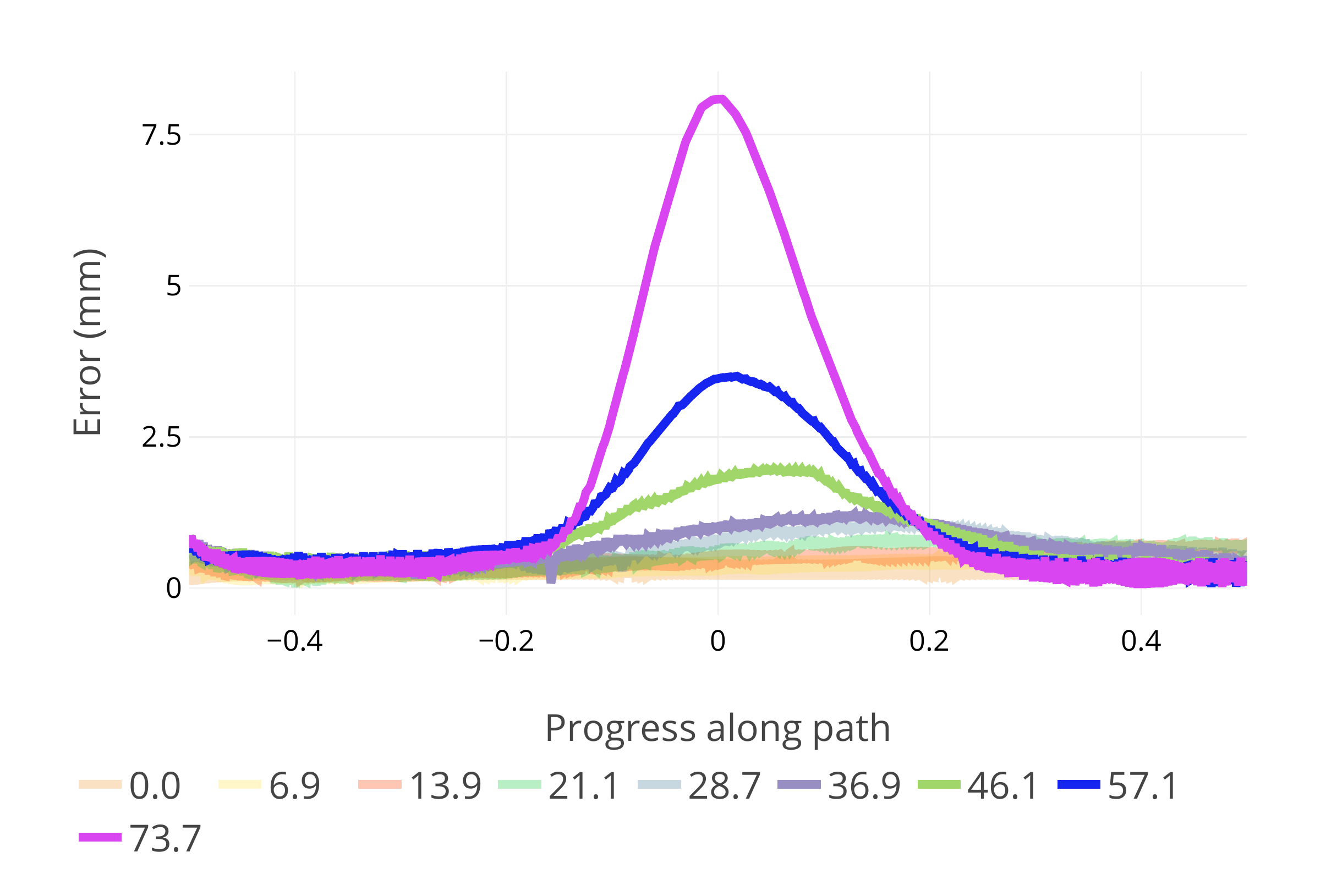}
    \end{subfigure}
    \caption{Curvature dependent error. Insets \protect\subref{subfig:angle_error_a}-\protect\subref{subfig:angle_error_c} show increasingly sharp cornered references (dotted lines) and the resulting pen trajectory (in simulation).  x-axis is normalized for cord length, so that all angles can be directly compared. Bottom: error over 9 different levels of curvature (in degrees).}
    \label{fig:angle_error}
\end{figure}

Finally, during our experiments we noticed a tendency to cut corners, well illustrated in the case of the sinusoidal (see \figref{fig:single_user_a}, \ref{fig:qualitative_results_e}). To unpack this issue further, we performed an experiment in simulation, with the user model from Eq. \ref{eq:user_sim}, tracing references with increasingly sharp angles (see \figref{fig:angle_error}). The plot clearly shows an increase in error with increase in curvature. It has been shown that humans trade-off speed and accuracy in tracing tasks and that they slow down when tracing high-curvature paths \cite{accot1997beyond}. Currently our implementation does not take curvature of the reference into account but it would be straightforward to penalize the progress $\theta$ along the reference according to its curvature.

\section{Limitations \& Future Work}
There are, of course, some limitations to our current approach. First, the maximum speed of the linear stage we used is relatively slow. This causes users to slow down their drawing speed. This could be overcome via a faster bi-axial linear stage, which are commercially available but expensive. A potentially more interesting and scalable direction would be to extend the proposed magnet model towards a grid of electromagnets. However, the interaction between several overlapping EM fields with a moving permanent magnet are non-trivial to model and would require significant research. 

Once the hardware-induced speed limitation is overcome, efficient closed-loop control approaches become an interesting direction for future work, since faster pen motion would also tighten the latency and accuracy budget. In the context of sensing it would be interesting to incorporate a mechanism to reconstruct the tilt of the pen. This could be achieved for example via an accelerometer built into the pen or via a grid of hall-sensors underneath the surface. Information on the pen tilt could then be combined with the angle dependent formulation of our EM model (see Appendix \ref{sc:ap.dipole.eq}). Furthermore, we believe there are many research opportunities in combining our approach with other sketch and ink beautification approaches (e.g., \cite{simo2016learning,simo2018mastering,xing2015autocomplete}). Particularly interesting would be to leverage fully predictive models (e.g., \cite{Aksan:2018:DeepWriting}) in order to overcome the need for a known reference. 

Another interesting direction of research is connected to the observation that different users perceive the feedback at different strength levels. \hl{This could be due to grip strength, or movement from the shoulder rather than wrist.} A stronger (i.e., bigger) electromagnet could increase the dynamic range. However, this would have to be carefully counterbalanced with weight and heat dissipation concerns as well as with a loss in accuracy towards the center of the electromagnet (the force goes to zero as $\mathbf{r_d}\rightarrow0$). 

We believe that it could be an interesting direction for future work to combine our approach with different types of haptic feedback, either environment mounted or body-worn. Moreover, we have so far focused our attention towards drawing applications. However, electromagnetic feedback in combination with spatial actuation maybe interesting in other settings. For example, a magnet mounted to a robotic arm could deliver contact-less feedback in VR scenarios. It would also be interesting to investigate how to best exploit the system capabilities in the context of motor memory and learning.

\section{Conclusions}
We have proposed \systemname{}, a system that delivers dynamically adjustable guidance in drawing and sketching tasks. We have detailed our hardware setup and discussed a novel model of the electromagnetic interactions in the system. The proposed model can be evaluated analytically and is hence suitable for iterative, real-time optimization approaches. We have furthermore demonstrated that the assumptions of dipole magnets and an upright pen lead only to a small approximation error. However, we have included a angle dependent formulation that maybe used in future work, where pen-tilt information is available. 

We have also discussed a novel optimization scheme based on the MPCC framework that leverages the EM model, in order to optimize the system states and its inputs over a receding horizon via solving a stochastic optimal control problem at each timestep. Our formulation has been designed to provide dynamically adjustable forces and automatically adjusts magnet position and strength. 

Our experiments have shown that the \hl{proposed} hardware-software solution is effective in improving accuracy and in guiding users in a variety of drawing tasks, without taking away agency and control from the user. We believe this is an interesting first step towards many exciting applications of electromagnetic haptic feedback. In order to foster future research we will release all hardware schematics and software source code to the public. 

\begin{acks}
This work was supported in part the Hasler Foundation (Switzerland), ERC (OPTINT StG-2016-717054) NSF (IIS-CAREER 1652515 and OAC:1835712), a gift from nTopology, and a gift from Adobe Research. We thank all participants for taking part in our experiments.
\end{acks}

\bibliographystyle{ACM-Reference-Format}
\bibliography{bibliography.bib}

%%% -*-BibTeX-*-
%%% Do NOT edit. File created by BibTeX with style
%%% ACM-Reference-Format-Journals [18-Jan-2012].

\begin{thebibliography}{45}

%%% ====================================================================
%%% NOTE TO THE USER: you can override these defaults by providing
%%% customized versions of any of these macros before the \bibliography
%%% command.  Each of them MUST provide its own final punctuation,
%%% except for \shownote{}, \showDOI{}, and \showURL{}.  The latter two
%%% do not use final punctuation, in order to avoid confusing it with
%%% the Web address.
%%%
%%% To suppress output of a particular field, define its macro to expand
%%% to an empty string, or better, \unskip, like this:
%%%
%%% \newcommand{\showDOI}[1]{\unskip}   % LaTeX syntax
%%%
%%% \def \showDOI #1{\unskip}           % plain TeX syntax
%%%
%%% ====================================================================

\ifx \showCODEN    \undefined \def \showCODEN     #1{\unskip}     \fi
\ifx \showDOI      \undefined \def \showDOI       #1{#1}\fi
\ifx \showISBNx    \undefined \def \showISBNx     #1{\unskip}     \fi
\ifx \showISBNxiii \undefined \def \showISBNxiii  #1{\unskip}     \fi
\ifx \showISSN     \undefined \def \showISSN      #1{\unskip}     \fi
\ifx \showLCCN     \undefined \def \showLCCN      #1{\unskip}     \fi
\ifx \shownote     \undefined \def \shownote      #1{#1}          \fi
\ifx \showarticletitle \undefined \def \showarticletitle #1{#1}   \fi
\ifx \showURL      \undefined \def \showURL       {\relax}        \fi
% The following commands are used for tagged output and should be
% invisible to TeX
\providecommand\bibfield[2]{#2}
\providecommand\bibinfo[2]{#2}
\providecommand\natexlab[1]{#1}
\providecommand\showeprint[2][]{arXiv:#2}

\bibitem[\protect\citeauthoryear{Accot and Zhai}{Accot and Zhai}{1997}]%
        {accot1997beyond}
\bibfield{author}{\bibinfo{person}{Johnny Accot} {and} \bibinfo{person}{Shumin
  Zhai}.} \bibinfo{year}{1997}\natexlab{}.
\newblock \showarticletitle{Beyond Fitts' law: models for trajectory-based HCI
  tasks}. In \bibinfo{booktitle}{\emph{Proceedings of the ACM SIGCHI Conference
  on Human factors in computing systems}}. ACM, \bibinfo{pages}{295--302}.
\newblock


\bibitem[\protect\citeauthoryear{Aguiar, Hespanha, and Kokotovic}{Aguiar
  et~al\mbox{.}}{2008}]%
        {AGUIAR2008}
\bibfield{author}{\bibinfo{person}{A.~Pedro Aguiar}, \bibinfo{person}{Joao~P.
  Hespanha}, {and} \bibinfo{person}{Petar~V. Kokotovic}.}
  \bibinfo{year}{2008}\natexlab{}.
\newblock \showarticletitle{Performance limitations in reference tracking and
  path following for nonlinear systems}.
\newblock \bibinfo{journal}{\emph{Automatica}} \bibinfo{volume}{44},
  \bibinfo{number}{3} (\bibinfo{year}{2008}), \bibinfo{pages}{598 -- 610}.
\newblock
\showISSN{0005-1098}
\urldef\tempurl%
\url{https://doi.org/10.1016/j.automatica.2007.06.030}
\showDOI{\tempurl}


\bibitem[\protect\citeauthoryear{Aksan, Pece, and Hilliges}{Aksan
  et~al\mbox{.}}{2018}]%
        {Aksan:2018:DeepWriting}
\bibfield{author}{\bibinfo{person}{Emre Aksan}, \bibinfo{person}{Fabrizio
  Pece}, {and} \bibinfo{person}{Otmar Hilliges}.}
  \bibinfo{year}{2018}\natexlab{}.
\newblock \showarticletitle{{DeepWriting: Making Digital Ink Editable via Deep
  Generative Modeling}}. In \bibinfo{booktitle}{\emph{SIGCHI Conference on
  Human Factors in Computing Systems}} \emph{(\bibinfo{series}{CHI '18})}.
  \bibinfo{publisher}{ACM}, \bibinfo{address}{New York, NY, USA}.
\newblock


\bibitem[\protect\citeauthoryear{{\AA}str{\"o}m and
  H{\"a}gglund}{{\AA}str{\"o}m and H{\"a}gglund}{1995}]%
        {aastrom1995pid}
\bibfield{author}{\bibinfo{person}{Karl~Johan {\AA}str{\"o}m} {and}
  \bibinfo{person}{Tore H{\"a}gglund}.} \bibinfo{year}{1995}\natexlab{}.
\newblock \bibinfo{booktitle}{\emph{PID controllers: theory, design, and
  tuning}}. Vol.~\bibinfo{volume}{2}.
\newblock \bibinfo{publisher}{Instrument society of America Research Triangle
  Park, NC}.
\newblock


\bibitem[\protect\citeauthoryear{Cho, Bianchi, Marquardt, and
  Bianchi-Berthouze}{Cho et~al\mbox{.}}{2016}]%
        {cho2016realpen}
\bibfield{author}{\bibinfo{person}{Youngjun Cho}, \bibinfo{person}{Andrea
  Bianchi}, \bibinfo{person}{Nicolai Marquardt}, {and} \bibinfo{person}{Nadia
  Bianchi-Berthouze}.} \bibinfo{year}{2016}\natexlab{}.
\newblock \showarticletitle{RealPen: Providing Realism in Handwriting Tasks on
  Touch Surfaces using Auditory-Tactile Feedback}. In
  \bibinfo{booktitle}{\emph{Proceedings of the 29th Annual Symposium on User
  Interface Software and Technology}}. ACM, \bibinfo{pages}{195--205}.
\newblock


\bibitem[\protect\citeauthoryear{Da~Silva, Abe, and Popović}{Da~Silva
  et~al\mbox{.}}{2008}]%
        {dasilva:2008:mpc}
\bibfield{author}{\bibinfo{person}{M. Da~Silva}, \bibinfo{person}{Y. Abe},
  {and} \bibinfo{person}{J. Popović}.} \bibinfo{year}{2008}\natexlab{}.
\newblock \showarticletitle{Simulation of Human Motion Data using Short-Horizon
  Model-Predictive Control}.
\newblock \bibinfo{journal}{\emph{Computer Graphics Forum}}
  \bibinfo{volume}{27}, \bibinfo{number}{2} (\bibinfo{year}{2008}),
  \bibinfo{pages}{371--380}.
\newblock
\urldef\tempurl%
\url{https://doi.org/10.1111/j.1467-8659.2008.01134.x}
\showDOI{\tempurl}
\showeprint{https://onlinelibrary.wiley.com/doi/pdf/10.1111/j.1467-8659.2008.01134.x}


\bibitem[\protect\citeauthoryear{Domahidi and Jerez}{Domahidi and
  Jerez}{2014}]%
        {domahidi2014forces}
\bibfield{author}{\bibinfo{person}{Alexander Domahidi} {and}
  \bibinfo{person}{Juan Jerez}.} \bibinfo{year}{2014}\natexlab{}.
\newblock \bibinfo{title}{FORCES Professional. embotech GmbH (http://embotech.
  com/FORCES-Pro)}.
\newblock
\newblock


\bibitem[\protect\citeauthoryear{Faulwasser, Kern, and Findeisen}{Faulwasser
  et~al\mbox{.}}{2009}]%
        {Faulwasser:2009}
\bibfield{author}{\bibinfo{person}{T. Faulwasser}, \bibinfo{person}{B. Kern},
  {and} \bibinfo{person}{R. Findeisen}.} \bibinfo{year}{2009}\natexlab{}.
\newblock \showarticletitle{Model predictive path-following for constrained
  nonlinear systems}. In \bibinfo{booktitle}{\emph{Proceedings of the 48h IEEE
  Conference on Decision and Control (CDC)}}. \bibinfo{pages}{8642--8647}.
\newblock
\showISSN{0191-2216}
\urldef\tempurl%
\url{https://doi.org/10.1109/CDC.2009.5399744}
\showDOI{\tempurl}


\bibitem[\protect\citeauthoryear{Favreau, Lafarge, and Bousseau}{Favreau
  et~al\mbox{.}}{2016}]%
        {favreau2016fidelity}
\bibfield{author}{\bibinfo{person}{Jean-Dominique Favreau},
  \bibinfo{person}{Florent Lafarge}, {and} \bibinfo{person}{Adrien Bousseau}.}
  \bibinfo{year}{2016}\natexlab{}.
\newblock \showarticletitle{Fidelity vs. simplicity: a global approach to line
  drawing vectorization}.
\newblock \bibinfo{journal}{\emph{ACM Transactions on Graphics (TOG)}}
  \bibinfo{volume}{35}, \bibinfo{number}{4} (\bibinfo{year}{2016}),
  \bibinfo{pages}{120}.
\newblock


\bibitem[\protect\citeauthoryear{Gebhardt, Stevsic, and Hilliges}{Gebhardt
  et~al\mbox{.}}{2018}]%
        {Gebhardt:2018}
\bibfield{author}{\bibinfo{person}{Christoph Gebhardt}, \bibinfo{person}{Stefan
  Stevsic}, {and} \bibinfo{person}{Otmar Hilliges}.}
  \bibinfo{year}{2018}\natexlab{}.
\newblock \showarticletitle{{Optimizing for Aesthetically Pleasing Quadrotor
  Camera Motion}}.
\newblock \bibinfo{journal}{\emph{ACM Transactions on Graphics (Proceedings of
  ACM SIGGRAPH)}} \bibinfo{volume}{37}, \bibinfo{number}{4}, Article
  \bibinfo{articleno}{90} (\bibinfo{year}{2018}), \bibinfo{numpages}{11}~pages.
\newblock


\bibitem[\protect\citeauthoryear{Gibbs}{Gibbs}{2011}]%
        {gibbs2011advanced}
\bibfield{author}{\bibinfo{person}{Bruce~P Gibbs}.}
  \bibinfo{year}{2011}\natexlab{}.
\newblock \bibinfo{booktitle}{\emph{Advanced Kalman filtering, least-squares
  and modeling: a practical handbook}}.
\newblock \bibinfo{publisher}{John Wiley \& Sons}.
\newblock


\bibitem[\protect\citeauthoryear{Hilaire and Tombre}{Hilaire and
  Tombre}{2006}]%
        {hilaire2006robust}
\bibfield{author}{\bibinfo{person}{Xavier Hilaire} {and} \bibinfo{person}{Karl
  Tombre}.} \bibinfo{year}{2006}\natexlab{}.
\newblock \showarticletitle{Robust and accurate vectorization of line
  drawings}.
\newblock \bibinfo{journal}{\emph{IEEE Transactions on Pattern Analysis and
  Machine Intelligence}} \bibinfo{volume}{28}, \bibinfo{number}{6}
  (\bibinfo{year}{2006}), \bibinfo{pages}{890--904}.
\newblock


\bibitem[\protect\citeauthoryear{Kim, Kim, Lee, Pak, Sohn, Lee, and Lee}{Kim
  et~al\mbox{.}}{2008}]%
        {kim2008digital}
\bibfield{author}{\bibinfo{person}{Hyunjung Kim}, \bibinfo{person}{Seoktae
  Kim}, \bibinfo{person}{Boram Lee}, \bibinfo{person}{Jinhee Pak},
  \bibinfo{person}{Minjung Sohn}, \bibinfo{person}{Geehyuk Lee}, {and}
  \bibinfo{person}{Woohun Lee}.} \bibinfo{year}{2008}\natexlab{}.
\newblock \showarticletitle{Digital rubbing: playful and intuitive interaction
  technique for transferring a graphic image onto paper with pen-based
  computing}. In \bibinfo{booktitle}{\emph{CHI'08 Extended Abstracts on Human
  Factors in Computing Systems}}. ACM, \bibinfo{pages}{2337--2342}.
\newblock


\bibitem[\protect\citeauthoryear{Kolmogorov}{Kolmogorov}{1933}]%
        {kolmogorov1933sulla}
\bibfield{author}{\bibinfo{person}{Andrey Kolmogorov}.}
  \bibinfo{year}{1933}\natexlab{}.
\newblock \showarticletitle{Sulla determinazione empirica di una lgge di
  distribuzione}.
\newblock \bibinfo{journal}{\emph{Inst. Ital. Attuari, Giorn.}}
  \bibinfo{volume}{4} (\bibinfo{year}{1933}), \bibinfo{pages}{83--91}.
\newblock


\bibitem[\protect\citeauthoryear{Kyung, Lee, and Park}{Kyung
  et~al\mbox{.}}{2008}]%
        {kyung2008haptic}
\bibfield{author}{\bibinfo{person}{Ki-Uk Kyung}, \bibinfo{person}{Jun-Young
  Lee}, {and} \bibinfo{person}{Junseok Park}.} \bibinfo{year}{2008}\natexlab{}.
\newblock \showarticletitle{Haptic stylus and empirical studies on braille,
  button, and texture display}.
\newblock \bibinfo{journal}{\emph{BioMed Research International}}
  \bibinfo{volume}{2008} (\bibinfo{year}{2008}).
\newblock


\bibitem[\protect\citeauthoryear{Kyung, Lee, and Srinivasan}{Kyung
  et~al\mbox{.}}{2009}]%
        {kyung2009precise}
\bibfield{author}{\bibinfo{person}{Ki-Uk Kyung}, \bibinfo{person}{Jun-Young
  Lee}, {and} \bibinfo{person}{Mandayam~A Srinivasan}.}
  \bibinfo{year}{2009}\natexlab{}.
\newblock \showarticletitle{Precise manipulation of GUI on a touch screen with
  haptic cues}. In \bibinfo{booktitle}{\emph{EuroHaptics conference, 2009 and
  Symposium on Haptic Interfaces for Virtual Environment and Teleoperator
  Systems. World Haptics 2009. Third Joint}}. IEEE, \bibinfo{pages}{202--207}.
\newblock


\bibitem[\protect\citeauthoryear{Lam, Manzie, and Good}{Lam
  et~al\mbox{.}}{2010}]%
        {lam2010model}
\bibfield{author}{\bibinfo{person}{Denise Lam}, \bibinfo{person}{Chris Manzie},
  {and} \bibinfo{person}{Malcolm Good}.} \bibinfo{year}{2010}\natexlab{}.
\newblock \showarticletitle{Model predictive contouring control}. In
  \bibinfo{booktitle}{\emph{Decision and Control (CDC), 2010 49th IEEE
  Conference on}}. IEEE, \bibinfo{pages}{6137--6142}.
\newblock


\bibitem[\protect\citeauthoryear{Lam, Manzie, and Good}{Lam
  et~al\mbox{.}}{2013}]%
        {lam2013model}
\bibfield{author}{\bibinfo{person}{Denise Lam}, \bibinfo{person}{Chris Manzie},
  {and} \bibinfo{person}{Malcolm~C Good}.} \bibinfo{year}{2013}\natexlab{}.
\newblock \showarticletitle{Model predictive contouring control for biaxial
  systems}.
\newblock \bibinfo{journal}{\emph{IEEE Transactions on Control Systems
  Technology}} \bibinfo{volume}{21}, \bibinfo{number}{2}
  (\bibinfo{year}{2013}), \bibinfo{pages}{552--559}.
\newblock


\bibitem[\protect\citeauthoryear{Lee, Dietz, Leigh, Yerazunis, and Hudson}{Lee
  et~al\mbox{.}}{2004}]%
        {lee2004haptic}
\bibfield{author}{\bibinfo{person}{Johnny~C Lee}, \bibinfo{person}{Paul~H
  Dietz}, \bibinfo{person}{Darren Leigh}, \bibinfo{person}{William~S
  Yerazunis}, {and} \bibinfo{person}{Scott~E Hudson}.}
  \bibinfo{year}{2004}\natexlab{}.
\newblock \showarticletitle{Haptic pen: a tactile feedback stylus for touch
  screens}. In \bibinfo{booktitle}{\emph{Proceedings of the 17th annual ACM
  symposium on User interface software and technology}}. ACM,
  \bibinfo{pages}{291--294}.
\newblock


\bibitem[\protect\citeauthoryear{Lee, Zitnick, and Cohen}{Lee
  et~al\mbox{.}}{2011}]%
        {lee2011shadowdraw}
\bibfield{author}{\bibinfo{person}{Yong~Jae Lee}, \bibinfo{person}{C~Lawrence
  Zitnick}, {and} \bibinfo{person}{Michael~F Cohen}.}
  \bibinfo{year}{2011}\natexlab{}.
\newblock \showarticletitle{Shadowdraw: real-time user guidance for freehand
  drawing}. In \bibinfo{booktitle}{\emph{ACM Transactions on Graphics (TOG)}},
  Vol.~\bibinfo{volume}{30}. ACM, \bibinfo{pages}{27}.
\newblock


\bibitem[\protect\citeauthoryear{Limpaecher, Feltman, Treuille, and
  Cohen}{Limpaecher et~al\mbox{.}}{2013}]%
        {limpaecher2013real}
\bibfield{author}{\bibinfo{person}{Alex Limpaecher}, \bibinfo{person}{Nicolas
  Feltman}, \bibinfo{person}{Adrien Treuille}, {and} \bibinfo{person}{Michael
  Cohen}.} \bibinfo{year}{2013}\natexlab{}.
\newblock \showarticletitle{Real-time drawing assistance through
  crowdsourcing}.
\newblock \bibinfo{journal}{\emph{ACM Transactions on Graphics (TOG)}}
  \bibinfo{volume}{32}, \bibinfo{number}{4} (\bibinfo{year}{2013}),
  \bibinfo{pages}{54}.
\newblock


\bibitem[\protect\citeauthoryear{Liniger, Domahidi, and Morari}{Liniger
  et~al\mbox{.}}{2014}]%
        {Liniger2014}
\bibfield{author}{\bibinfo{person}{Alexander Liniger},
  \bibinfo{person}{Alexander Domahidi}, {and} \bibinfo{person}{Manfred
  Morari}.} \bibinfo{year}{2014}\natexlab{}.
\newblock \showarticletitle{{Optimization-based autonomous racing of 1:43 scale
  RC cars}}.
\newblock \bibinfo{journal}{\emph{Optimal Control Applications and Methods}}
  (\bibinfo{year}{2014}).
\newblock
\showISSN{01432087}
\urldef\tempurl%
\url{https://doi.org/10.1002/oca.2123}
\showDOI{\tempurl}


\bibitem[\protect\citeauthoryear{Liu, Rosales, and Sheffer}{Liu
  et~al\mbox{.}}{2018}]%
        {liu2018strokeaggregator}
\bibfield{author}{\bibinfo{person}{Chenxi Liu}, \bibinfo{person}{Enrique
  Rosales}, {and} \bibinfo{person}{Alla Sheffer}.}
  \bibinfo{year}{2018}\natexlab{}.
\newblock \showarticletitle{StrokeAggregator: consolidating raw sketches into
  artist-intended curve drawings}.
\newblock \bibinfo{journal}{\emph{ACM Transactions on Graphics (TOG)}}
  \bibinfo{volume}{37}, \bibinfo{number}{4} (\bibinfo{year}{2018}),
  \bibinfo{pages}{97}.
\newblock


\bibitem[\protect\citeauthoryear{Mueller and D'Andrea}{Mueller and
  D'Andrea}{2013}]%
        {Mueller2013}
\bibfield{author}{\bibinfo{person}{M.W. Mueller} {and} \bibinfo{person}{R.
  D'Andrea}.} \bibinfo{year}{2013}\natexlab{}.
\newblock \showarticletitle{{A model predictive controller for quadrocopter
  state interception}}. In \bibinfo{booktitle}{\emph{Proceedings of the
  European Control Conference (ECC), 2013}}. \bibinfo{pages}{1383--1389}.
\newblock
\showISBNx{9783952417348}
\urldef\tempurl%
\url{http://ieeexplore.ieee.org/xpls/abs\_all.jsp?arnumber=6669415}
\showURL{%
\tempurl}


\bibitem[\protect\citeauthoryear{Mullins, Mawson, and Nahavandi}{Mullins
  et~al\mbox{.}}{2005}]%
        {mullins2005haptic}
\bibfield{author}{\bibinfo{person}{James Mullins}, \bibinfo{person}{Christopher
  Mawson}, {and} \bibinfo{person}{Saeid Nahavandi}.}
  \bibinfo{year}{2005}\natexlab{}.
\newblock \showarticletitle{Haptic handwriting aid for training and
  rehabilitation}. In \bibinfo{booktitle}{\emph{Systems, Man and Cybernetics,
  2005 IEEE International Conference on}}, Vol.~\bibinfo{volume}{3}. IEEE,
  \bibinfo{pages}{2690--2694}.
\newblock


\bibitem[\protect\citeauthoryear{N{\"a}geli, Meier, Domahidi, Alonso-Mora, and
  Hilliges}{N{\"a}geli et~al\mbox{.}}{2017}]%
        {Naegeli:2017:MultiDroneCine}
\bibfield{author}{\bibinfo{person}{Tobias N{\"a}geli}, \bibinfo{person}{Lukas
  Meier}, \bibinfo{person}{Alexander Domahidi}, \bibinfo{person}{Javier
  Alonso-Mora}, {and} \bibinfo{person}{Otmar Hilliges}.}
  \bibinfo{year}{2017}\natexlab{}.
\newblock \showarticletitle{Real-time Planning for Automated Multi-View Drone
  Cinematography}.
\newblock \bibinfo{journal}{\emph{ACM Transactions on Graphics (Proceedings of
  ACM SIGGRAPH)}}.
\newblock


\bibitem[\protect\citeauthoryear{Pangaro, Maynes-Aminzade, and Ishii}{Pangaro
  et~al\mbox{.}}{2002}]%
        {pangaro2002actuated}
\bibfield{author}{\bibinfo{person}{Gian Pangaro}, \bibinfo{person}{Dan
  Maynes-Aminzade}, {and} \bibinfo{person}{Hiroshi Ishii}.}
  \bibinfo{year}{2002}\natexlab{}.
\newblock \showarticletitle{The actuated workbench: computer-controlled
  actuation in tabletop tangible interfaces}. In
  \bibinfo{booktitle}{\emph{Proceedings of the 15th annual ACM symposium on
  User interface software and technology}}. ACM, \bibinfo{pages}{181--190}.
\newblock


\bibitem[\protect\citeauthoryear{Peng, Zoran, and Guimbreti{\`e}re}{Peng
  et~al\mbox{.}}{2015}]%
        {peng2015d}
\bibfield{author}{\bibinfo{person}{Huaishu Peng}, \bibinfo{person}{Amit Zoran},
  {and} \bibinfo{person}{Fran{\c{c}}ois~V Guimbreti{\`e}re}.}
  \bibinfo{year}{2015}\natexlab{}.
\newblock \showarticletitle{D-Coil: A Hands-on Approach to Digital 3D Models
  Design}. In \bibinfo{booktitle}{\emph{Proceedings of the 33rd Annual ACM
  Conference on Human Factors in Computing Systems}}. ACM,
  \bibinfo{pages}{1807--1815}.
\newblock


\bibitem[\protect\citeauthoryear{Portillo, Avizzano, Raspolli, and
  Bergamasco}{Portillo et~al\mbox{.}}{2005}]%
        {portillo2005haptic}
\bibfield{author}{\bibinfo{person}{O Portillo}, \bibinfo{person}{Carlo~Alberto
  Avizzano}, \bibinfo{person}{Mirko Raspolli}, {and} \bibinfo{person}{Massimo
  Bergamasco}.} \bibinfo{year}{2005}\natexlab{}.
\newblock \showarticletitle{Haptic desktop for assisted handwriting and
  drawing}. In \bibinfo{booktitle}{\emph{ROMAN 2005. IEEE International
  Workshop on Robot and Human Interactive Communication, 2005.}} IEEE,
  \bibinfo{pages}{512--517}.
\newblock


\bibitem[\protect\citeauthoryear{Poupyrev, Okabe, and Maruyama}{Poupyrev
  et~al\mbox{.}}{2004}]%
        {poupyrev2004haptic}
\bibfield{author}{\bibinfo{person}{Ivan Poupyrev}, \bibinfo{person}{Makoto
  Okabe}, {and} \bibinfo{person}{Shigeaki Maruyama}.}
  \bibinfo{year}{2004}\natexlab{}.
\newblock \showarticletitle{Haptic feedback for pen computing: directions and
  strategies}. In \bibinfo{booktitle}{\emph{CHI'04 extended abstracts on Human
  factors in computing systems}}. ACM, \bibinfo{pages}{1309--1312}.
\newblock


\bibitem[\protect\citeauthoryear{Rockafellar and Wets}{Rockafellar and
  Wets}{2009}]%
        {rockafellar2009variational}
\bibfield{author}{\bibinfo{person}{R~Tyrrell Rockafellar} {and}
  \bibinfo{person}{Roger J-B Wets}.} \bibinfo{year}{2009}\natexlab{}.
\newblock \bibinfo{booktitle}{\emph{Variational analysis}}.
  Vol.~\bibinfo{volume}{317}.
\newblock \bibinfo{publisher}{Springer Science \& Business Media}.
\newblock


\bibitem[\protect\citeauthoryear{Simo-Serra, Iizuka, and Ishikawa}{Simo-Serra
  et~al\mbox{.}}{2018a}]%
        {simo2018mastering}
\bibfield{author}{\bibinfo{person}{Edgar Simo-Serra}, \bibinfo{person}{Satoshi
  Iizuka}, {and} \bibinfo{person}{Hiroshi Ishikawa}.}
  \bibinfo{year}{2018}\natexlab{a}.
\newblock \showarticletitle{Mastering sketching: adversarial augmentation for
  structured prediction}.
\newblock \bibinfo{journal}{\emph{ACM Transactions on Graphics (TOG)}}
  \bibinfo{volume}{37}, \bibinfo{number}{1} (\bibinfo{year}{2018}),
  \bibinfo{pages}{11}.
\newblock


\bibitem[\protect\citeauthoryear{Simo-Serra, Iizuka, and Ishikawa}{Simo-Serra
  et~al\mbox{.}}{2018b}]%
        {simo2018real}
\bibfield{author}{\bibinfo{person}{Edgar Simo-Serra}, \bibinfo{person}{Satoshi
  Iizuka}, {and} \bibinfo{person}{Hiroshi Ishikawa}.}
  \bibinfo{year}{2018}\natexlab{b}.
\newblock \showarticletitle{Real-time data-driven interactive rough sketch
  inking}.
\newblock \bibinfo{journal}{\emph{ACM Transactions on Graphics (TOG)}}
  \bibinfo{volume}{37}, \bibinfo{number}{4} (\bibinfo{year}{2018}),
  \bibinfo{pages}{98}.
\newblock


\bibitem[\protect\citeauthoryear{Simo-Serra, Iizuka, Sasaki, and
  Ishikawa}{Simo-Serra et~al\mbox{.}}{2016}]%
        {simo2016learning}
\bibfield{author}{\bibinfo{person}{Edgar Simo-Serra}, \bibinfo{person}{Satoshi
  Iizuka}, \bibinfo{person}{Kazuma Sasaki}, {and} \bibinfo{person}{Hiroshi
  Ishikawa}.} \bibinfo{year}{2016}\natexlab{}.
\newblock \showarticletitle{Learning to simplify: fully convolutional networks
  for rough sketch cleanup}.
\newblock \bibinfo{journal}{\emph{ACM Transactions on Graphics (TOG)}}
  \bibinfo{volume}{35}, \bibinfo{number}{4} (\bibinfo{year}{2016}),
  \bibinfo{pages}{121}.
\newblock


\bibitem[\protect\citeauthoryear{Su, Li, Wang, and Fu}{Su
  et~al\mbox{.}}{2014}]%
        {su2014ez}
\bibfield{author}{\bibinfo{person}{Qingkun Su}, \bibinfo{person}{Wing Ho~Andy
  Li}, \bibinfo{person}{Jue Wang}, {and} \bibinfo{person}{Hongbo Fu}.}
  \bibinfo{year}{2014}\natexlab{}.
\newblock \showarticletitle{EZ-sketching: three-level optimization for
  error-tolerant image tracing}.
\newblock \bibinfo{journal}{\emph{ACM Transactions on Graphics}}
  \bibinfo{volume}{33}, \bibinfo{number}{4} (\bibinfo{year}{2014}).
\newblock


\bibitem[\protect\citeauthoryear{Weiss, Wacharamanotham, Voelker, and
  Borchers}{Weiss et~al\mbox{.}}{2011}]%
        {weiss2011fingerflux}
\bibfield{author}{\bibinfo{person}{Malte Weiss}, \bibinfo{person}{Chat
  Wacharamanotham}, \bibinfo{person}{Simon Voelker}, {and} \bibinfo{person}{Jan
  Borchers}.} \bibinfo{year}{2011}\natexlab{}.
\newblock \showarticletitle{FingerFlux: near-surface haptic feedback on
  tabletops}. In \bibinfo{booktitle}{\emph{Proceedings of the 24th annual ACM
  symposium on User interface software and technology}}. ACM,
  \bibinfo{pages}{615--620}.
\newblock


\bibitem[\protect\citeauthoryear{Withana, Kondo, Makino, Kakehi, Sugimoto, and
  Inami}{Withana et~al\mbox{.}}{2010}]%
        {withana2010impact}
\bibfield{author}{\bibinfo{person}{Anusha Withana}, \bibinfo{person}{Makoto
  Kondo}, \bibinfo{person}{Yasutoshi Makino}, \bibinfo{person}{Gota Kakehi},
  \bibinfo{person}{Maki Sugimoto}, {and} \bibinfo{person}{Masahiko Inami}.}
  \bibinfo{year}{2010}\natexlab{}.
\newblock \showarticletitle{ImpAct: Immersive haptic stylus to enable direct
  touch and manipulation for surface computing}.
\newblock \bibinfo{journal}{\emph{Computers in Entertainment (CIE)}}
  \bibinfo{volume}{8}, \bibinfo{number}{2} (\bibinfo{year}{2010}),
  \bibinfo{pages}{9}.
\newblock


\bibitem[\protect\citeauthoryear{Xing, Wei, Shiratori, and Yatani}{Xing
  et~al\mbox{.}}{2015}]%
        {xing2015autocomplete}
\bibfield{author}{\bibinfo{person}{Jun Xing}, \bibinfo{person}{Li-Yi Wei},
  \bibinfo{person}{Takaaki Shiratori}, {and} \bibinfo{person}{Koji Yatani}.}
  \bibinfo{year}{2015}\natexlab{}.
\newblock \showarticletitle{Autocomplete hand-drawn animations}.
\newblock \bibinfo{journal}{\emph{ACM Transactions on Graphics (TOG)}}
  \bibinfo{volume}{34}, \bibinfo{number}{6} (\bibinfo{year}{2015}),
  \bibinfo{pages}{169}.
\newblock


\bibitem[\protect\citeauthoryear{Yamaoka and Kakehi}{Yamaoka and
  Kakehi}{2013}]%
        {yamaoka2013depend}
\bibfield{author}{\bibinfo{person}{Junichi Yamaoka} {and}
  \bibinfo{person}{Yasuaki Kakehi}.} \bibinfo{year}{2013}\natexlab{}.
\newblock \showarticletitle{dePENd: augmented handwriting system using
  ferromagnetism of a ballpoint pen}. In \bibinfo{booktitle}{\emph{Proceedings
  of the 26th annual ACM symposium on User interface software and technology}}.
  ACM, \bibinfo{pages}{203--210}.
\newblock


\bibitem[\protect\citeauthoryear{Yang, Bischof, and Boulanger}{Yang
  et~al\mbox{.}}{2008}]%
        {yang2008validating}
\bibfield{author}{\bibinfo{person}{Xing-Dong Yang}, \bibinfo{person}{Walter~F
  Bischof}, {and} \bibinfo{person}{Pierre Boulanger}.}
  \bibinfo{year}{2008}\natexlab{}.
\newblock \showarticletitle{Validating the performance of haptic motor skill
  training}. In \bibinfo{booktitle}{\emph{2008 Symposium on Haptic Interfaces
  for Virtual Environment and Teleoperator Systems}}. IEEE,
  \bibinfo{pages}{129--135}.
\newblock


\bibitem[\protect\citeauthoryear{Yoshida, Noma, and Hosaka}{Yoshida
  et~al\mbox{.}}{2006}]%
        {yoshida2006proactive}
\bibfield{author}{\bibinfo{person}{Shunsuke Yoshida}, \bibinfo{person}{Haruo
  Noma}, {and} \bibinfo{person}{Kenichi Hosaka}.}
  \bibinfo{year}{2006}\natexlab{}.
\newblock \showarticletitle{Proactive desk II: Development of a new
  multi-object haptic display using a linear induction motor}. In
  \bibinfo{booktitle}{\emph{Virtual Reality Conference, 2006}}. IEEE,
  \bibinfo{pages}{269--272}.
\newblock


\bibitem[\protect\citeauthoryear{Yung, Landecker, and Villani}{Yung
  et~al\mbox{.}}{1998}]%
        {yung1998analytic}
\bibfield{author}{\bibinfo{person}{Kar~W Yung}, \bibinfo{person}{Peter~B
  Landecker}, {and} \bibinfo{person}{Daniel~D Villani}.}
  \bibinfo{year}{1998}\natexlab{}.
\newblock \showarticletitle{An analytic solution for the force between two
  magnetic dipoles}.
\newblock \bibinfo{journal}{\emph{Physical Separation in Science and
  Engineering}} \bibinfo{volume}{9}, \bibinfo{number}{1}
  (\bibinfo{year}{1998}), \bibinfo{pages}{39--52}.
\newblock


\bibitem[\protect\citeauthoryear{Zitnick}{Zitnick}{2013}]%
        {zitnick2013handwriting}
\bibfield{author}{\bibinfo{person}{C~Lawrence Zitnick}.}
  \bibinfo{year}{2013}\natexlab{}.
\newblock \showarticletitle{Handwriting beautification using token means}.
\newblock \bibinfo{journal}{\emph{ACM Transactions on Graphics (TOG)}}
  \bibinfo{volume}{32}, \bibinfo{number}{4} (\bibinfo{year}{2013}),
  \bibinfo{pages}{53}.
\newblock


\bibitem[\protect\citeauthoryear{Zoran and Paradiso}{Zoran and
  Paradiso}{2013}]%
        {zoran2013freed}
\bibfield{author}{\bibinfo{person}{Amit Zoran} {and} \bibinfo{person}{Joseph~A
  Paradiso}.} \bibinfo{year}{2013}\natexlab{}.
\newblock \showarticletitle{FreeD: a freehand digital sculpting tool}. In
  \bibinfo{booktitle}{\emph{Proceedings of the SIGCHI Conference on Human
  Factors in Computing Systems}}. ACM, \bibinfo{pages}{2613--2616}.
\newblock


\bibitem[\protect\citeauthoryear{Zoran, Shilkrot, Goyal, Maes, and
  Paradiso}{Zoran et~al\mbox{.}}{2014}]%
        {zoran2014wise}
\bibfield{author}{\bibinfo{person}{Amit Zoran}, \bibinfo{person}{Roy Shilkrot},
  \bibinfo{person}{Pragun Goyal}, \bibinfo{person}{Pattie Maes}, {and}
  \bibinfo{person}{Joseph~A Paradiso}.} \bibinfo{year}{2014}\natexlab{}.
\newblock \showarticletitle{The wise chisel: The rise of the smart handheld
  tool}.
\newblock \bibinfo{journal}{\emph{IEEE Pervasive Computing}}
  \bibinfo{volume}{13}, \bibinfo{number}{3} (\bibinfo{year}{2014}),
  \bibinfo{pages}{48--57}.
\newblock


\end{thebibliography}
\appendix

\section{Notation and Variables} \label{sc:ap.notation}

We summarize the notations used in this paper in Table \ref{tab:notations}. Please refer to Table \ref{tab:var} for all weights, constants and parameters used in our implementation.

\begin{table}[!htb]
  \caption{List of notations used on this work.}
  \label{tab:notations}
  \begin{tabular}{ll}
    \toprule
    Symbol&Description\\
    \midrule
    $\mathbf{x}_t$ & State of the system at time $t$ \\
    $\mathbf{u}_t$ & Input to the system at time $t$ \\
    $\mathcal{C}_{(\cdot)}$ & Cost term for optimization problem\\
    $w_{(\cdot)}$ & Weight for the optimization cost terms\\
    $J_k$ & Cost function ($=\sum w_i\mathcal{C}_i$) for the $k$-time prediction\\
    $\bm{\chi}$ & Set of state constraints\\
    $\bm{\zeta}$ & Set of input constraints\\
    $dt$ & Sampling time\\
    $\mmBold$ & Magnetic dipole of the electromagnet\\
    $\mpBold$ & Magnetic dipole of the pen magnet \\
    $\mpBoldt$ & Approx. of $\mpBold$, with only $\ez$ component and $\ctheta \simeq 1$\\
    $\RmagtopenBold$ & 3D Vector distance from $\mmBold$ to $\mpBold$\\
    $\RmagtopenBoldt$ & 3D Vector distance from $\mmBold$ to $\mpBoldt$\\
    $\posp$ & Pen tip position \\
    $\posm$ & Center of electromagnet, projected into paper plane\\
    $\mathbf{r_d}$ & in-plane vector $\mathbf{r_d} = \posm - \posp$\\
    $\ed$ & unity vector that goes from $\ed = \mathbf{r_d} / \norm{\mathbf{r_d}}$\\
    $d$ & in-plane separation between $\posp$ and $\posm$, i.e $d=\norm{\mathbf{r_d}}$\\
    $h$ & $= h_m + h_p$, i.e vertical distance between $\mpBold$ and $\mpBoldt$ \\
    $\alpha$ & intensity of the electromagnet (PWM input)\\
    $\mathbf{F_a}$ & Actuation force on the pen, from $\posp$ to $\posm$\\
    $\theta$ & Control parameter to progress along the desired path\\
    $\posst$ & Current desired point in the path\\
    $\mathbf{r_{\theta}}$ & in-plane vector $\mathbf{r_{\theta}} = \posst - \posm$\\
    $\mathbf{F_{\theta}}$ & Desired force on the pen, from $\posp$ to $\posst$\\
    $\mathbf{n}$ & The normalized tangent at $\mathbf{s}(\theta)$ \\
  \bottomrule
\end{tabular}
\end{table}

\begin{table}[!htb]
  \caption{List of variables and optimization weights used in this work.}
  \label{tab:var}
  \begin{tabular}{c|cccccc}
    \toprule
    Variable&$N$&$w_l$&$w_c$&$w_\theta$&$w_{\dot{\theta}}$&$w_f$\\
    \midrule
    Value&10&1.5&1.5&10.&0.1&10.\\
   \bottomrule
   \end{tabular}
  \vspace{10pt}
  
  \emph{cont.}\\
  \begin{tabular}{c|cccccc}
    \toprule
    Variable&$w_d$&$w_\alpha$&$c$&$w_v$&$w_m$&$v_C$\\
    \midrule
    Value&0.05&7.&5.&1.&1.&0.1 \\
   \bottomrule
\end{tabular}
\end{table}

\section{Electromagnet dipole equivalent} \label{sc:ap.dipole.eq}%

Here we describe the experimental validation of the dipole model for our electromagnet, that allows us to compute the force that the electromagnet exerts onto the permanent magnet on the pen. This interaction is minimal at distance $h$, when the electromagnet and pen are located directly above each other (See Figure \ref{fig:em_model}).
The magnetic field generated by a dipole $\mmBold$ (electromagnet) at the position of dipole $\mpBold$ (pen) can be written as,
\begin{equation}
  \mathbf{B_m} (\RmagtopenBold,\mmBold) = \frac{\mu_0}{4\pi} \left( \frac{3\RmagtopenBold \left( \mmBold \cdot \RmagtopenBold \right) }{\Rmagtopen^5} - \frac{\mmBold}{\Rmagtopen^3} \right ) \label{eq:ap.B2}
\end{equation}
\noindent where the vector $\RmagtopenBold$ is the vector that goes from $\mmBold$ to $\mpBold$ (see \figref{fig:dipole_dipole}). 
The magnetic field $\mathbf{B_m}$ is well described in a cylindrical system centered on the dipole and with the z-axis aligned on the direction of $\mmBold$. Taking only the $z$ component on Eq. \ref{eq:ap.B2} and using the definitions of $\RmagtopenBold$ (Eq. \ref{eq:r21b}) and $\mmBold$ (Eq. \ref{eq:m2}) we arrive at,

\begin{eqnarray}
B_{m,z}(d) &=& \frac{\mu_0 \alpha m_m}{4\pi} \left( \frac{2 h^2 - d^2}{\left( d^2 + h^2\right)^{\frac{5}{}2}} \right) \label{eq:apB2z}
\end{eqnarray}

\begin{figure}[tb]
\centering
\includegraphics[width=\columnwidth]{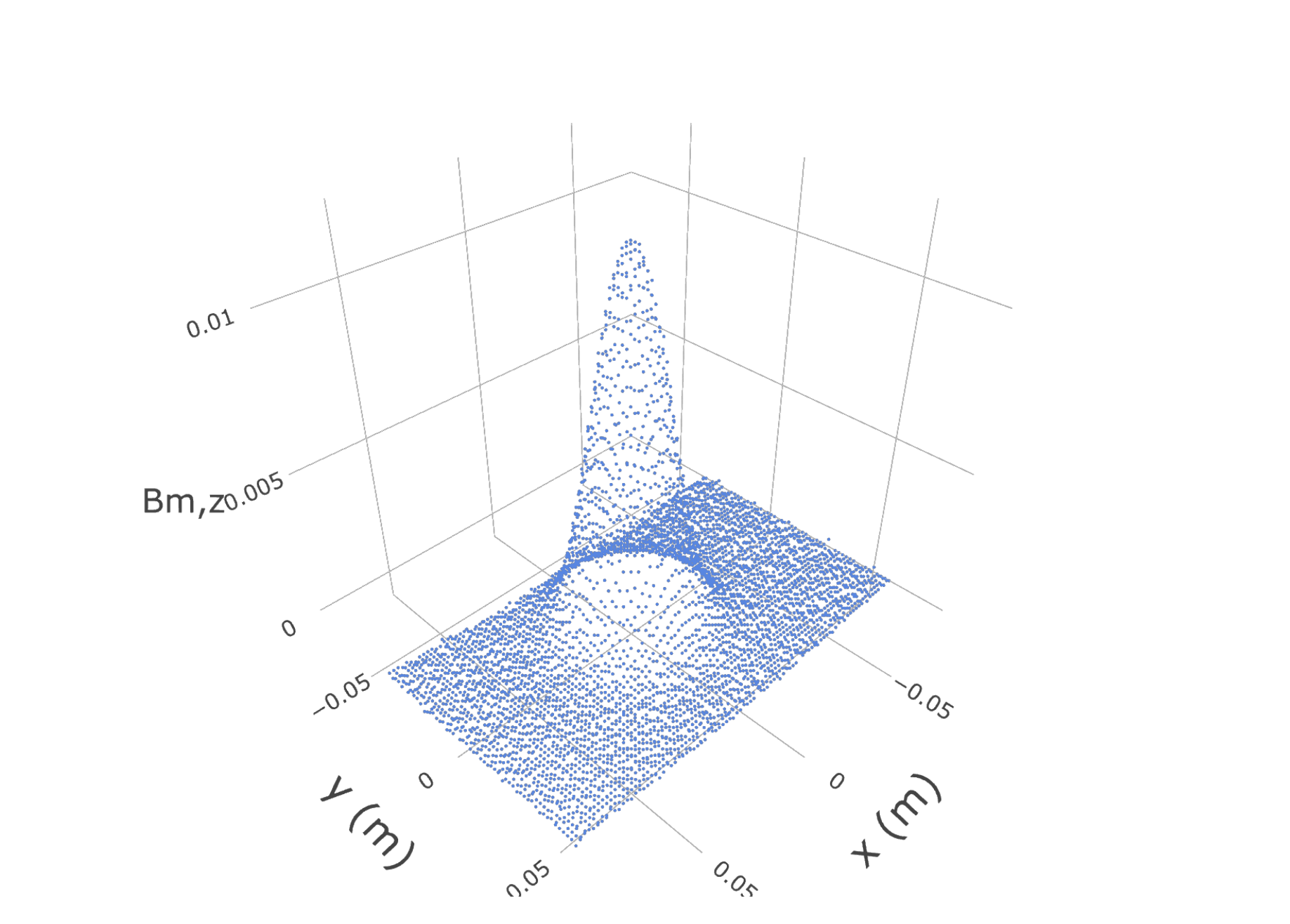} \\
\includegraphics[width=\columnwidth]{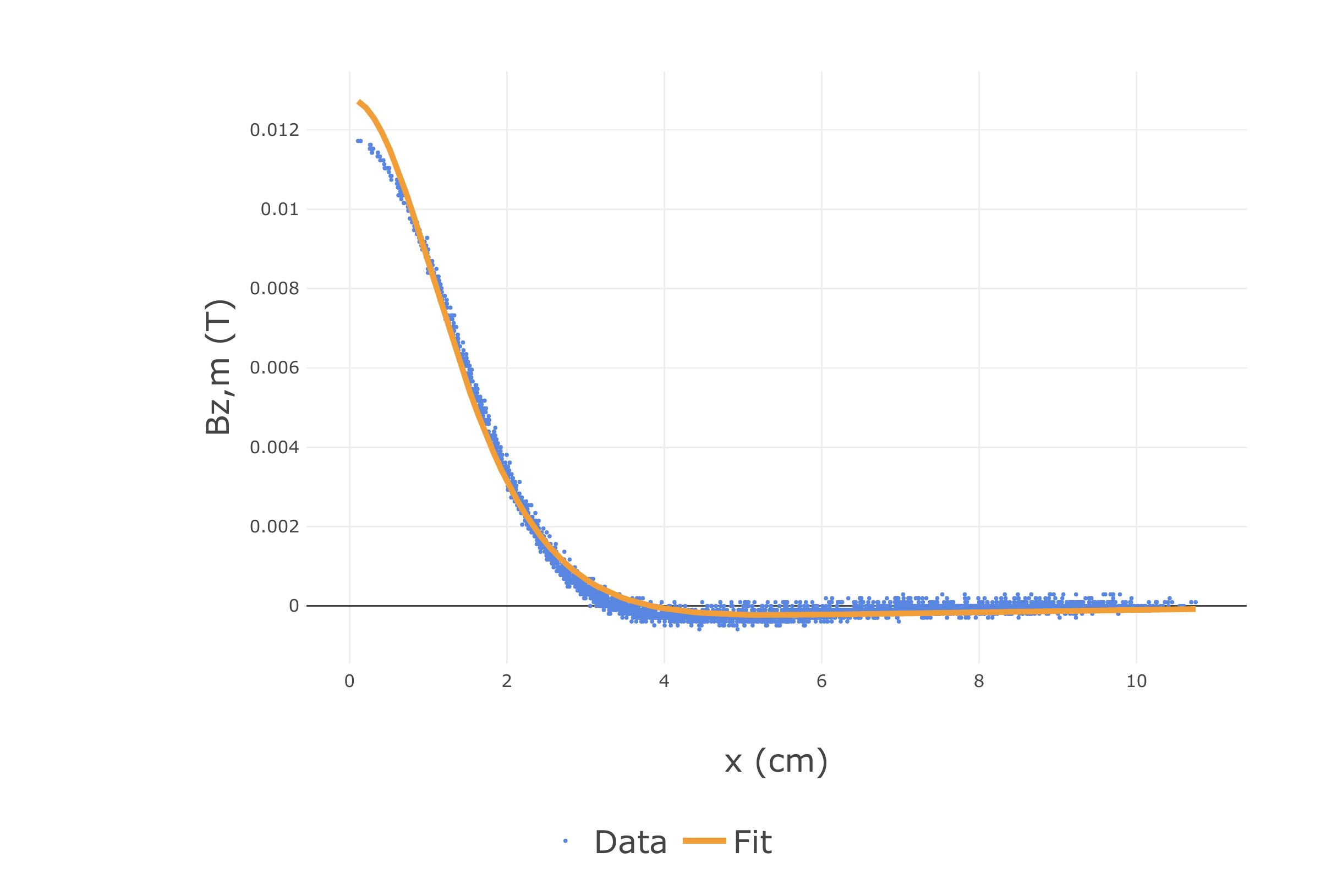}
\caption{3D overview of all data points. The x,y axis are position and the z-axis is $B_2^{(z)}$}
\label{fig:3D-hall-fit}
\end{figure}

We measure the z-component of the magnetic field generated by the electromagnet to compare it with the dipole prediction of Eq. \ref{eq:apB2z}. We use a hall sensor (Allegro A1324, sensitivity is 5 mV/G)\footnote{\url{http://www.farnell.com/datasheets/1538021.pdf}} to measure the z-magnetic flux at a fix height $h_m$, where the magnet of the pen would be. Setting the electromagnet to $\alpha =1$ and moving it in a grid we attain multiple readings of the hall sensor for different electromagnet positions $\posm$. We present the obtained magnetic field plotted in Figure \ref{fig:3D-hall-fit}, top.

Due to symmetry over the z-axis we expect for $B_{m,z}$, we re-plot all points as a function of distance $d_s = \norm{\mathbf{p_s}-\posm}$, with $\mathbf{p_s}=(0,0)$ the in-plane position of the hall sensor. In turn, Eq. \ref{eq:apB2z} can be expressed in the form,
\begin{equation}
   B_{m,z}(d_s) = C_1 \frac{2 C_2^2 - d_s^2}{\left( d_s^2 + C_2^2\right)^{5/2}} \label{eq:apB2_fit}
\end{equation}
\noindent where we have defined two parameters used for the fitting,
\begin{eqnarray}
C_1 &=& \frac{\mu_0 \alpha m_m}{4\pi} \label{eq:C1}\\
C_2 &=& h \label{eq:C2}
\end{eqnarray}

The bottom plot of Figure \ref{fig:3D-hall-fit} shows the measured data for magnetic flux $B_{m,z}(d_s)$ and the fitting to Eq. \ref{eq:apB2_fit}, from which we obtained $C_1 = -1.276 \ 10^{-07}$ and $C_2 = 2.713 \ 10^{-02}$. By replacing these values in equations \ref{eq:C1} and \ref{eq:C2}, we observe that our system can be described by the values $m_m = 1.286$ [A m$^2$] and $h = 2.71$ [cm]. We want to emphasize the excellent agreement in Figure \ref{fig:3D-hall-fit} between the experimental values and the proposed dipole model for the electromagnet. However, we should note that the experimental points show a flattening of $B_{m,z}(x)$ for values of $x < 3$ [mm], that translates into smaller values of forces in that region, as $\mathbf{F_a} \propto \nabla{B_{m,z}}$. This experimental behaviour may explain the position dispersion results we report in Section \ref{sc:results-pos-dispersion}.

\section{Angle Aware Dipole-dipole Model} 
\label{ap:sc.EM.model}

In this section we describe the derivation of the dipole-dipole model for the in-plane actuation force, in the case of considering a pen tilt $\angt$ of the pen. Please refer to the schematic Figure \ref{fig:dipole_dipole} for vector notations we use in this section. The coordinate system is given by,
\begin{eqnarray}
 \ed &=& \frac{\posm - \posp}{||\posm - \posp||} \\
 \ez &=& [0,0,1]^T \\
 \et &=& \ed \times \ez
\end{eqnarray}
\noindent with $\ed$ the in-paper-plane distance from the pen contact point to the electromagnet center projection, $\ez$ the vertical out-of-plane direction and $\et$ the orthogonal vector to the former two.

The dipole-dipole expression for the force acting on $\mpBold$ due to $\mmBold$ and separated by $\RmagtopenBold$ is given by Eq. \ref{eq:F21-dip}), repeated here:
\begin{multline}
   \mathbf{F_p} = {\dfrac  {3\mu _{0}}{4\pi \Rmagtopen^{5}}}
   \left [ \left(\langle\mpBold,\RmagtopenBold\rangle \right) \mmBold + 
   \left(\langle\mmBold,\RmagtopenBold\rangle\right) \mpBold \right . +
   \\
   \left(\langle\mpBold,\mmBold\rangle\right) \RmagtopenBold - 
    \left . {\dfrac{5\left(\langle\mpBold,\RmagtopenBold\rangle\right)
    \left(\langle\mmBold,\RmagtopenBold\rangle\right)}{\Rmagtopen^{2}}} \RmagtopenBold \right ] \ , \label{eq:ap.F21-dip}
\end{multline}

The two dipoles and the vector distance between them can be expressed in the proposed coordinate system as,
\begin{eqnarray}
 \mmBold &=& \alpha \ m_m \ \ez \label{ap.mm}\\
 \mpBold &=& - (m_p \stheta \cphi) \ \ed \nonumber \\
          && + (m_p \stheta \sphi) \ \et \nonumber \\
          && + (m_p \ctheta) \ \ez \label{ap.mp} \\
 \RmagtopenBold &=& - (d+h_p \stheta \cphi) \ \ed \nonumber \\
          && + (h_p \stheta \sphi) \ \et \nonumber \\
          && + (h - (1-\ctheta) h_p) \ \ez \label{ap.rmp}
\end{eqnarray}
\noindent and the three scalar products of equation \ref{eq:ap.F21-dip},
\begin{eqnarray}
 \langle\mmBold,\RmagtopenBold\rangle&=& \alpha \ m_m [h-(1-\ctheta)h_p] \label{eq:ap.mm.r.1}\\
 \langle\mpBold,\RmagtopenBold\rangle&=&mp \ [-\stheta \cphi (d + h_p \stheta \cphi) \nonumber\\
 &&+\stheta^2 \sphi^2 h_p +\ctheta (h - h_p(1-\ctheta))] \ \  \label{eq:ap.mp.r.1} \\
 \langle\mmBold,\mpBold\rangle&=& \alpha m_m m_p \ctheta \label{ap:mm.mp.1}
\end{eqnarray}

We continue the deduction of $\mathbf{F_p}$ by substituting Eq. \ref{ap.mm}---\ref{ap:mm.mp.1} into the main expression Eq. \ref{eq:ap.F21-dip}. However, by following that path we wouldn't necessarily attain information on how strong the actuation force depends on the tilting angles $\angt$ and $\angp$. Here we take a different path. Based on the geometry of our system, we consider the cases where the pen is tilted by only a small angle ($\angt < 30 ^o$). We introduce this small-angle approximation by keeping only the first order terms in $\angt$,
\begin{eqnarray}
 \stheta &\approx& \angt \ \ \ \ \text{(with} \ \angt \ \text{in \ radians)} \label{ap.appsin} \\
 \ctheta &\approx& 1 \label{ap.appcos}
\end{eqnarray}
\noindent As an indication of what this approximation means, for an angle $\angt = 30 ^\circ$, the difference between using $\stheta$ or $\ctheta$ or their approximations forms (Eq. \ref{ap.appsin} and \ref{ap.appcos}) is 5\% and 15 \%, respectively. Under the small-$\angt$ approximation, the dipoles' vectors are,
\begin{eqnarray}
 \mmBold &=& \alpha \ m_m \ \ez \label{ap.mm2}\\
 \mpBold &\simeq& -m_p \angt \cphi \ed + m_p \angt \sphi \et + m_p \ \ez \label{ap.mp2}
\end{eqnarray}
\noindent and the distance between dipoles,
\begin{equation}
\RmagtopenBold \simeq -(d+h_p \angt \cphi) \ \ed + h_p \angt \sphi \ \et + h \ \ez
\end{equation}
\noindent with the length of that distance, at first order on $\angt$,
\begin{equation}
    \Rmagtopen \simeq d^2 + h^2 + 2 d h_p \angt \cphi \label{eq:ap.rapp}
\end{equation}

In turn, the scalar products (Eq. \ref{eq:ap.mm.r.1}---\ref{ap:mm.mp.1}) can be written as,
\begin{eqnarray}
 \langle\mmBold,\RmagtopenBold\rangle &\simeq& \alpha \ m_m h \label{eq:ap.mm.r.2}\\
 \langle\mpBold,\RmagtopenBold\rangle &\simeq& mp \ [-\angt \cphi d + h] \label{eq:ap.mp.r.2} \\
 \langle\mmBold,\mpBold\rangle &\simeq&\alpha \ m_m m_p \label{ap:mm.mp.2}
\end{eqnarray}

We can now substitute these expressions into the main force equation \ref{eq:ap.F21-dip}. As we do in Section \ref{sc:em_model}, we consider only the terms that contribute to the component $\ed$ of the force. Keeping only these terms that contain $\angt$ up to the first order,
 \begin{eqnarray}
 \mathbf{F_{p}^{(d)}} =&& \frac{3\mu _{0} \alpha m_m m_p}{4\pi \Rmagtopen^{5}} \left[-d + \frac{5 d h^2}{\Rmagtopen^2} - h \angt \cphi - h_p \angt \cphi + \right.\nonumber \\
 && \left. + \frac{5 h^2 h_p \angt \cphi}{\Rmagtopen^2} - \frac{5 h d^2 \angt \cphi}{\Rmagtopen^2} \right] \ \ed \\
 =&& \frac{3\mu _{0} \alpha m_m m_p}{4\pi (h^2+d^2)^{5/2}} \left[ \frac{-d (d^2+h^2) +5 d h^2}{(h^2+d^2)} + \right. \nonumber \\
 && \left. \angt \cphi \left( -h -h_p + \frac{5(h^2 h_p - h d^2)}{(h^2+d^2)} - \frac{5 d^2 h^2 h_p}{(h^2+d^2)^2}  \right) \right] \ed \nonumber \\
 && \label{eq:ap.Fp2} \\
  \mathbf{F_{p}^{(d)}} =&& \alpha \ F_0 \ \left[ \ f_0(d) \ + \ \angt \cphi \ f_1(d) \ \right] \ \ed \label{eq:ap.Fd}
 \end{eqnarray}
 \noindent where we define, 
\begin{eqnarray}
 F_0 &=&  \frac{3 \ \mu_0 \ m_p \ m_em}{4 \ \pi \ h^4} \ . \label{eq:ap.F0}\\
 f_0(d) &=& \frac{d \left(4 - \frac{d^2}{h^2}\right)}{h \left(1 + \frac{d^2}{h^2}\right)^\frac{7}{2}} \label{ap.f_0}\\
 f_1(d) &=& \frac{1 + \frac{h_p}{h}}{\left(1 + \frac{d^2}{h^2} \right)^\frac{5}{2}} 
  + \frac{5 \left(\frac{h_p}{h} + \frac{d^2}{h^2}\right)}{\left(1 + \frac{d^2}{h^2} \right)^\frac{7}{2}} 
  - \frac{5 \left(\frac{h_p}{h}\right) \left(\frac{d^2}{h^2}\right)}{\left(1 + \frac{d^2}{h^2} \right)^\frac{9}{2}} 
\end{eqnarray}

Note that by considering the case $\angt = 0$ in Eq. \ref{eq:ap.Fd}, we recover what we obtain in the Sec. \ref{sc:em_model} for $\mathbf{F_a}$. That means that the equation for $\mathbf{F_{p}^{(d)}}$ we obtained in this Appendix subsumes the cases of the pen being tilted by a small angles, and it can be used in future EM actuated systems which may be able to track $\angt$ and $\angp$.

\end{document}